\ifx\texorpdfstring\undefined\newcommand\texorpdfstring[2]{{#1}}\fi
\PassOptionsToPackage{  colorlinks,
  linkcolor = blue!60!black,
  urlcolor = blue!60!black,
  citecolor = blue!60!black,
  hypertexnames=false}{hyperref}
\documentclass[aps,prd,10pt,twocolumn,superscriptaddress,preprintnumbers,floatfix,nofootinbib,notitlepage,showkeys,showpacs]{revtex4-2}
\makeatletter
\newcommand\showtitleinbib{{\escapechar=`\\ \immediate\write\@auxout{%
\csname citation{REVTEX42Control}\endcsname^^J%
\csname citation{apsrev42Control}\endcsname
}}}
\makeatother

\usepackage{orcidlink}
\usepackage{multirow}
\usepackage{siunitx}
\usepackage{bbold}
\usepackage{braket}
\newcommand\me[1]{\left\langle#1\right\rangle}

\renewcommand\>\rangle

\usepackage{microtype}
\usepackage[
  colorlinks,
  linkcolor = blue!60!black,
  urlcolor = blue!60!black,
  citecolor = blue!60!black,
  hypertexnames = false
]{hyperref}

\usepackage{tikz}
\usetikzlibrary{decorations.markings, arrows}

\usepackage{graphicx,times}
\usepackage{latexsym}
\usepackage{mathtools}

\usepackage{amsmath,amssymb,amsbsy,amsfonts}
\usepackage{array}
\usepackage{bm}
\usepackage{graphics}
\usepackage{mathrsfs}
\usepackage{xcolor}
\usepackage{cancel}
\usepackage[normalem]{ulem}

\usepackage[capitalise]{cleveref}

\usepackage{xcolor}
\usepackage{makecell}
\newcommand\TopRule{\Xhline{0.08em}}

\newcommand\MidRule{\Xhline{0.03em}}
\newcommand\BotRule{\Xhline{0.08em}}

\setcellgapes{2pt}

\newcommand{\bfn}{{\mathbf{n}}}

\newcommand{\Id}{{\mathbb{I}}}

\newcommand{\Wc}{{\cal W}}

\newcommand{\ua}{{\mathord\uparrow}}
\newcommand{\da}{{\mathord\downarrow}}

\newcommand\Heff{H_{\text{eff}}}

\newcommand\Order{O}

\usepackage{relsize}

\usepackage[acronyms, nohypertypes={acronym}, nopostdot, style=super, nonumberlist, toc]{glossaries}
\glsenableentrycount
\setacronymstyle{long-sc-short}

\newcommand\ac[1]{\gls{#1}}
\newcommand\acp[1]{\glspl{#1}}

\newacronym{WF}{wf}{Wilson-Fisher}
\newacronym{AF}{af}{asymptotically free}

\newacronym{RG}{rg}{renormalization group}

\newacronym{QEC}{qec}{quantum error correction}

\newacronym[longplural={conformal field theories}]{CFT}{cft}{conformal field theory}
\newacronym[longplural={lattice field theories}]{LFT}{lft}{lattice field theory}
\newacronym[longplural={effective field theories}]{EFT}{eft}{effective field theory}
\newacronym[longplural={quantum field theories}]{QFT}{qft}{quantum field theory}

\newacronym{JLP}{jlp}{Jordan-Lee-Preskill}

\newacronym{BBN}{bbn}{big bang nucleosynthesis}

\newacronym{LEC}{lec}{low-energy constant}

\newacronym{QCD}{qcd}{quantum chromodynamics}
\newacronym{MC}{mc}{Monte Carlo}

\newacronym{IR}{ir}{infrared}
\newacronym{UV}{uv}{ultraviolet}

\newacronym{QED}{qed}{quantum electrodynamics}
\newacronym{SNR}{snr}{signal-to-noise ratio}

\newacronym{NLSM}{nlsm}{nonlinear sigma model}

\newacronym{CL}{cl}{Complex Langevin}

\newacronym{CSA}{csa}{Cartan subalgebra}

\newacronym{SSB}{ssb}{spontaneous symmetry breaking}

\newacronym{AFQMC}{afqmc}{auxiliary field quantum Monte Carlo}
\newacronym{iHMC}{ihmc}{imaginary-mass Hybrid Monte Carlo}

\newacronym{MCMC}{mcmc}{Markov Chain Monte Carlo}
\newacronym{WZW}{wzw}{Wess-Zumino-Witten}

\newacronym{QEA}{qea}{Qubit-Embedding Algebra}
\newacronym{LSM}{lsm}{Lieb-Schultz-Mattis}

\newcommand\ketbraop[3][]{\mathopen{\ket{#2}}#1\mathclose{\bra{#3}}}
\newcommand\braketop[3][]{\mathopen{\bra{#2}}#1\mathclose{\ket{#3}}}
\newcommand\link[1]{\langle#1\rangle}
\newcommand\ketu{\ket{\ua}}

\newcommand\ketd{\ket{\da}}

\newcommand\kete{\ket{0}}

\newcommand\ketbrauu{\ketbraop\ua\ua}
\newcommand\ketbraud{\ketbraop\ua\da}
\newcommand\ketbraue{\ketbraop\ua0}
\newcommand\ketbradu{\ketbraop\da\ua}
\newcommand\ketbradd{\ketbraop\da\da}

\newcommand\ketbraed{\ketbraop0\da}
\newcommand\ketbraee{\ketbraop00}

\newcommand\trr[1]{{\color{orange}#1}}
\newcommand\sch[1]{{\color{red}#1}}
\newcommand\tbh[1]{{\color{green!70!black}#1}}

\newcommand\footnotepunct[2]{{\edef\tmp{\spacefactor=\the\spacefactor\relax#2}%
    \toks0={#1}%
    \hbox to 0pt{\tmp
      \edef\tmp{\noexpand\footnote{\the\toks0}%
        \spacefactor=\the\spacefactor\relax}%
      \hss\expandafter}%
    \tmp}}

\allowdisplaybreaks

\begin{document}

\title{Topological terms with qubit regularization and relativistic quantum circuits}
\preprint{LA-UR-22-27102}
\preprint{MITP-23-013}
\author{Tanmoy Bhattacharya\,\orcidlink{0000-0002-1060-652X}}
\email{tanmoy@lanl.gov}
\affiliation{Los Alamos National Laboratory, Los Alamos, New Mexico 87545, USA}
\author{Shailesh Chandrasekharan\,\orcidlink{0000-0002-3711-4998}}
\email{sch27@duke.edu}
\affiliation{Department of Physics, Box 90305, Duke University, Durham, North Carolina 27708, USA}
\author{Rajan Gupta\,\orcidlink{0000-0003-1784-3058}}
\email{rg@lanl.gov}
\affiliation{Los Alamos National Laboratory, Los Alamos, New Mexico 87545, USA}
\author{Thomas R.~Richardson\,\orcidlink{0000-0001-6314-7518}}
\email{richardt@uni-mainz.de}
\affiliation{Department of Physics, Box 90305, Duke University, Durham, North Carolina 27708, USA}
\affiliation{Institut f\"ur Kernphysik and PRISMA$^+$ Cluster of Excellence, Johannes Gutenberg-Universit\"at, 55128 Mainz, Germany}
\author{Hersh Singh\,\orcidlink{0000-0002-2002-6959}}
\email{hershs@fnal.gov}
\affiliation{Fermi National Accelerator Laboratory, Batavia, Illinois, 60510, USA}

\begin{abstract}
Qubit regularization provides a rich framework to explore quantum field theories. The freedom to choose how the important symmetries of the theory are embedded in the qubit regularization scheme allows us to construct new lattice models with rich phase diagrams. Some of the phases can contain topological terms which lead to critical phases. In this work we introduce and study the SU(3)-F qubit regularization scheme to embed the SO(3) spin-symmetry. We argue that qubit models in this regularization scheme contain several phases including a critical phase which describes the $k=1$ \ac{WZW} \ac{CFT} at long distances, and two massive phases one of which is trvially gapped and the other which breaks the lattice translation symmetry. We construct a simple space-time Euclidean lattice model with a single coupling $U$ and study it using the Monte Carlo method. We show the model has a critical phase at small $U$ and a trivially massive phase at large $U$ with a first order transition separating the two. Another feature of our model is that it is symmetric under space-time rotations, which means the temporal and spatial lattice spacing are connected to each other. The unitary time evolution operator obtained by a Wick rotation of the transfer matrix of our model can help us compute the physics of the $k=1$ \ac{WZW} \ac{CFT} in real time without the need for tuning the temporal lattice spacing to zero. We use this idea to introduce the concept of a relativistic quantum circuit on a discrete space-time lattice. 
\end{abstract}

\maketitle

\section{Introduction}
\label{sec1}

Quantum computers have the potential to revolutionize our ability to study quantum systems~\cite{Arute:2019zxq}. Due to this promise, formulating quantum field theories of interest in high energy and nuclear physics so that they can be studied using a quantum computer has recently emerged as a new area of research in both particle and nuclear physics~\cite{Bauer:2022hpo}. The literature on how we can use quantum computers to study a variety of quantum phenomena including applications to quantum field theories has exploded in the past few years~\cite{Jordan:2011ci,Casanova:2011wh,Casanova:2012zz,PhysRevX.6.031007,Macridin:2018oli,PhysRevLett.123.090501,Raychowdhury:2018osk,Roggero:2018hrn,Alexandru:2019nsa,Davoudi:2019bhy,Davoudi:2020yln,Ciavarella:2021nmj,Hall:2021rbv,Meurice:2021pvj}. A common theme that guides many studies is to be able to solve strongly interacting quantum field theories like quantum chromodynamics and compute experimentally useful quantities using a quantum computer. The only non-perturbative approach that is currently available to do this is to use the lattice regularization. Monte Carlo algorithms allow us in principle to study static properties of the theory using the Euclidean path integral, if it is free of sign problems~\cite{Troyer:2004ge}. In such studies one usually begins with a lattice Lagrangian on a space-time lattice. However, to study the theory in real time one must instead start with a lattice Hamiltonian. The Hamiltonian approach is also very natural for quantum computers since they can evolve quantum states through a unitary time evolution~\cite{Jordan:2011ne,Wiese:2013uua,Banuls:2019bmf,Bender:2020jgr}. Unfortunately, traditional lattice Hamiltonians of bosonic systems are constructed with lattice operators that act on an infinite dimensional local Hilbert space. To construct a theory suitable for a quantum computer one must reformulate the theory so that the local Hilbert space is finite dimensional. We can then obtain the original field theory as the long-distance physics of the lattice theory at an appropriate quantum critical point; this is called qubit regularization~\cite{Singh:2019uwd}.

Qubit regularization of a quantum field theory can be accomplished in several ways~\cite{Hackett:2018cel,Alexandru:2019nsa,Carena:2022kpg,Ji:2022qvr}. However, from a renormalization perspective it would be useful to preserve as many important symmetries of the quantum field theory as possible. One simple approach is to begin with the traditional lattice regularized quantum field theory and truncate the local Hilbert space to a
subspace forming a low-dimensional, often reducible, representation of the relevant symmetry group.  The commutation relations between the `field' operators on this space are constrained by the symmetry, but are not uniquely specified by them---to close the algebra, one needs to specify additional relations.  Along with these extra relations needed to specify the theory, the operators form a representation of a bigger algebra, called the \ac{QEA}, that contains the original symmetry algebra as a subalgebra.
Thus, the choice of the \ac{QEA} and the corresponding representation we choose to realize it,
together define the qubit-regularization scheme completely. It was recently shown how one can systematically develop such truncation schemes that preserve the symmetry algebra exactly, and this process reveals various possible \acp{QEA} along with their representations~\cite{Liu:2021tef}. However, one can also construct non-traditional \acp{QEA}  by simply embedding the symmetry algebra in other groups by hand, without motivating it as a truncation from some traditional lattice Hamiltonian. This freedom of qubit regularization was exploited in the D-theory approach long ago~\cite{Chandrasekharan:1996ih,PhysRevD.60.094502,Beard:2004jr,Brower:2003vy,Wiese:2006kp}.

The freedom to choose the \ac{QEA} and its representation while defining the qubit regularization scheme suggests that we may be able to access several quantum critical points using qubit regularized Hamiltonians than with traditional lattice Hamiltonians. Each quantum critical point may describe a different continuum quantum field theory in its vicinity. Some of these may contain topological terms while others may have emergent symmetries. It is possible that some of them may not even be relativistic. It is also possible that qubit models constructed using one \ac{QEA} may contain a quantum critical point that is not present in models constructed with another \ac{QEA}. Likewise, models based the same \ac{QEA} but realized through different representations may contain different quantum critical points. Finally, we cannot rule out the possibility that some qubit regularization schemes simply do not have any quantum critical points. In short the possibilities of what we will find is quite rich and a fertile ground for research.

In this paper we explain the above richness within the context of qubit regularizing the SO(3) spin symmetry.
After discussing some well known examples, we explore a new qubit regularization scheme which promises to have a rich phase diagram containing several known phases within a single model, thus providing a framework to study possible new quantum critical points. In particular we show the presence of a critical phase where the long distance physics is the $k=1$ \ac{WZW} conformal field theory. As a part of our study we also address another important challenge for quantum computation of relativistic quantum field theories. Since quantum computation forces us to work in the lattice Hamiltonian framework which breaks the symmetry between space and time, we need to search for quantum critical points which are relativistic in order to recover this symmetry. However, since quantum circuits are usually implemented by coupling local qubits using the Trotter approximation, there will be errors from temporal lattice spacing. Thus, studies of relativistic quantum field theories will be faced with the challenge of tuning to the critical point while at the same time making the Trotter errors small. While these two limits are independent in the Hamiltonian approach, we
develop a space-time lattice formulations that connect them. For example, in the traditional Euclidean Lagrangian lattice formulations, the Lorentz symmetry in the continuum is tied to Euclidean lattice models that are invariant under discrete space-time rotations. If this can also be implemented in a qubit regularized theory, the usual Wick rotation into real time should in principle be sufficient to obtain the correct real time correlation functions near quantum critical points. In this work we explicitly construct a Euclidean space-time lattice field theory that is invariant under discrete space-time rotations. The time evolution operator in real time is then obtained by a simple Wick rotation. Thus we obtain a quantum circuit on a space-time lattice which we define as a {\em relativistic quantum circuit}. We argue that our circuit should help us study the $k=1$ \ac{WZW} conformal field theory in real time without the need to take the Trotter errors to zero.

Our paper is organized as follows. In \cref{sec2} we explain the richness of the qubit regularization schemes within the context of the $O(3)$ \ac{NLSM}. We explain what is known and introduce a new regularization scheme in this paper which we argue has several phases. In \cref{sec3} we introduce a simple class of model within the new scheme and show how we can construct a space-time symmetric Euclidean lattice field theory for it, and find that the model is free of sign problem only a limited range of parameters. We present results from a Monte Carlo study of our model in this range of parameters in \cref{sec4}. We show the presence of a quantum phase transition between a critical phase, which we argue to be in the same universality as the $k=1$ \ac{WZW} \ac{CFT}, and the massive phase with no spontaneous symmetry breaking. Unfortunately, our results seem to suggest that the phase transition is first order phase transition. We discuss the implications for this and the possibility of finding the quantum critical point in the extended parameter space. In \cref{sec5} we construct the relativistic quantum circuit of our model and discuss the possibility of studying the real time physics of the the $k=1$ \ac{WZW} \ac{CFT} using it. Finally in \cref{sec6} we present our conclusions.

\section{Examples of SO(3) Qubit Embedding Algebras}
\label{sec2}

To understand the richness that qubit regularization brings to the table, let us consider the example of embedding the SO(3) symmetry of a quantum field theory in various qubit regularization schemes. To be concrete we can assume we are interested in reproducing the physics of the (1+1)-dimensional $O(3)$ \ac{NLSM} described by the Euclidean action
\begin{align}
S[\bfn] = \int\!\! d^2x\ \Big(\frac{1}{2g^2} \partial_\mu\bfn \cdot \partial_\mu \bfn 
+ i\frac{\theta}{4\pi} \bfn \cdot (\partial_\mu \bfn \ \times\ \partial_\nu\bfn) \Big).
\label{eq:cmodel+theta}
\end{align}
where we have included the topological $\theta$ term for completeness. The SO(3) symmetry plays an important role in the physics of the model so we would like to explore various ways of preserving it in the qubit regularization scheme. The traditional lattice Hamiltonian preserves this symmetry by introducing a quantum particle on the surface of a unit sphere on each lattice site. This then leads to the lattice quantum fields $\phi^i$ ($i=1,2,3$) and $L^a$ ($a=1,2,3$) at each lattice site, where $\phi^i$ is the ``position'' of this particle
constrained to lie on the unit sphere,
\begin{align}
    \sum_i \phi^i \phi^i &= 1, \label{eq:phinorm}
\end{align}
$L^a$ is the ``orbital angular momentum''  which generates rotations on the sphere. The fields $\phi^i$ and $L^a$ satisfy the commutation relations
\begin{subequations}
\begin{align}
[L^a,L^b]\ &=\ i\epsilon^{abc} L^c, \\   
[L^a,\phi^i]\ &=\ i\epsilon^{aij} \phi^j.
\end{align}
\label{eq:symalg}%
\end{subequations}
In addition, the traditional model imposes the commutation relation
\begin{align}
    [\phi^i,\phi^j] &= 0. \label{eq:phicommute}
\end{align}
\cref{eq:phinorm,eq:symalg,eq:phicommute} together force the local Hilbert of the traditional model to be infinite dimensional, which can be identified as the direct sum of all integer-spin representations of the SO(3) algebra,
\begin{align}
\mathbb{V}_{\rm trad} \ =\ \bigoplus_{\ell = 0}^\infty \ \mathbb{V}_\ell. 
\end{align}
However, from the perspective of the SO(3) symmetry, only \cref{eq:symalg} is essential, and we refer to it as the symmetry algebra.
A qubit regularization scheme takes advantage of this freedom and allows us to work in finite-dimensional local Hilbert spaces by sacrificing \cref{eq:phinorm,eq:phicommute}.
%
For example, we can work with the simplest qubit-regularization scheme by truncating the local Hilbert space to a single $\ell$ value,
\begin{align}
\mathbb{V}_{\rm Q}  = \mathbb{V}_{\ell}
\label{eq:qr0}
\end{align}
which gives us a quantum spin-$\ell$ Hilbert space on each lattice site and the operator algebra of \(\phi^i\) and \(L^a\) are closed into the \(\ell\) representation of SO(3) by imposing
\begin{align}
\phi^i = 0\label{eq:ellscheme}
\end{align}
and sacrificing \cref{eq:phinorm} \cite{Liu2022}.
We call this the SO(3)-\(\ell\) scheme and say that the SO(3) symmetry is embedded into itself. It is well known that models in this scheme can have several interesting phases and quantum critical points~\cite{PhysRevLett.59.799,Affleck:1987ch,PhysRevB.55.8295,PhysRevB.102.014447}. Another simple example is obtained by by keeping only the $\ell=0$ and $\ell=1$ sectors of the original Hilbert space,
\begin{align}
\mathbb{V}_{\rm Q} = \mathbb{V}_0 \oplus \mathbb{V}_1
\label{eq:qr1}
\end{align}
which would be four dimensional. As was argued in Ref.~\cite{Liu:2021tef}, this truncation naturally lets us define $4\times 4$ matrices of operators $\phi^i_Q$ and $L^a_Q$ that satisfy the symmetry algebra in \cref{eq:symalg}, which can be closed by imposing
\begin{align}
[\phi^i_Q,\phi^j_Q] = \frac{1}{3} i\epsilon^{ija}L^a_Q.
\end{align}
Thus, in this case, the \ac{QEA} is the SO(4) algebra realized through the four-dimensional fundamental representation. We will refer to this as SO(4)-F (`F' for fundamental) qubit regularization scheme for the SO(3) symmetry.
It was shown recently how the physics 
of \cref{eq:cmodel+theta} at $\theta=0$ can emerge as a continuum of this simple four-dimensional qubit regularization scheme~\cite{Bhattacharya:2020gpm}. 

From the perspective of the SO(3) symmetry, one can in principle also begin with other choices of \ac{QEA}, that do not begin with the traditional lattice Hamiltonian, but are motivated purely based on symmetries. For example, we could choose the spinorial representations of SO(3) involving $\ell=1/2,3/2,\dotsc$. In fact one can argue that these representations arise naturally if we assume the fictitious quantum particle that lives on each lattice site is charged and is moving on the surface of the sphere but in the presence of a magnetic monopole sitting at the origin~\cite{Wu:1976ge}. Can a qubit regularized theory constructed using these spinorial representations of the SO(3) symmetry lead to new quantum critical points not easily accessible through those constructed with integer ones? We already know through the work of Haldane, Affleck and others that the long distance physics of antiferromagnetic quantum spin-half Heisenberg ladders with an even number of legs (which, viewed as one-dimensional lattice instead of as a ladder, are naturally described by local Hilbert spaces with integer $\ell$s) are usually massive while the ones with odd number of legs (which naturally describe local Hilbert spaces with half-integer $\ell$s) are critical. This difference is entirely due to topological terms that are induced in the low energy theory. 
In a seminal work, Haldane showed how the Berry phase of spin-$S$ representation on each lattice site can conspire to create the $\theta$-term in the physics of antiferromagnetic Heisenberg chains, which has also been generalized to spin ladders
\cite{Haldane:1982rj, sierra_nonlinear_1996, sierra_application_1997, martin-delgado_phase_1996}. The Berry phases can mimic the affect of the magnetic monopole that we referred to above and allow for half-integer representations of angular momentum. When the number of legs is odd, we obtain the physics of \cref{eq:cmodel+theta} at $\theta=\pi$, which at long distances is described by the $k=1$ \ac{WZW} conformal field theory.
Recently it was shown how both the \ac{UV} and \ac{IR} physics of \cref{eq:cmodel+theta} can be obtained from spin-half ladders using the D-theory idea~\cite{PhysRevLett.129.022003} at arbitrary $\theta\neq 0, \pi$. The idea of inducing topological terms by choosing unconventional \acp{QEA} within the qubit regularization framework has also been extended to (1+1)-dimensional $CP(N-1)$ and Grassmannian models~\cite{Beard:2004jr, nguyen_lattice_2023a}.  

From the perspective of qubit regularization, the simplest spinorial Hilbert space would be through the SU(2)-F qubit regularization scheme defined on the Hilbert space with $\ell=1/2$,
\begin{align}
\mathbb V_Q = \mathbb{V}_{\ell=1/2} \,.
\end{align}
In this scheme, quantum field operators will be spin-half generators $S^a$ which live on each lattice site and replace the $L^a$ of the traditional model and \(\phi^i\equiv0\) as in the SO(3)-\(\ell\) scheme.
Again the symmetry algebra in \cref{eq:symalg} is satisfied and the algebra is closed, so that the \ac{QEA} is the SO(3) algebra itself. The most common qubit models that emerge in this regularization scheme are the quantum spin-half chains and there is extensive literature on them. One of the main results comes from the \ac{LSM} theorem which states that the ground states of these models can only be one of two types: a gapless critical phase and a massive phase with a broken translation symmetry which we will call the dimer phase~\cite{LIEB1961407}. The quantum critical point that separates these phases describes the $k=1$ \ac{WZW} model which, as we mentioned above, is the long distance physics of \cref{eq:cmodel+theta} at $\theta=\pi$. This quantum critical point has been studied in the literature directly in the spin model~\cite{PhysRevB.54.R9612} and via a fermionic realization~\cite{Liu:2020ygc}. Thus, even the simplest spinorial qubit regularization allows us to study a quantum critical point whose vicinity describes a quantum field theory with topological terms with the SU(2) symmetry of the spin chain enhanced to
$\left[{\rm SU(2)}\times {\rm SU(2)}\right]/\mathbb{Z}_2$ at the quantum critical point.

In this work we explore a different qubit regularization scheme to embed the SO(3) symmetry. In this scheme we define
\begin{align}
\mathbb V_Q = \mathbb V_0 \oplus \mathbb{V}_{\ell=1/2},
\end{align}
which is a three-dimensional local Hilbert space on each lattice site spanned by the basis states $\kete, \ketu, \ketd$ where $|0\>$ is a singlet under SU(2) transformations (referred to as the Fock vacuum state), while the remaining two transform as a doublet. In this scheme, in addition to the spin operators $S^a$, which are now $3\times 3$ block diagonal matrices in the two angular momentum sectors, it is natural to introduce four additional operators $c^\dagger_\ua = \ketbraop{\ua}{ 0}$ and $c^\dagger_\da = \ketbraop{\da}{ 0}$ and their Hermitian conjugates. These are a doublet of hard-core boson creation and annihilation operators. Thus, $S^a$, $c^\dagger_i$,and $c_i$ are the seven operators which we will consider as the qubit regularized quantum fields of the theory. Notice that the original $\phi_i$ of the traditional field operators have been replaced by $c^\dagger_\alpha, c_\alpha\ (\alpha=\ua,\da)$ in this regularization scheme. Closing the algebra of these seven operators generates one more operator $Y$, which is diagonal and traceless, and the qubit embedding algebra turns out to be SU(3), realized in the fundamental representation.
Hence we will call this as the SU(3)-F qubit regularization scheme.

\begin{figure*}[ht]
\begin{center}
\includegraphics[width=0.4\textwidth]{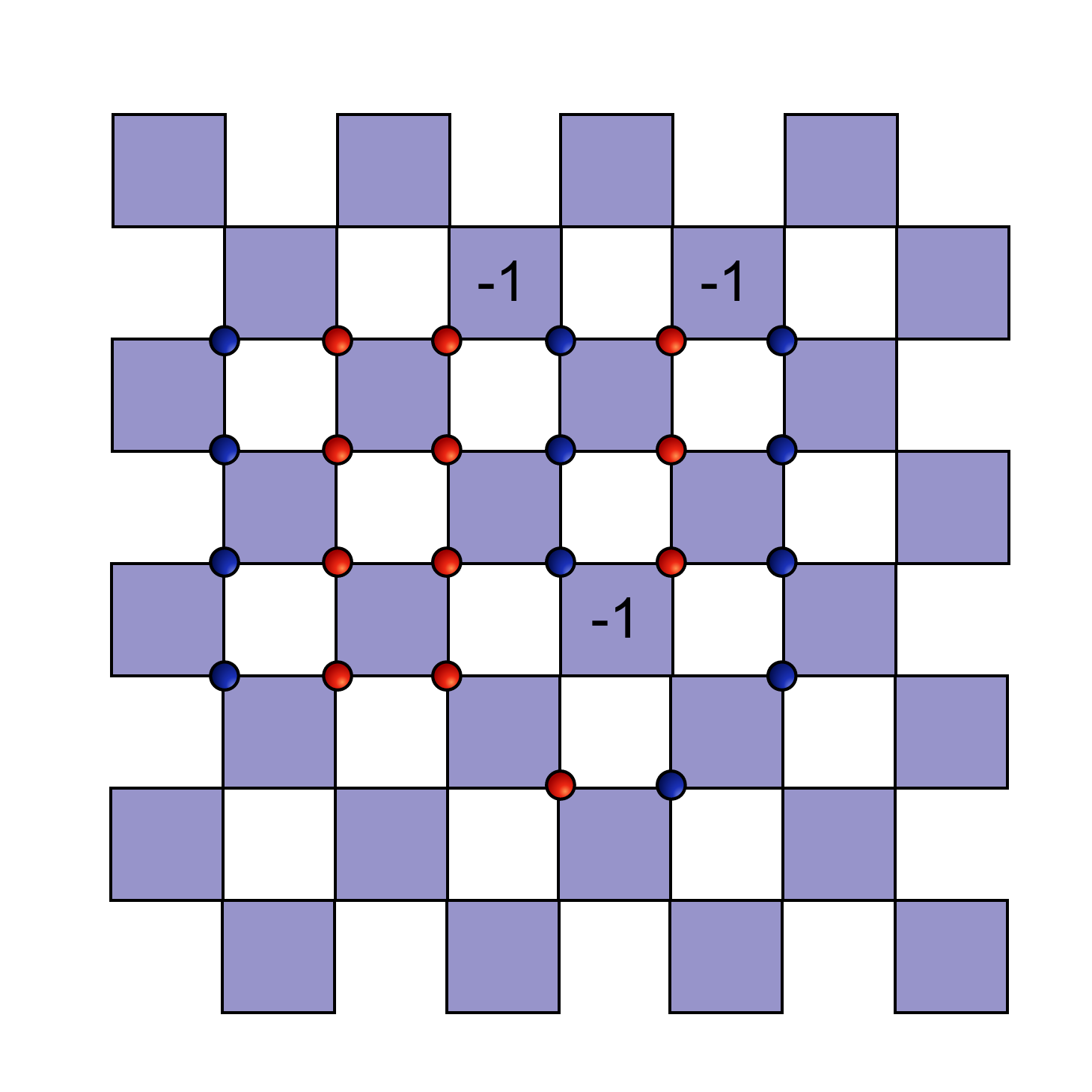}
\includegraphics[width=0.52\textwidth]{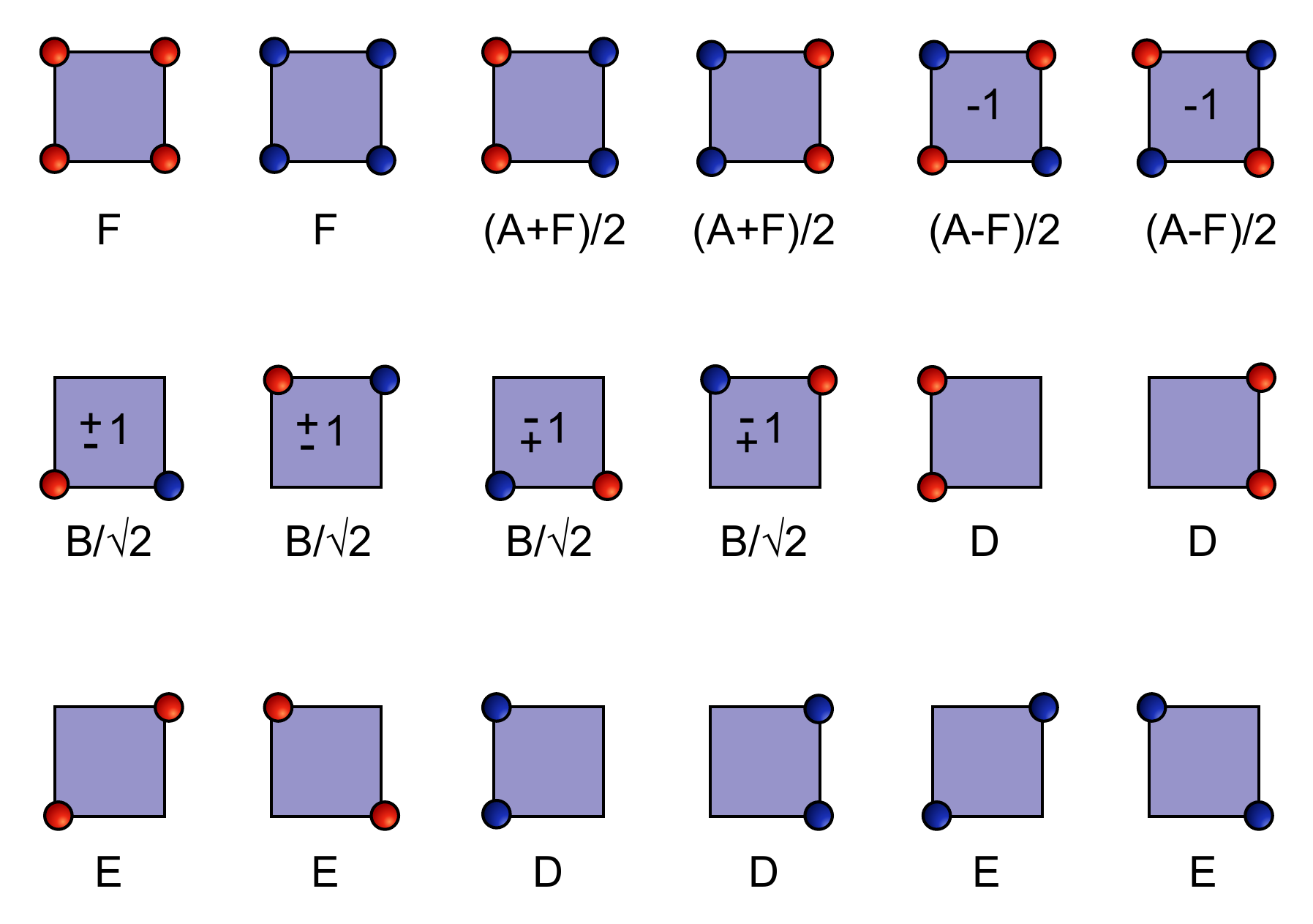}
\end{center}
\caption{\label{fig:sconf} The left figure shows an illustration of an SU(3) configuration $c$ on a Euclidean space-time lattice, that contributes to the partition function of our model. Each site contains one of the three projection operators, $\ketbrauu$ (a red circle) or $\ketbradd$ (blue circle) or  $\ketbraee$ (empty site). The model is defined through transfer matrix elements $e^{-i\varepsilon H_j}$ associated to each shaded plaquette. Along with the projections on the sites we can see that there is a matrix element associated to each plaquette whose weights are denoted by ${\cal W}_p(c)$ and shown in the right figure. Due to the sign associated with the singlet state $\ket{s}$ defined in \cref{eq:singlet}, the weights are negative on some plaquettes. This extra negative sign due to a $\ket{\ua}$ or a $\bra{\ua}$ on an even site is indicated in the figure inside the box; the left hand figure assumes that the leftmost site is an even site, whereas in the right hand figure, for the boxes that have \(\pm1\) or \(\mp1\), the upper and lower signs apply when the left site is an even and odd site, respectively.
}
\end{figure*}

Inclusion of the $\kete$ states in the Hilbert space changes things in an interesting way as compared to the SU(2)-F scheme, because the \ac{LSM}-theorem is no longer applicable.  In particular, using a Hamiltonian that lowers the energy of the Fock vacuum states allows us to construct models with a trivially massive phase where no symmetries are broken spontaneously. On the other hand by suppressing these $\kete$ states, we can explore the two other phases predicted by the \ac{LSM}-theorem\tbh{:} the critical and dimer phases of the spin-half chain. Thus, the SU(3)-F scheme of embedding the SO(3) symmetry promises to contain a rich set of quantum critical points. One of the motivations of our work is to search for a quantum critical point connected to the trivially gapped phase where we can recover asymptotic freedom. We know that this is not guaranteed since we know of a different quantum critical point where the \ac{UV} physics is governed by two decoupled spin-half chains~\cite{Bhattacharya:2020gpm}. In the next section we construct a model where it is easy to argue for the existence of a quantum phase transition between the critical phase of the spin-half chain and the trivially massive phase. We want to explore if this quantum phase transition is second order and if so, which quantum critical point controls the \ac{UV} physics of the massive phase.

\section{A SU(3)-F Qubit Regularized Model}
\label{sec3}

Our model within the SU(3)-F qubit-regularization scheme is constructed on a one dimensional periodic spatial lattice whose sites are labeled with $j=0,1,\dotsc,L_x-1$, where we will assume $L_x$ is even. On each site $j$ the three dimensional Hilbert space is spanned by the SU(3) fundamental basis vectors $\kete$, $\ketu$, and $\ketd$.
In order to construct the lattice Hamiltonian that is invariant under global SO(3) transformations realized in the spinorial representation, it is useful to define the nine-dimensional nearest-neighbor unentangled states $\ket{00}, \ket{\ua 0}, \ket{\da 0}, \ket{\uparrow\uparrow}, \ket{\ua\da}$ and so on.
These are defined independently for each bond connecting neighboring sites $j$ and $j+1$ with the first element of the ket denoting the state on site $j$ and the second the state on site $j+1$, and we suppress the site label \(j\) to simplify notation. In addition to these unentangled states it is useful to also define two entangled states
\begin{subequations}
\begin{align}
\ket{s} &= \pm
\big(\ket{\ua \da} - 
\ket{\da \ua}\big)/
\sqrt{2} 
\label{eq:singlet}
\\
\ket{t} &=  \hphantom{\pm}
\big(\ket{\ua \da} + 
\ket{\da \ua}\big)/
\sqrt{2}
\label{eq:triplet}
\end{align}
\end{subequations}
on each bond where the sign in \(\ket{s}\) is chosen such that \(\ket{\ua}\) gets a negative sign iff it is on an even site. With this notation and convention, consider the nearest neighbor Hamiltonian $H = \sum_j H_j$ where
$H_j \ =\ \ H^{(1)}_j + H^{(2)}_j$ is a sum of two types of nearest neighbor interaction terms that couple the sites $j$ and $j+1$. These two terms are given by
\begin{subequations}
\begin{align}
H^{(1)}_j\ =\ & \ (\alpha+\delta)\ \ketbraop{s}{s} + (\alpha-\delta)\ \ketbraop{00}{00} \nonumber \\
& - \gamma \ (\ketbraop{00}{s} + \ketbraop{s}{00}) \\
H^{(2)}_j \ =\ & \ \eta \ 
(\ketbraop{0\ua}{0\ua} + \ketbraop{\ua 0}{ \ua 0} + \ketbraop{0\da}{0\da} + \ketbraop{\da 0}{ \da 0}) \nonumber \\
& - \kappa \ (\ketbraop{0\ua}{\ua 0} 
+ \ketbraop{\ua 0}{0\ua} + \ketbraop{0\da}{\da 0} 
+ \ketbraop{\da 0}{0\da}).
\label{eq:hamiltonian}
\end{align}
\end{subequations}
Since the nearest neighbor Hilbert space is nine dimensional, the most general two-site Hamiltonian is a $9\times 9$ matrix. If we preserve the SO(3) symmetry the Hamiltonian needs to be a unit matrix in the triplet space. We set the energy of this triplet space to be zero for later convenience. However the SO(3) symmetry allows the $\ket{s}$ and $\ket{00}$ to mix. This means two spin-half particles in the singlet state can annihilate into the Fock vaccuum. Here we choose this mixing Hamiltonian to be given by $H^{(1)}_j$. The overall energy of this combined space is $\alpha$ while the splitting between $\ket{s}$ and $\ket{00}$ is given by $2\delta$ and the mixing energy scale is \(\gamma\). In addition there are four spinorial states involving $\ket{0\ua}$, $\ket{0\da}$, $\ket{\ua 0}$, $\ket{\da 0}$. The Hamiltonian in this subspace is $H^{(2)}_j$.  This space has an overall energy \(\eta\), but spin-half particles can hop to nearest neighbor sites across the spinorial link with strength \(\kappa\). Thus, \(H^{(1)}_j+H^{(2)}_j\) is the most general SO(3) invariant Hamiltonian involving nearest neighbor sites that has translation and parity symmetries.

 When $\alpha=\gamma=\kappa=0$, $\eta \to \infty$, and $\delta \rightarrow -\infty$ we expect our model to mimic the anti-ferromagnetic spin-half chain since singlets will have the least weight. Hence we expect our model to be critical in the vicinity of these parameters and the long distance physics will be described by \cref{eq:cmodel+theta} at $\theta=\pi$, or equivalently the $k=1$ \ac{WZW} \ac{CFT} \cite{Shankar:1989ee}. On the other hand when $\alpha=\gamma=\kappa=\eta=0$ and $\delta \rightarrow \infty$, there will be a phase where our model will be trivially gapped with a unique ground state dominated by $\kete$ states. This massive phase should naturally be described by \cref{eq:cmodel+theta} but at $\theta=0$. However, we should note that this connection is only true within an effective field theory framework since without a quantum critical point the lattice theory can have several higher dimensional irrelevant operators whose physics will be non-universal. Thus, the natural next question is whether there is a second-order quantum critical points connecting the massive phase to another phase. For example, can a second-order phase transition exist directly between the $\theta=0$ and $\theta=\pi$ phases? This was one of the prime motivations to study the SU(3)-F model.

In order to construct a Euclidean path integral method to study a possible phase transition in our model we first split the Hamiltonian into two terms $H_e = \sum_{j\in \text{even}} H_j$ and $H_o = \sum_{j\in \text{odd}} H_j$ and construct the Euclidean time evolution operator for our model in imaginary time on a space-time lattice through the expression
\begin{align}
U_E(\tau) \ =\ \Big(e^{-H_o\varepsilon}e^{-H_e\varepsilon}\Big)^{L_t/2},
\label{eq:Ete}
\end{align}
where $\varepsilon$ is a real parameter that plays the role of the temporal lattice spacing. The Euclidean time $\tau = (L_t/2) \varepsilon$ usually plays the role of the inverse temperature when we study equilibrium thermodynamics. The discretized partition function at a temperature $\tau = 1/T$ is defined through the expression $Z = \mathrm{Tr}(U_E(\tau))$. The Euclidean lattice field theory that emerges is a statistical mechanics of the SU(3) configurations on a checker-board lattice as illustrated on the left side in \cref{fig:sconf}. At each space-time lattice site we imagine inserting the complete basis of states which is a sum of  $\ketbrauu$ (red circle), $\ketbradd$ (blue circle), and 
$\ketbraee$ (empty site). We will denote each such configurations as $c$. 

The partition function of the model can be written as a sum over Boltzmann weights ${\cal W}(c)$, 
\begin{align}
Z \ =\ \sum_{c} \ {\cal W}(c) 
\label{eq:spinpf}
\end{align}
where each weight is obtained as a product of local weights ${\cal W}_p(c)$ which are the matrix elements of $e^{-\varepsilon H_j}$ in the described SU(3) basis. 
\begin{align}
& e^{-\varepsilon H_j} \ =\  A \ketbraop{s}{s} + B (\ketbraop{00}{s} + \ketbraop{s}{00}) + C \ketbraop{00}{00} \nonumber \\
& + D (\ketbraop{0\ua}{0\ua} + \ketbraop{\ua 0}{\ua 0} + \ketbraop{0\da}{0\da} + \ketbraop{\da 0}{\da 0}) \nonumber \\
& + E (\ketbraop{0\ua}{\ua 0} + \ketbraop{\ua 0}{0 \ua} + \ketbraop{0\da}{\da 0} + \ketbraop{\da 0}{0 \da}) \nonumber \\
& + F(\ketbraop{\ua \ua}{\ua \ua} + \ketbraop{\da \da}{\da \da} + \ketbraop{t}{ t}).
\label{eq:tmate}
\end{align}
The non-zero matrix elements are shown diagrammatically on the right side of \cref{fig:sconf}, except for the element 
$\braketop[e^{-\varepsilon H_j}]{00}{00}$ (i.e., the completely empty plaquette), which has the weight $C$. The weights $A$--$F$ are given by
\begin{subequations}
\begin{align}
A &= e^{-\varepsilon\alpha}
\Big(\cosh(\varepsilon\lambda)-(\delta/\lambda)\sinh(\varepsilon\lambda)\Big) \\
B &= e^{-\varepsilon\alpha}
(\gamma/\lambda)\sinh(\varepsilon\lambda) \\
C &= e^{-\varepsilon\alpha}
\Big(\cosh(\varepsilon\lambda)+(\delta/\lambda)\sinh(\varepsilon\lambda)\Big) \\
D &= e^{-\varepsilon\eta}\cosh(\varepsilon\kappa) \\
E &= e^{-\varepsilon\eta}\sinh(\varepsilon\kappa) \\
F &= 1
\end{align}
\end{subequations}
with the definition that $\lambda=\sqrt{\delta^2+\gamma^2}$. When $\gamma=\eta=\kappa = 0$, it is easy to verify that $A=e^{-\varepsilon(\alpha+\delta)}$, $C=e^{-\varepsilon(\alpha-\delta)}$, $B=E=0$, $D=F=1$. When $A$ dominates we obtain the physics of the spin-chain and when $C$ dominates we get the trivial phase. Here we wish to study the transition between these two phases using a Monte Carlo method.

Unfortunately, there are configurations whose Boltzmann weights are negative. One such configuration is shown in \cref{fig:sconf}. As explained in the cation of the figure, the negative signs arise due to the sign associated with the definition of the singlet state $\ket{s}$ in \cref{eq:singlet}. We could choose to remove this sign by a change of 
%
%
the sign of the basis state $\ket{\ua} \rightarrow -\ket{\ua}$ on all even spatial sites. This is essentially a gauge choice in the complete basis states that we choose to expand the partition function. This 
change moves the negative sign to the hopping term of the $\ket\ua$ state. This choice is useful to understand why the sign problem is eliminated when $\kappa = 0$ (i.e., when the single particle hopping term is eliminated),
but it makes the action of the SO(3) symmetry different on the basis chosen at odd and even sites, making it less manifest.  In our choice that makes the symmetry manifest, the absence of sign problem whe \(\kappa=0\) is explained by the observation that negative sign plaquettes in configurations without spinorial hops always come in pairs.

\begin{figure}[ht]
\begin{center}
\includegraphics[width=0.3\textwidth]{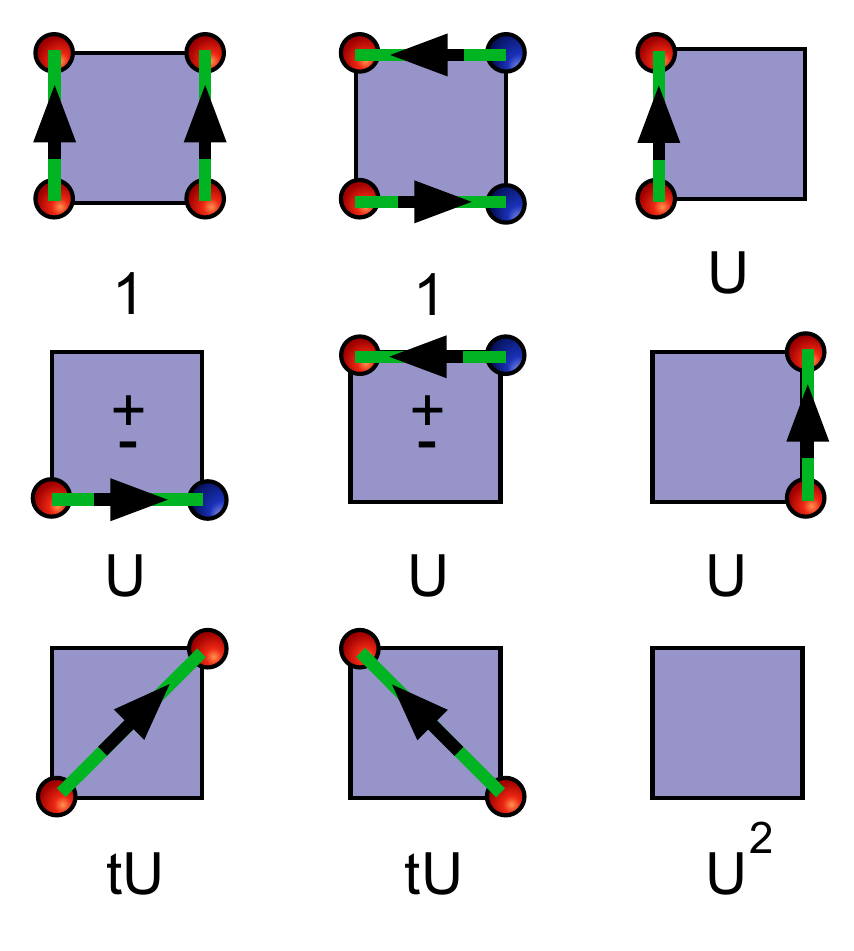}
\end{center}
\caption{\label{fig:bndwt} Weights $\Omega_p(b)$ of plaquette configurations are shown, along with the equivalent `bond' description. In some cases an extra sign is needed: this is indicated inside the box, the upper sign to be used when the red spin is on an even site. In each case, flipping the orientation of any of the bonds on the plaquette simultaneously with the `color' of the spins at its ends gives a plaquette with the same absolute value of the weight, with a sign change for every flip of an horizontal arrow. All other plaquette configurations have zero weight.  Note that to convert a bond description to the corresponding plaquette description, we need to assign blue circles on downward pointing bonds and red circles on upward pointing bonds. 
}
\end{figure}

\begin{figure*}[ht]
\begin{center}
\includegraphics[width=0.46\textwidth]{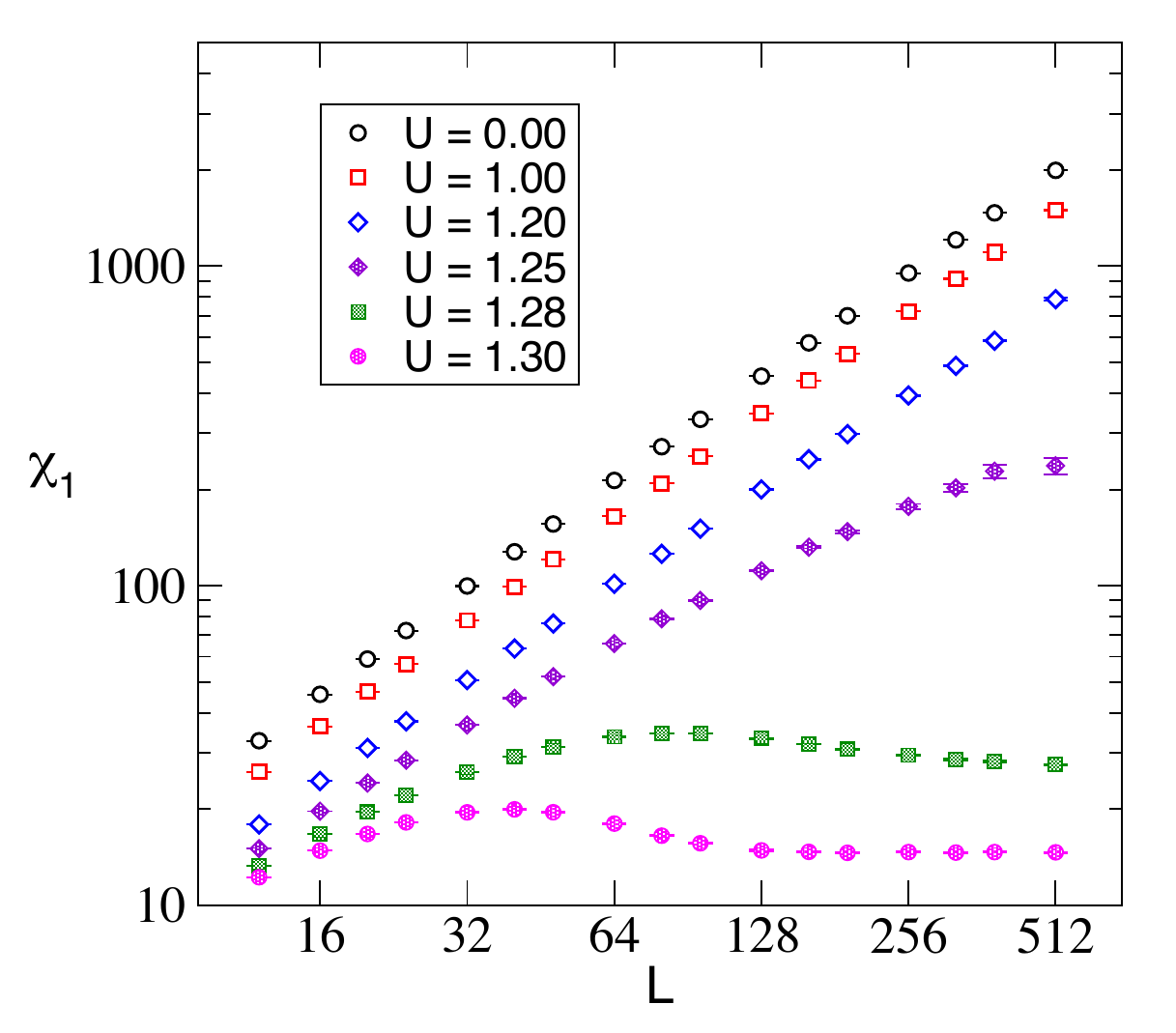}
\includegraphics[width=0.49\textwidth]{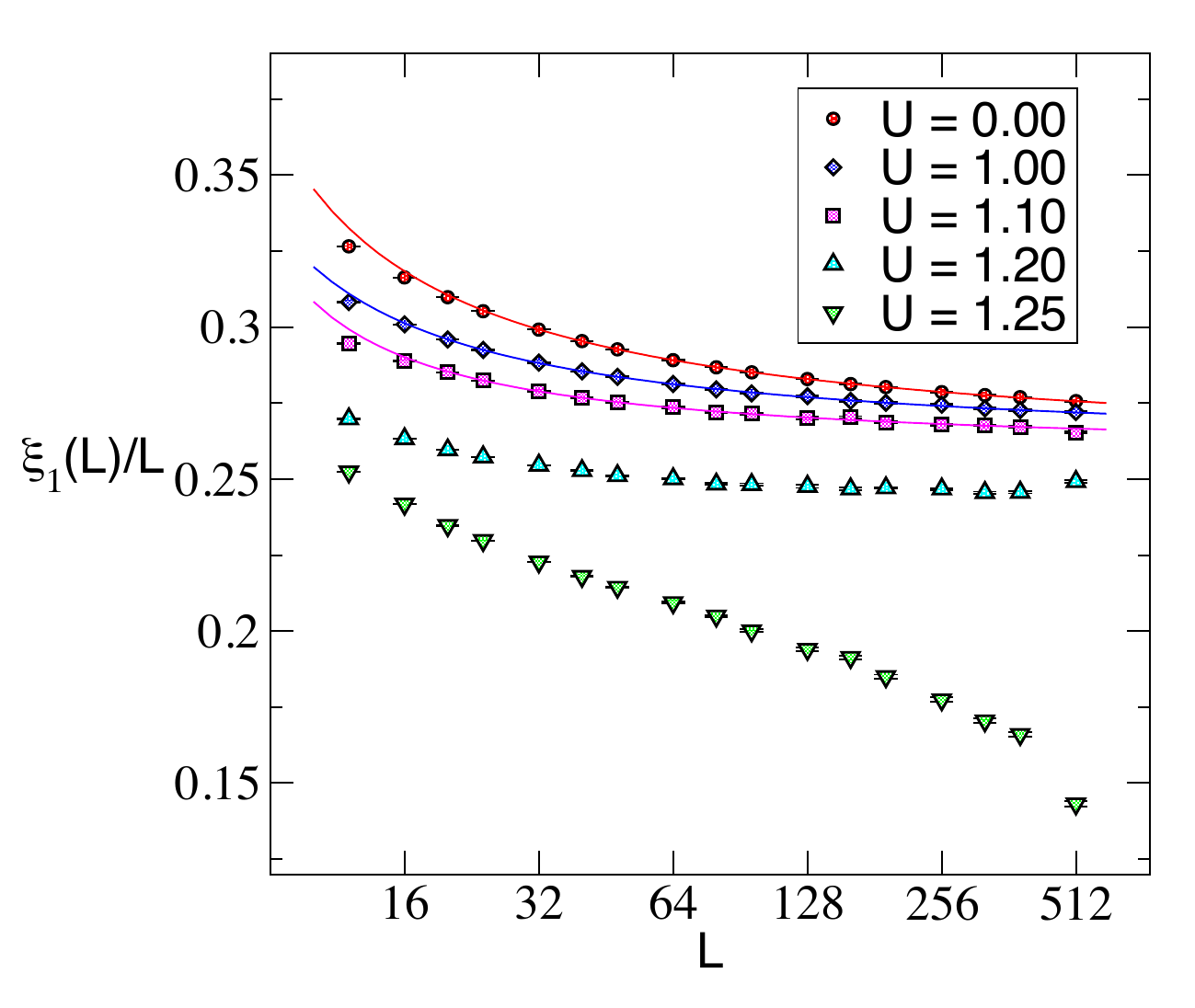}
\end{center}
\caption{\label{fig:susxi1} The left figure shows a plot of $\chi_1$ as a function of $L$. We observe that for $U\leq 1.2$, $\chi_1$ diverges almost linearly, suggesting that our model is critical. This is also confirmed in the right figure, which shows $\xi_1(L)/L$ approaching a constant for large $L$ for these couplings. For $U \geq 1.28$ we are clearly in a massive phase since now $\chi$ begins to saturate for large values of $L$. At $U=1.25$ the saturation in $\chi_1$ is only visible at the largest lattice sizes, but the fact that the correlation length is no longer growing with $L$ is much clearer in the right plot. The solid lines in the right plot are fits to \cref{eq:marginal} which takes into account the presence of a marginally irrelevant operator in the critical phase.}
\end{figure*}

\section{Monte Carlo Results}
\label{sec4}

Before we construct the Monte Carlo method, let us first argue that if we set $A=3$, \(F=1\), $B/\sqrt{2}=D=U$, $E=t U$ and $C=U^2$ the partition function \cref{eq:spinpf} can also be written as a sum over oriented-bond configurations $b$, 
\begin{align}
Z \ =\ \sum_{b} \ \Omega(b)
\label{eq:bondpf}
\end{align}
where the weight $\Omega(b)$ can be written as a product over plaquette weights $\Omega_p(b)$. The bond configurations are constructed by introducing oriented bond variables locally on each plaquette for every fixed spin configuration, such that a sum over $\Omega_p(b)$ of all bond configurations introduced results in the original plaquette weight $W_p(c)$. {Plaquette configurations with both spins and bonds along with their weights ${\Omega}_p(b)$, are shown in \cref{fig:bndwt}.}
Notice that though some spin configurations correspond to multiple bond configurations\footnotepunct{The middle two plaquettes of the top row of \cref{fig:sconf} correspond to two bond descriptions each: either with two horizontal arrows, or with two vertical arrows.}, for every oriented-bond configuration $b$ we can identify a unique spin configuration $c$. 
To identify $c$ from a given $b$ the rule is that if the bonds are oriented upwards in time it is always associated with the $\ket{\ua}$ state and if it is oriented downwards in time it is associated with $\ket{\da}$. If the bond points in the spatial direction, the spin-flips as we hop to the neighboring spatial site.

With the explicit spin labels thus deleted, an interesting feature of the bond configuration appears: for every bond configuration $b$ we can perform a $\pi/2$ rotation to get another bond configuration $b_R$ such that $|\Omega(b_R)| = |\Omega(b)|$. 
%
This, in turn, leads to a space-time rotation symmetry of the model 
since after every $\pi/2$ rotation the bond configuration on each plaquette is rotated and up to a sign gives another bond configuration with the same weight. When we set $\kappa=0$, all configurations have a positive weight\footnote{This happens because, as explained previously, in this case, the negative weight plaquettes come in pairs} and the partition function of the model is symmetric under space-time rotations. In the remaining part of this paper we will focus on this $\kappa=0$ model which depends on a single parameter $U$, which can be associated with the weight of an empty site (or the Fock vacuum state $\ket{0}$) which is clear by noting that each site belongs to two plaquettes and every plaquette gets a weight $\sqrt{U}$ from each empty site (see \cref{fig:bndwt}). For $U=0$, no Fock vacuum states are allowed and our model is a Euclidean space-time lattice model of the spin-half chain, with the additional property that it is also invariant under space-time rotations. Hence, we expect our model to be critical for small $U$ which we will demonstrate below using our Monte Carlo methods. {\em We would like to emphasize here that we have constructed a space-time symmetric qubit regularized lattice field theory to study the $O(3)$ \ac{NLSM} at $\theta=\pi$. The qubit regularization was able to naturally induce the $\theta$-term.}

As $U$ increases to some critical value, we expect to the vacuum states to dominate and the lattice model will not longer be critical. For sufficiently large values of $U$ the theory will be gapped trivially. This suggests for sufficiently large values of $U$ we expect our lattice model describes some lattice regularization of the $O(3)$ \ac{NLSM} at $\theta=0$. The difficult question which we cannot answer easily is whether the phase transition between the two phases will be second order or first order. For this we have to perform Monte Carlo calculations. 

\begin{figure*}[ht]
\begin{center}
\includegraphics[width=0.48\textwidth]{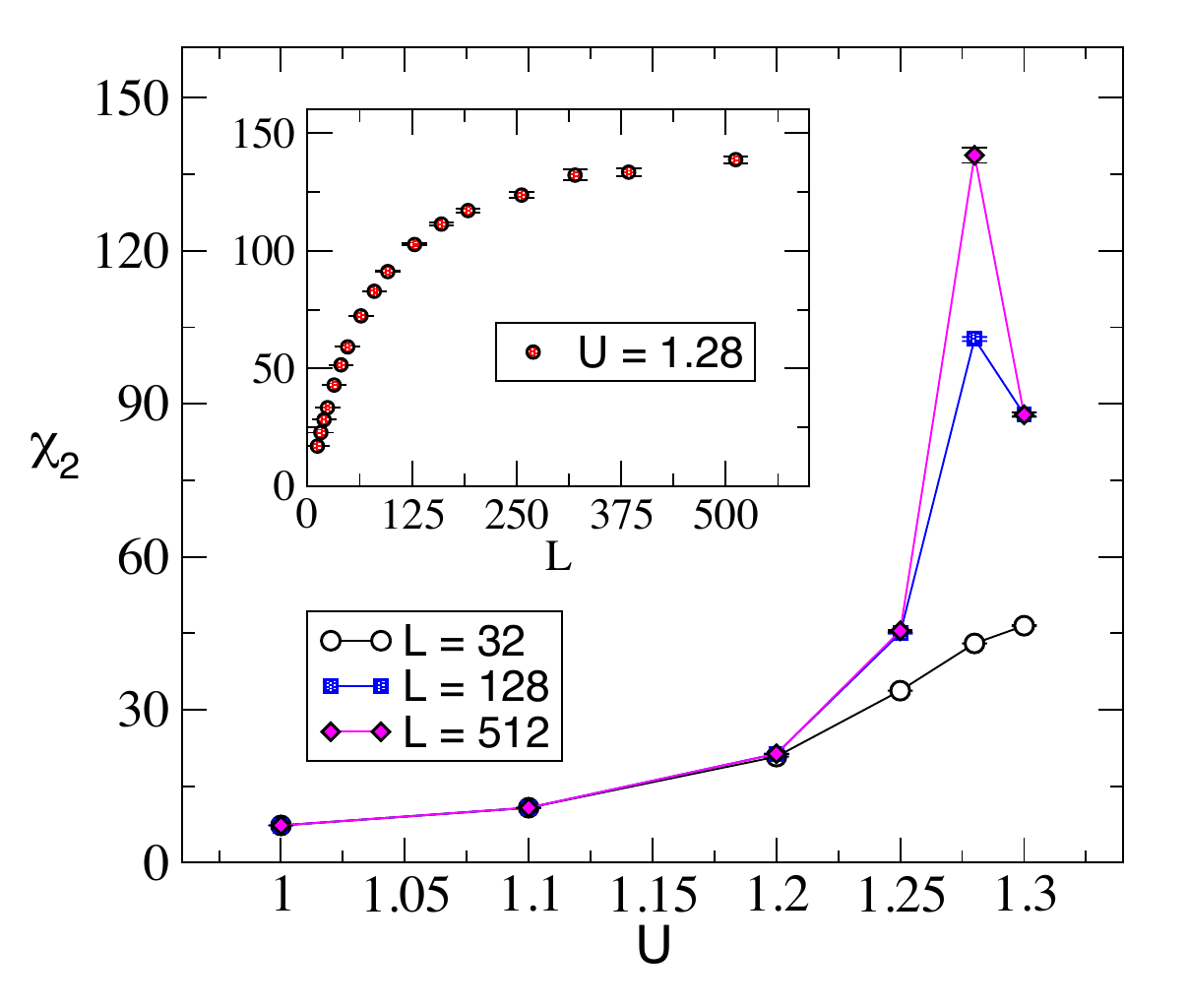}
\includegraphics[width=0.46\textwidth]{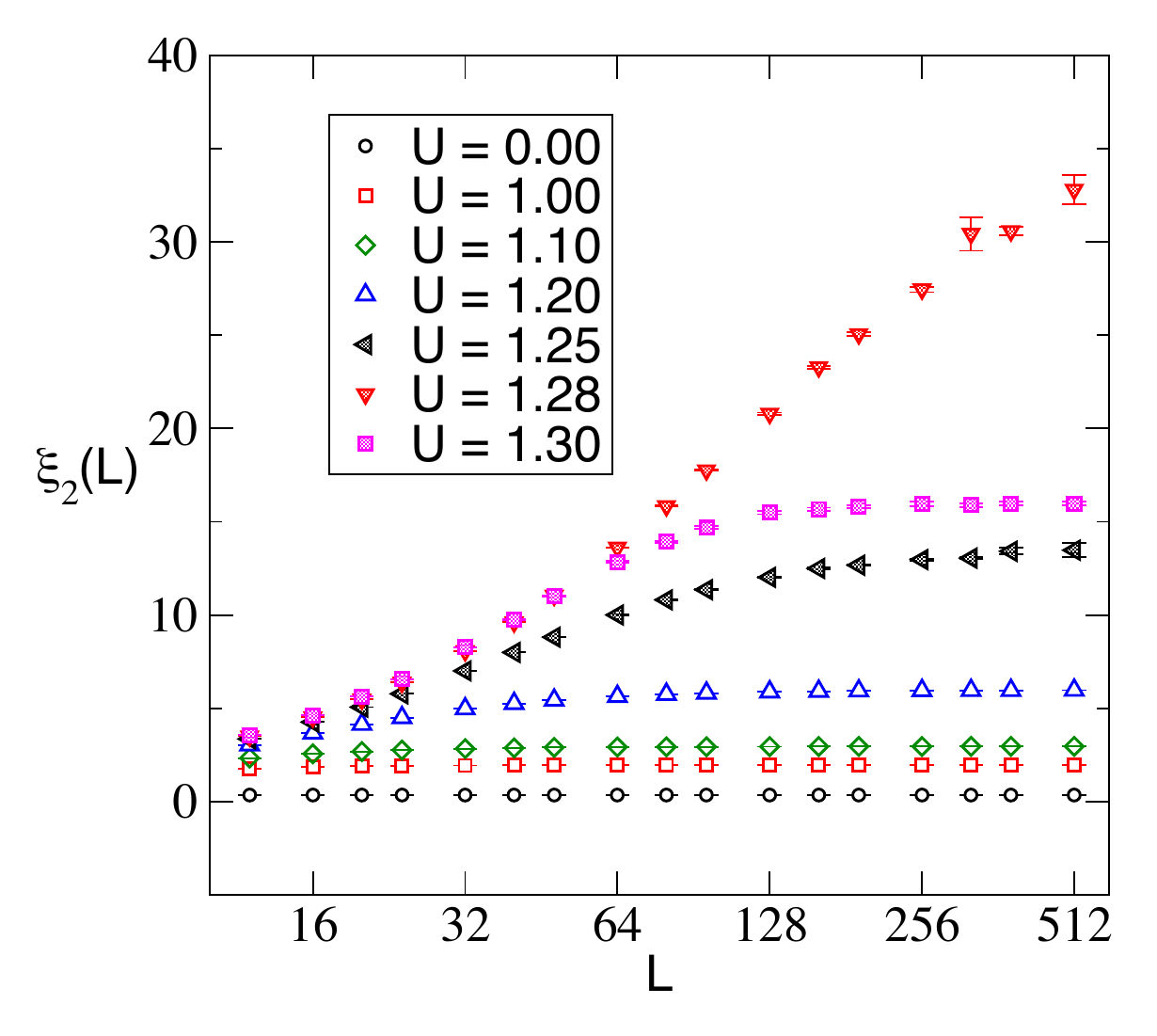}
\end{center}
\caption{\label{fig:susxi2}
In the left figure we plot $\chi_2$ as a function of $U$ for three different values of $L$. This plot suggests that the $G_2$ correlation function is never critical for any value of $U$. Although $\chi_2$ appears to be growing with $L$ at $U=1.28$, this rise is not due to criticality. We clarify this in the inset of the figure where we show $L$ dependence of $\chi_2$ at $U=1.28$. In the right figure we plot the $L$ dependence of $\xi_2(L)$ and again find that there is no real critical behavior although we do see an anomalously large correlation length at $U=1.28$, consistent with the left figure.}
\end{figure*}

Our Monte Carlo method is based on an extension of worm algorithms we have developed for similar problems before \cite{Adams:2003cca,Chandrasekharan:2006tz}. We have measured several observables. First is the density of vaccuum states (empty sites) which we will define as 
\begin{align}
\rho_0 \ =\ \frac{1}{L_x L_t}\frac{U}{Z} \frac{\partial Z}{\partial U}.
\label{eq:rho0}
\end{align}
where $L_x L_t$ is the total number of space-time lattice sites. Since our model is SU(2) invariant and in our path integral we can track the worldlines of the $z$-component of the spin, we can measure the current-current correlation function
\begin{align}
G_c(x,t) \ = \ \frac{1}{Z} \sum_{c}\ J^z_{x,t} \ J^z_{0,0}\  {\cal W}(c)
\end{align}
where $J^z_{x,t} = \ketbrauu - \ketbradd$ is the spin charge at the space-time site $(x,t)$. Due to rotational invariance of our model there is also a corresponding conserved spatial current that can be defined at every lattice site. Using $J^z_{x,t}$ and the corresponding spatial current, we can define a temporal winding charge $Q^t_w(c)$ and a spatial winding charge $Q^x_w(c)$ of the spin world lines in each configuration $b$. Using this we measure the current-current susceptibility $\rho^x_w$ defined as
\begin{align}
\rho^x_w &= \frac{L_x}{ZL_t} \sum_{c} (Q^a_w(c))^2 {\cal W}(c), \\
\rho^t_w &= \frac{L_t}{ZL_x} \sum_{c} (Q^a_w(c))^2 {\cal W}(c).
\label{eq:currsus}
\end{align}
On a square lattice, due to space-time rotational symmetry, both of these susceptibilities are the same and our Monte Carlo method measures $\rho_w = (\rho^x_w+\rho^t_w)/2$.

We have also measured properties of two different correlation functions using a worm algorithm. One is the anti-ferromagnetic spin-spin correlation function defined as
\begin{align}
G_1(x,t) \ = \ \frac{1}{Z} \sum_{c}\ (-1)^xS^1_{x,t} \ S^1_{0,0}\ {\cal W}(c)
\label{eq:G1}
\end{align}
where $S^1_{x,t} = \ketbraud + \ketbradu$ on the space-time lattice site $(x,t)$. The other correlation function we compute is the spin-half particle creation-annihilation correlation function defined as
\begin{align}
G_2(x,t) \ = \ \frac{1}{Z} \sum_{c}\ O^\dagger_{x,t}\ O_{0,0}\ {\cal W}(c),
\label{eq:G2}
\end{align}
where $O=((-1)^{x+1}\ketbraue + \ketbraed)$ based on our algorithm. Since $O$ and $O^\dagger$ do not commute we have to define $G_2(0,0)$ carefully. Our algorithm computes 
\begin{align}
G_2(0,0) \ = \ \frac{1}{Z} \sum_{c} 
\{O_{0,0}, O^\dagger_{0,0}\} {\cal W}(c),
\label{eq:G2a}
\end{align}
where $\{,\}$ is the anti-commutator.
With these definitions of the correlation functions we can define their susceptibilities as
\begin{align}
\chi_i \ =\ \sum_{x,t} G_i(x,t).
\label{eq:Chi}
\end{align}
To compute the finite size correlation lengths we also compute
\begin{subequations}
\begin{align}
F_i^x \ &=\ \sum_{x,t} G_i(x,t)\cos(2\pi x/L)
\label{eq:Fx}
\\
F_i^t \ &=\ \sum_{x,t} G_i(x,t)\cos(2\pi t/L)
\label{eq:Ft}
\end{align}
\end{subequations}
where on a square lattice we can check that $F_i^x=F^i_t$ due to space-time rotation symmetry. Using these we can define the second moment correlation length using the usual definition~\cite{Caracciolo:1992nh}
\begin{align}
\xi_i(L) \ =\ \frac{1}{2\sin(\pi/L)}\sqrt{\frac{\chi_i}{F^x_i}-1}.
\label{eq:fsscl}
\end{align}
The behavior of these correlation lengths as a function of $L$ can help us distinguish between the critical phase from the massive phase. {In the appendix we provide exact formulae for these observables on small lattices and compare them with Monte Carlo results.}

\begin{figure*}[ht]
\begin{center}
\includegraphics[width=0.48\textwidth]{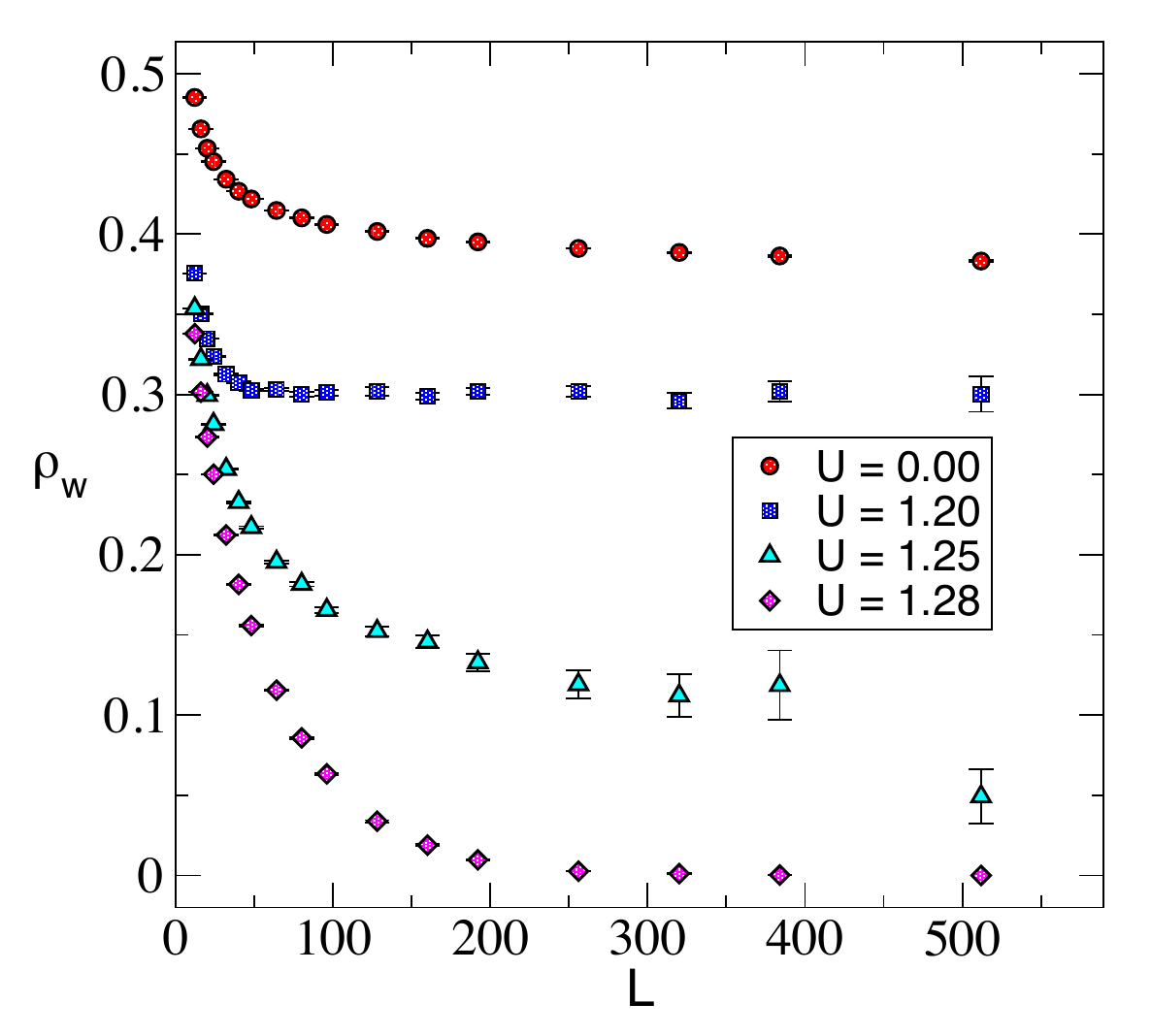}
\includegraphics[width=0.48\textwidth]{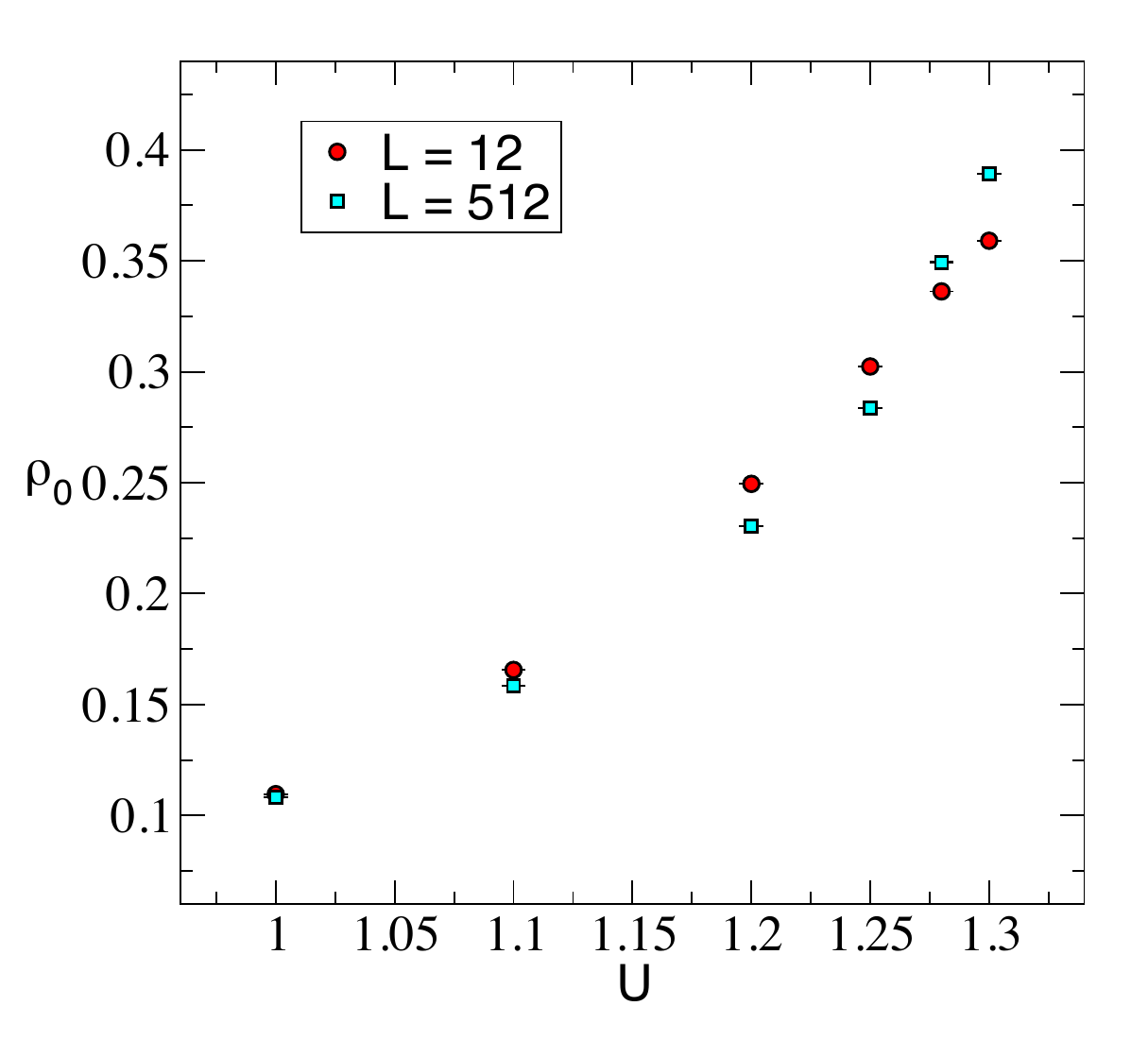}
\end{center}
\caption{\label{fig:wsusmono} In the left figure we plot the $\rho_w$ as a function of $L$ for various values of $U$. Again there is a dramatic change in the behavior between $U=1.2$ and $1.25$. On the right we plot density of the vacuum states $\rho_0$ as a function of the coupling $U$. As a local density that is not an order parameter, there is no significant change in it at the phase transition. However we note that the phase transition occurs when $\rho_0 \approx 0.25$. }
\end{figure*}

We have performed Monte Carlo calculations starting on lattices with $L=12,16,20$ and then doubling each of them several times until we reach $L=512$. Thus, in total we have studied nineteen different lattice sizes. For each of these lattices we have done calculations with $U=0,1$,$1.1$,$1.2$,$1.25$,$1.28$,$1.3$. We will argue below that our model is in a critical phase for $U\leq 1.20$ and in a massive phase for $U \geq 1.25$, suggesting that the critical coupling where the phase transition occurs is in the range $1.20 < U_c < 1.25$. We have not tried to determine the transition more precisely because we believe the transition is first order as we explain below. 

\begin{table}[b]
\centering
\setlength{\tabcolsep}{4pt}
\makegapedcells
\begin{tabular}{r|c|c|c|c}
\TopRule
\multicolumn{1}{c|}{$U$} & \multicolumn{1}{c|}{$c$} & \multicolumn{1}{c|}{$a$} &  \multicolumn{1}{c|}{$b$} & \multicolumn{1}{c}{$\chi^2$/DOF}\\
\MidRule 
0.0 & 0.2566(3) & 0.37(1) & - 1.23(3) & 0.90\\
1.0 & 0.2600(8) & 0.22(2) & -1.4(1) & 0.97 \\
1.1 & 0.2564(10) & 0.19(2) & -1.3(2) & 2.14 \\
1.2 & 0.2430(5) & 0.047(6) & -2.49(10) & 9.26\\
\BotRule
\end{tabular}
\caption{\label{tab:fits1} Results of fitting $\xi_1(L)/L$ to the form given in \cref{eq:marginal} at various values of $U$. Notice the fits are excellent for $U=0.0$ and $1.0$ but begin to become worse for larger values of $U$ suggesting the massive phase begins to have a big impact to the scaling behavior.}
\end{table}

At $U=0$ we expect the theory to be critical since our model is just a space-time symmetric Euclidean lattice field theory of a spin-half chain which we know is critical. When $U$ becomes non-zero but remains small, the vacuum states ($\ketbraee$) appear in small isolated space-time plaquettes. We believe this only renormalizes the couplings of the anti-ferromagnetic chain. Of course a less likely possibility is that the patches mimic a relevant topological operator that destroys the critical behavior as soon as it is introduced. Our numerical work confirms the former scenario. In \cref{fig:susxi1} we plot $\chi_1$ and $\xi_1/L$ as a function of $L$ for various values of $U$. We observe that both these observables show clear evidence of the critical phase for $U\leq 1.2$. In the critical theory we expect $\xi_i(L) = c L$. Since we expect our model to mimic the spin-half chain in the critical phase we do expect the model to have a marginally irrelevant operator in the theory. The effects of this will be visible as logarithmic corrections to the scaling behavior. While we have not worked out theoretically what the correct finite size effects would be one simple possibility is
\begin{align}
\frac{\xi_i(L)}{L}\ =\ c\Big(1 + \frac{a}{\log L + b}\Big) .
\label{eq:marginal}
\end{align}
Our data fits well to this ansatz for $U=0.0$ and $U=1.0$ for $32\leq L \leq 512$. For higher values of $U$ the fits rapidly become bad. We give the values of the fit parameters in table \cref{tab:fits1}.

The evidence for the phase transition between the critical and massive phases can also be clearly observed through the winding number susceptibility $\rho_w$ as shown in \cref{fig:wsusmono}. In the critical phase we expect $\rho_w$ to reach a contant at large values of $L$, while in the massive phase it is expected to vanish exponentially. At both $U=0.0$ and $1.20$ we see that $\rho_w$ seems to be saturating as expected. While the exponential decay is clear at $U=1.28$ it seems we will need much larger lattices to see this at $U=1.25$. In the right of \cref{fig:wsusmono} we plot the density of the vacuum states as a function of $U$ and we see that the critical phase survives until this density reaches $0.25$.

\begin{figure*}[ht]
\begin{center}
\includegraphics[width=0.48\textwidth]{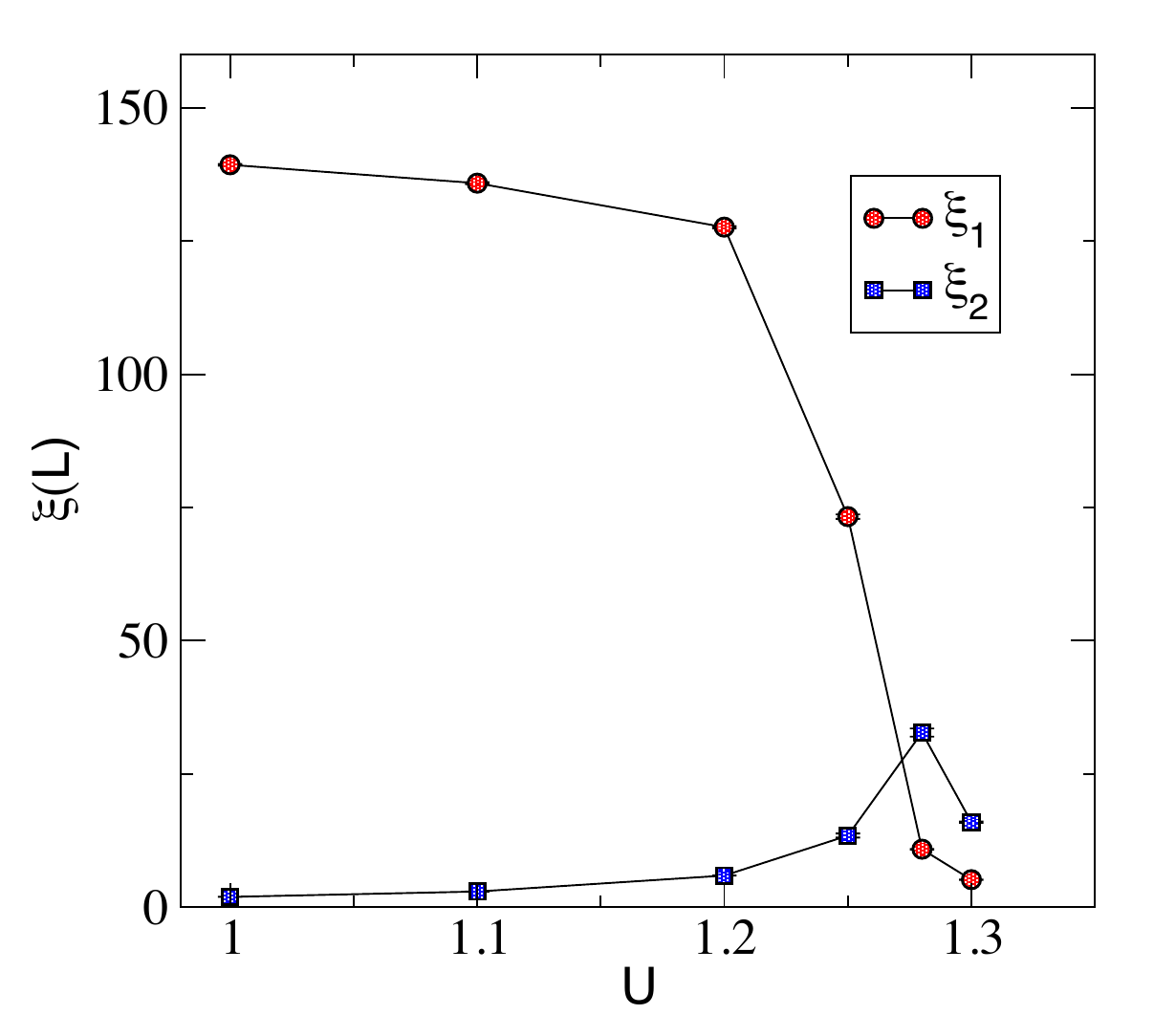}
\includegraphics[width=0.48\textwidth]{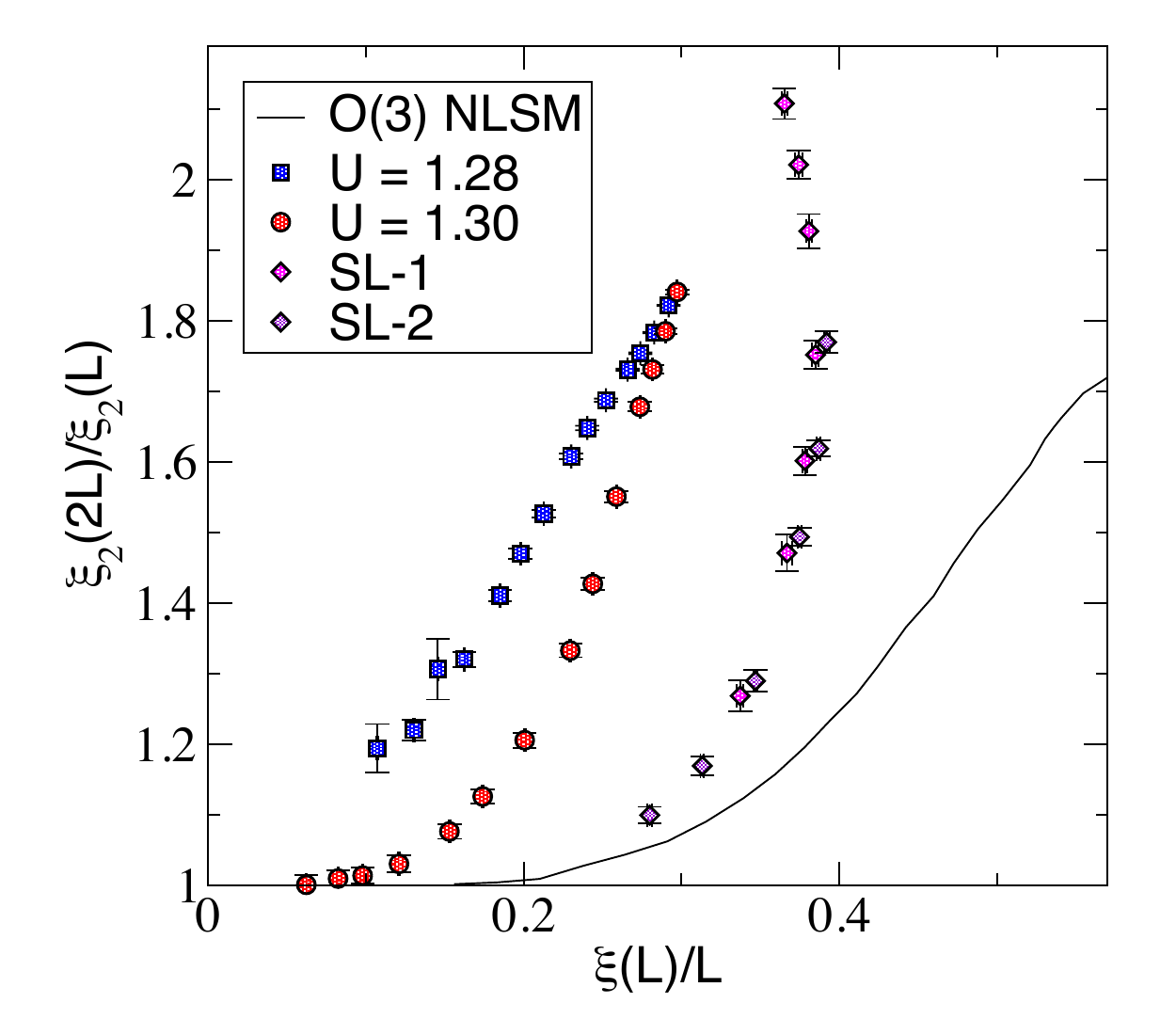}
\end{center}
\caption{\label{fig:xivsU} In the left figure we show the behavior of the two correlation lengths $\xi_1(L)$ and $\xi_2(L)$ for $L=512$ as a function of $U$ across the phase transition. The fact that $\xi_2(L)$ does not diverge across the phase transition provides strong evidence that the phase transition is first order. In the right figure we plot the step scaling function derived from $\xi_2(L)$ to study the \ac{UV} physics of the massive phase at $U=1.28$ and $U=1.3$. For reference we also plot the same function for the asymtotically free $O(3)$ \ac{NLSM} model described by \cref{eq:cmodel+theta} at $\theta=0$ (solid line) and data from an earlier study published in~\cite{Bhattacharya:2020gpm} for two weakly coupled spin-half chains (denoted as SL-1, SL-2 in the figure). The \ac{UV} physics of our model seems to be governed by the latter than the former. }
\end{figure*}

Let us now turn to the phase transition itself. In \cref{fig:xivsU} we plot the two correlation lengths $\xi_1(L)$ and $\xi_2(L)$ as a function of $U$ on the largest lattice sizes we have, i.e., $L=512$. There is a clear jump in $\xi_1(L)$ between the couplings $U=1.20$ and $1.25$. Along with the behavior of $\xi_1(L)/L$ in \cref{fig:susxi1} we can conclude that there is a phase transition between these two couplings. However, notice that $\xi_2(L)$ has only increased slightly between $U=1.20$ and $1.25$ and there is no indication of a divergence. This strongly suggests that the transition is first order. A close examination of the the behavior of the correlation lengths in the massive phase reveals some interesting peculiarities. At $U=1.25$ note that $\xi_1 > \xi_2$, while for $U\geq 1.28$ we notice a reversal and we find $\xi_1(L) < \xi_2(L)$. This is because $\xi_2(L)$ continues to increase between $U=1.25$ and $1.28$ while $\xi_1(L)$ continues to drop. Thus the massive phase in a small region from $1.25 < U < 1.28$ seems to be affected by the critical phase more dramatically than for large values of $U$. Perhaps some of this is an affect of the vicinity of the critical phase. We also believe the non-monotonic behavior observed in $\chi_1$ at $U=1.28$ and $1.30$ in \cref{fig:susxi1} could be another feature of the phase transition. Such behavior has been observed before near the vicinity of first order transitions~\cite{Chandrasekharan:2006tz}.

\section{Relativistic Quantum Circuits}
\label{sec5}

In conventional space-time lattice field theory since one begins with a Euclidean Lagrangian, the Wick rotation from imaginary to real time makes the action complex. On the other hand in the Hamiltonian formulation we can begin with a Euclidean space-time lattice field theory and perform Wick rotation by simply setting the Trotter time step $\varepsilon = i\Delta$ and one naturally gets a unitary time evolution. This can be seen from \cref{eq:Ete} where by making this change of variables and defining $\tau = i t$ with both $\Delta$ and $t=\Delta L_t$ as real parameters, \cref{eq:Ete} changes to
\begin{align}
U(t) = U_E(\tau \rightarrow it),    
\end{align}
which is nothing but a time evolution operator in real time. {In fact,} $U(t)$ {is the} analytic continuation of the imaginary time expression $U_E(\tau)$ to real time. {Obviously, as long as the whole formulation is on a discrete finite lattice, the Wick rotation and exponentiation operations commute.}

{The question is: }How does the continuum limit emerge from such a formulation. Clearly{,} we will need to tune the Hamiltonian to a quantum critical point{, and study appropriate\footnote{{The appropriate quantities are built from operators that do not couple to the high-energy `ultraviolet' modes.   Examples in space-time symmetric models include local operators smeared over spatial distances larger than this ultraviolet scale. This smearing produces corrections at times smaller than or comparable to the smearing scale, which, therefore, needs to be much smaller than the correlation length.}} quantities at a fixed physical interval, i.e., at a constant multiple of the (temporal) correlation length. At any fixed value of the parameters in the Hamiltonian away from the critical point, the spectrum has finite corrections, and hence the errors grow exponentially at large times. On the other hand, if we hold the physical interval fixed, the closer we go to the critical point, the smaller these errors become, even though the interval being fixed in physical units needs an increasing number of lattice steps.

This is fundamentally different than the usual approach of evolving the spatially discrete Hamiltonian with small temporal steps, and taking the limit of $\varepsilon \rightarrow 0$. In that approach, it is important to understand how the limits must be taken~\cite{Carena:2021ltu}}, {essentially due to a lack of symmetry between spatial correlations and temporal correlations. In contrast}, {because of our space-time symmetric approach, the temporal and spatial lattice spacings are automatically related, and w}hen such a theory is critical, like in the critical phase in our model or at a quantum critical point, we do not need to take the time discretization error to zero {separately}. 
This can naturally lead to efficient quantum-circuits for real-time evolution of the corresponding Minkowski field theory.

Note that we did not have to use the Trotter {step size} $\varepsilon$ in this analysis due to space-time symmetry. {Instead}, {the} Wick-rotated correlation function can be obtained by sequential application of an unitary given by Wick-rotating the Euclidean transfer matrix.  This unitary is completely local like the transfer matrix, {and} provides an approximation to the time evolution that is often distinct from what one would obtain by Trotterizing a Hamiltonian time-step. {In summary, t}his evolution has three remarkable properties: {(i) as we take the `continuum limit' by tuning the couplings to get larger spatial correlation lengths, the temporal correlation length stays equal to the spatial correlation length and does not have to be tuned separately; (ii) the difference between the lattice and continuum correlators are not proportional to the time duration in lattice units, which grows as the correlation length; and (iii) the discretization error for evolution in the} low-energy subspace is not proportional to the ultraviolet energy scale. The possibility of {the last} is known for local interactions~\cite{_ahino_lu_2021}, and {the control of errors specifically at long simulation times} has also been studied in a wide class of systems~\cite{Gu:2021hyo}, but our construction provides a particularly simple example of this for relativistic field theories for which a Euclidean qubit formulation {with space-time rotation symmetry} can be constructed.

\section{Conclusions}
\label{sec6}

In this work we introduced the SU(3)-F regularization scheme to embed the SO(3) spin-symmetry and constructed a Euclidean lattice field theory with a coupling $U$ that is symmetric under space-time rotations. Using the Monte Carlo method we showed that our lattice model has two phases, a critical phase for $U \leq 1.2 $ and a trivially massive phase for $U\geq 1.25$. We reached this conclusion my studying the behavior of the current-current correlation function $G_0(x,t)$, the anti-ferromagnetic spin-spin correlation function $G_1(x,t)$, and the spin creation-annihilation correlation function $G_2(x,t)$ through their susceptibilities. We also computed the two finite size correlation lengths $\xi_1(L)$ and $\xi_2(L)$ obtained from $G_1$ and $G_2$ respectively. Since the $U=0$ model is just a space-time symmetric realization of the spin-half chain, we believe the critical phase is describing the $O(3)$ \ac{NLSM} at $\theta=\pi$ or equivalently the $k=1$ \ac{WZW} \ac{CFT} with a marginally irrelevant coupling. The presence of such a coupling could be seen in our data through the behavior of $\xi_1(L)$ as a function of $L$, which we modeled using \cref{eq:marginal}. 

One of our motivations to study the model was the hope that the quantum phase transition between the critical phase and the massive phase would be second order. We could then explore if the \ac{UV} physics of the massive phase near the critical point is governed by the asymptotically free fixed point described by \cref{eq:cmodel+theta} at $\theta=0$ or by the fixed point of two decoupled spin-half chains. Unfortunately, our data suggests that the phase transition is first order. This is clear from \cref{fig:xivsU} where there is no evidence of a diverging $\xi_2$. Still if we study the \ac{UV} physics of the massive phase at $U=1.28$ and $1.30$ using $\xi_2(L)$, we notice that it is more similar to the \ac{UV} physics of the decoupled spin-half chains. We show this in the right plot of \cref{fig:xivsU}, where we plot the step scaling function obtained from $\xi_2(L)$. On small lattices the physics is dominated by the critical phase similar to what happens when two spin-half chains are weakly coupled. In the figure we also show the data published in~\cite{Bhattacharya:2020gpm} for two weakly coupled spin-half chains, labeled as SL-1 and SL-2. The solid curve is the step-scaling function of the asymptotocally free $O(3)$ \ac{NLSM}. Our model is clearly different.
A more extensive search of the model space within the SU(3)-F scheme may still reveal the existence of a true quantum critical connected to the massive phase. If it exists, its properties will be interesting to study.

Finally, in this paper we also argued how we could begin with our space-time symmetric Euclidean lattice model and perform a Wick rotation to real time. The Euclidean time propagation operator $U_E(\tau)$ is then modified to the time evolution operator in real time $U(t)$ which is naturally unitary. We defined this operator on a space-time lattice as a relativistic quantum circuit. One of the features of this quantum circuit is that there is no need to tune the temporal lattice spacing to zero separately, since it emerges naturally in the critical phase at long times measured in lattice units.

\section*{Acknowledgements}

S.C. would like to thank Ribhu Kaul, Hanqing Liu and Rui Xian Siew for helpful discussions on the subject. S.C would also like to thank the International Center for Theoretical Studies, Bengaluru for hospitality where part of this work was completed.
H.S. would like to thank Stephan Caspar for many useful conversations on related work. 
The material presented here is based on work supported by the U.S. Department of Energy, Office of Science --- High Energy Physics Contract KA2401032 (Triad National Security, LLC Contract Grant No. 89233218CNA000001) to Los Alamos National Laboratory. S.C. is supported by a Duke subcontract of this grant.
The algorithmic research of S.C. related to this work is supported in part by the U.S. Department of Energy, Office of Science, Nuclear Physics program under Award No. DE-FG02-05ER41368.
This work was supported in part by the Deutsche Forschungsgemeinschaft (DFG) through the Cluster of Excellence “Precision Physics, Fundamental Interactions, and Structure of Matter” (PRISMA${}^+$ EXC 2118/1) funded by the DFG within the German Excellence Strategy (Project ID 39083149).
The work of HS is funded
in part by the DOE QuantISED program through the theory consortium ``Intersections of QIS and Theoretical Particle Physics'' at Fermilab with Fermilab Subcontract No. 666484,
in part by the Institute for Nuclear Theory with US Department of Energy Grant No. DE-FG02-00ER41132,
and in part by U.S. Department of Energy, Office of Science, Office of Nuclear Physics, InQubator for Quantum Simulation (IQuS) under Award No. DOE (NP) DE-SC0020970.

\showtitleinbib
\bibliographystyle{apsrev4-2}
\bibliography{ref,topology,qc}

\begin{thebibliography}{54}%
\makeatletter
\providecommand \@ifxundefined [1]{%
 \@ifx{#1\undefined}
}%
\providecommand \@ifnum [1]{%
 \ifnum #1\expandafter \@firstoftwo
 \else \expandafter \@secondoftwo
 \fi
}%
\providecommand \@ifx [1]{%
 \ifx #1\expandafter \@firstoftwo
 \else \expandafter \@secondoftwo
 \fi
}%
\providecommand \natexlab [1]{#1}%
\providecommand \enquote  [1]{``#1''}%
\providecommand \bibnamefont  [1]{#1}%
\providecommand \bibfnamefont [1]{#1}%
\providecommand \citenamefont [1]{#1}%
\providecommand \href@noop [0]{\@secondoftwo}%
\providecommand \href [0]{\begingroup \@sanitize@url \@href}%
\providecommand \@href[1]{\@@startlink{#1}\@@href}%
\providecommand \@@href[1]{\endgroup#1\@@endlink}%
\providecommand \@sanitize@url [0]{\catcode `\\12\catcode `\$12\catcode
  `\&12\catcode `\#12\catcode `\^12\catcode `\_12\catcode `\%12\relax}%
\providecommand \@@startlink[1]{}%
\providecommand \@@endlink[0]{}%
\providecommand \url  [0]{\begingroup\@sanitize@url \@url }%
\providecommand \@url [1]{\endgroup\@href {#1}{\urlprefix }}%
\providecommand \urlprefix  [0]{URL }%
\providecommand \Eprint [0]{\href }%
\providecommand \doibase [0]{https://doi.org/}%
\providecommand \selectlanguage [0]{\@gobble}%
\providecommand \bibinfo  [0]{\@secondoftwo}%
\providecommand \bibfield  [0]{\@secondoftwo}%
\providecommand \translation [1]{[#1]}%
\providecommand \BibitemOpen [0]{}%
\providecommand \bibitemStop [0]{}%
\providecommand \bibitemNoStop [0]{.\EOS\space}%
\providecommand \EOS [0]{\spacefactor3000\relax}%
\providecommand \BibitemShut  [1]{\csname bibitem#1\endcsname}%
\let\auto@bib@innerbib\@empty
\bibitem [{\citenamefont {Arute}\ \emph {et~al.}(2019)\citenamefont {Arute},
  \citenamefont {Arya}, \citenamefont {Babbush}, \citenamefont {Bacon},
  \citenamefont {Bardin} \emph {et~al.}}]{Arute:2019zxq}%
  \BibitemOpen
  \bibfield  {author} {\bibinfo {author} {\bibfnamefont {F.}~\bibnamefont
  {Arute}}, \bibinfo {author} {\bibfnamefont {K.}~\bibnamefont {Arya}},
  \bibinfo {author} {\bibfnamefont {R.}~\bibnamefont {Babbush}}, \bibinfo
  {author} {\bibfnamefont {D.}~\bibnamefont {Bacon}}, \bibinfo {author}
  {\bibfnamefont {J.~C.}\ \bibnamefont {Bardin}}, \emph {et~al.},\ }\bibfield
  {title} {\bibinfo {title} {Quantum supremacy using a programmable
  superconducting processor},\ }\href
  {https://doi.org/10.1038/s41586-019-1666-5} {\bibfield  {journal} {\bibinfo
  {journal} {Nature}\ }\textbf {\bibinfo {volume} {574}},\ \bibinfo {pages}
  {505} (\bibinfo {year} {2019})},\ \Eprint {https://arxiv.org/abs/1910.11333}
  {arXiv:1910.11333 [quant-ph]} \BibitemShut {NoStop}%
\bibitem [{\citenamefont {Bauer}\ \emph {et~al.}(2022)\citenamefont {Bauer},
  \citenamefont {Davoudi}, \citenamefont {Balantekin}, \citenamefont
  {Bhattacharya}, \citenamefont {Carena} \emph {et~al.}}]{Bauer:2022hpo}%
  \BibitemOpen
  \bibfield  {author} {\bibinfo {author} {\bibfnamefont {C.~W.}\ \bibnamefont
  {Bauer}}, \bibinfo {author} {\bibfnamefont {Z.}~\bibnamefont {Davoudi}},
  \bibinfo {author} {\bibfnamefont {A.~B.}\ \bibnamefont {Balantekin}},
  \bibinfo {author} {\bibfnamefont {T.}~\bibnamefont {Bhattacharya}}, \bibinfo
  {author} {\bibfnamefont {M.}~\bibnamefont {Carena}}, \emph {et~al.},\
  }\bibfield  {title} {\bibinfo {title} {Quantum simulation for high energy
  physics},\ }\href@noop {} {\bibfield  {journal} {\bibinfo  {journal}
  {arXiv:2204.03381}\ } (\bibinfo {year} {2022})}\BibitemShut {NoStop}%
\bibitem [{\citenamefont {Jordan}\ \emph {et~al.}(2011)\citenamefont {Jordan},
  \citenamefont {Lee},\ and\ \citenamefont {Preskill}}]{Jordan:2011ci}%
  \BibitemOpen
  \bibfield  {author} {\bibinfo {author} {\bibfnamefont {S.~P.}\ \bibnamefont
  {Jordan}}, \bibinfo {author} {\bibfnamefont {K.~S.~M.}\ \bibnamefont {Lee}},\
  and\ \bibinfo {author} {\bibfnamefont {J.}~\bibnamefont {Preskill}},\
  }\href@noop {} {\bibinfo {title} {Quantum computation of scattering in scalar
  quantum field theories}} (\bibinfo {year} {2011}),\ \bibinfo {note} {[Quant.
  Inf. Comput.14,1014(2014)]},\ \Eprint {https://arxiv.org/abs/1112.4833}
  {arXiv:1112.4833 [hep-th]} \BibitemShut {NoStop}%
\bibitem [{\citenamefont {Casanova}\ \emph {et~al.}(2011)\citenamefont
  {Casanova}, \citenamefont {Lamata}, \citenamefont {Egusquiza}, \citenamefont
  {Gerritsma}, \citenamefont {Roos}, \citenamefont {Garcia-Ripoll},\ and\
  \citenamefont {Solano}}]{Casanova:2011wh}%
  \BibitemOpen
  \bibfield  {author} {\bibinfo {author} {\bibfnamefont {J.}~\bibnamefont
  {Casanova}}, \bibinfo {author} {\bibfnamefont {L.}~\bibnamefont {Lamata}},
  \bibinfo {author} {\bibfnamefont {I.~L.}\ \bibnamefont {Egusquiza}}, \bibinfo
  {author} {\bibfnamefont {R.}~\bibnamefont {Gerritsma}}, \bibinfo {author}
  {\bibfnamefont {C.~F.}\ \bibnamefont {Roos}}, \bibinfo {author}
  {\bibfnamefont {J.~J.}\ \bibnamefont {Garcia-Ripoll}},\ and\ \bibinfo
  {author} {\bibfnamefont {E.}~\bibnamefont {Solano}},\ }\bibfield  {title}
  {\bibinfo {title} {Quantum simulation of quantum field theories in trapped
  ions},\ }\href {https://doi.org/10.1103/PhysRevLett.107.260501} {\bibfield
  {journal} {\bibinfo  {journal} {Phys. Rev. Lett.}\ }\textbf {\bibinfo
  {volume} {107}},\ \bibinfo {pages} {260501} (\bibinfo {year} {2011})},\
  \Eprint {https://arxiv.org/abs/1107.5233} {arXiv:1107.5233 [quant-ph]}
  \BibitemShut {NoStop}%
\bibitem [{\citenamefont {Casanova}\ \emph {et~al.}(2012)\citenamefont
  {Casanova}, \citenamefont {Mezzacapo}, \citenamefont {Lamata},\ and\
  \citenamefont {Solano}}]{Casanova:2012zz}%
  \BibitemOpen
  \bibfield  {author} {\bibinfo {author} {\bibfnamefont {J.}~\bibnamefont
  {Casanova}}, \bibinfo {author} {\bibfnamefont {A.}~\bibnamefont {Mezzacapo}},
  \bibinfo {author} {\bibfnamefont {L.}~\bibnamefont {Lamata}},\ and\ \bibinfo
  {author} {\bibfnamefont {E.}~\bibnamefont {Solano}},\ }\bibfield  {title}
  {\bibinfo {title} {Quantum simulation of interacting {F}ermion lattice models
  in trapped ions},\ }\href {https://doi.org/10.1103/PhysRevLett.108.190502}
  {\bibfield  {journal} {\bibinfo  {journal} {Phys. Rev. Lett.}\ }\textbf
  {\bibinfo {volume} {108}},\ \bibinfo {pages} {190502} (\bibinfo {year}
  {2012})},\ \Eprint {https://arxiv.org/abs/1110.3730} {arXiv:1110.3730
  [quant-ph]} \BibitemShut {NoStop}%
\bibitem [{\citenamefont {O'Malley}\ \emph {et~al.}(2016)\citenamefont
  {O'Malley}, \citenamefont {Babbush}, \citenamefont {Kivlichan}, \citenamefont
  {Romero}, \citenamefont {McClean}, \citenamefont {Barends}, \citenamefont
  {Kelly}, \citenamefont {Roushan}, \citenamefont {Tranter}, \citenamefont
  {Ding}, \citenamefont {Campbell}, \citenamefont {Chen}, \citenamefont {Chen},
  \citenamefont {Chiaro}, \citenamefont {Dunsworth}, \citenamefont {Fowler},
  \citenamefont {Jeffrey}, \citenamefont {Lucero}, \citenamefont {Megrant},
  \citenamefont {Mutus}, \citenamefont {Neeley}, \citenamefont {Neill},
  \citenamefont {Quintana}, \citenamefont {Sank}, \citenamefont {Vainsencher},
  \citenamefont {Wenner}, \citenamefont {White}, \citenamefont {Coveney},
  \citenamefont {Love}, \citenamefont {Neven}, \citenamefont {Aspuru-Guzik},\
  and\ \citenamefont {Martinis}}]{PhysRevX.6.031007}%
  \BibitemOpen
  \bibfield  {author} {\bibinfo {author} {\bibfnamefont {P.~J.~J.}\
  \bibnamefont {O'Malley}}, \bibinfo {author} {\bibfnamefont {R.}~\bibnamefont
  {Babbush}}, \bibinfo {author} {\bibfnamefont {I.~D.}\ \bibnamefont
  {Kivlichan}}, \bibinfo {author} {\bibfnamefont {J.}~\bibnamefont {Romero}},
  \bibinfo {author} {\bibfnamefont {J.~R.}\ \bibnamefont {McClean}}, \bibinfo
  {author} {\bibfnamefont {R.}~\bibnamefont {Barends}}, \bibinfo {author}
  {\bibfnamefont {J.}~\bibnamefont {Kelly}}, \bibinfo {author} {\bibfnamefont
  {P.}~\bibnamefont {Roushan}}, \bibinfo {author} {\bibfnamefont
  {A.}~\bibnamefont {Tranter}}, \bibinfo {author} {\bibfnamefont
  {N.}~\bibnamefont {Ding}}, \bibinfo {author} {\bibfnamefont {B.}~\bibnamefont
  {Campbell}}, \bibinfo {author} {\bibfnamefont {Y.}~\bibnamefont {Chen}},
  \bibinfo {author} {\bibfnamefont {Z.}~\bibnamefont {Chen}}, \bibinfo {author}
  {\bibfnamefont {B.}~\bibnamefont {Chiaro}}, \bibinfo {author} {\bibfnamefont
  {A.}~\bibnamefont {Dunsworth}}, \bibinfo {author} {\bibfnamefont {A.~G.}\
  \bibnamefont {Fowler}}, \bibinfo {author} {\bibfnamefont {E.}~\bibnamefont
  {Jeffrey}}, \bibinfo {author} {\bibfnamefont {E.}~\bibnamefont {Lucero}},
  \bibinfo {author} {\bibfnamefont {A.}~\bibnamefont {Megrant}}, \bibinfo
  {author} {\bibfnamefont {J.~Y.}\ \bibnamefont {Mutus}}, \bibinfo {author}
  {\bibfnamefont {M.}~\bibnamefont {Neeley}}, \bibinfo {author} {\bibfnamefont
  {C.}~\bibnamefont {Neill}}, \bibinfo {author} {\bibfnamefont
  {C.}~\bibnamefont {Quintana}}, \bibinfo {author} {\bibfnamefont
  {D.}~\bibnamefont {Sank}}, \bibinfo {author} {\bibfnamefont {A.}~\bibnamefont
  {Vainsencher}}, \bibinfo {author} {\bibfnamefont {J.}~\bibnamefont {Wenner}},
  \bibinfo {author} {\bibfnamefont {T.~C.}\ \bibnamefont {White}}, \bibinfo
  {author} {\bibfnamefont {P.~V.}\ \bibnamefont {Coveney}}, \bibinfo {author}
  {\bibfnamefont {P.~J.}\ \bibnamefont {Love}}, \bibinfo {author}
  {\bibfnamefont {H.}~\bibnamefont {Neven}}, \bibinfo {author} {\bibfnamefont
  {A.}~\bibnamefont {Aspuru-Guzik}},\ and\ \bibinfo {author} {\bibfnamefont
  {J.~M.}\ \bibnamefont {Martinis}},\ }\bibfield  {title} {\bibinfo {title}
  {Scalable quantum simulation of molecular energies},\ }\href
  {https://doi.org/10.1103/PhysRevX.6.031007} {\bibfield  {journal} {\bibinfo
  {journal} {Phys. Rev. X}\ }\textbf {\bibinfo {volume} {6}},\ \bibinfo {pages}
  {031007} (\bibinfo {year} {2016})}\BibitemShut {NoStop}%
\bibitem [{\citenamefont {Macridin}\ \emph {et~al.}(2018)\citenamefont
  {Macridin}, \citenamefont {Spentzouris}, \citenamefont {Amundson},\ and\
  \citenamefont {Harnik}}]{Macridin:2018oli}%
  \BibitemOpen
  \bibfield  {author} {\bibinfo {author} {\bibfnamefont {A.}~\bibnamefont
  {Macridin}}, \bibinfo {author} {\bibfnamefont {P.}~\bibnamefont
  {Spentzouris}}, \bibinfo {author} {\bibfnamefont {J.}~\bibnamefont
  {Amundson}},\ and\ \bibinfo {author} {\bibfnamefont {R.}~\bibnamefont
  {Harnik}},\ }\bibfield  {title} {\bibinfo {title} {Digital quantum
  computation of {F}ermion-{B}oson interacting systems},\ }\href
  {https://doi.org/10.1103/PhysRevA.98.042312} {\bibfield  {journal} {\bibinfo
  {journal} {Phys. Rev.}\ }\textbf {\bibinfo {volume} {A98}},\ \bibinfo {pages}
  {042312} (\bibinfo {year} {2018})},\ \Eprint
  {https://arxiv.org/abs/1805.09928} {arXiv:1805.09928 [quant-ph]} \BibitemShut
  {NoStop}%
\bibitem [{\citenamefont {Alexandru}\ \emph
  {et~al.}(2019{\natexlab{a}})\citenamefont {Alexandru}, \citenamefont
  {Bedaque}, \citenamefont {Lamm},\ and\ \citenamefont
  {Lawrence}}]{PhysRevLett.123.090501}%
  \BibitemOpen
  \bibfield  {author} {\bibinfo {author} {\bibfnamefont {A.}~\bibnamefont
  {Alexandru}}, \bibinfo {author} {\bibfnamefont {P.~F.}\ \bibnamefont
  {Bedaque}}, \bibinfo {author} {\bibfnamefont {H.}~\bibnamefont {Lamm}},\ and\
  \bibinfo {author} {\bibfnamefont {S.}~\bibnamefont {Lawrence}} (\bibinfo
  {collaboration} {NuQS Collaboration}),\ }\bibfield  {title} {\bibinfo {title}
  {$\ensuremath{\sigma}$ models on quantum computers},\ }\href
  {https://doi.org/10.1103/PhysRevLett.123.090501} {\bibfield  {journal}
  {\bibinfo  {journal} {Phys. Rev. Lett.}\ }\textbf {\bibinfo {volume} {123}},\
  \bibinfo {pages} {090501} (\bibinfo {year} {2019}{\natexlab{a}})}\BibitemShut
  {NoStop}%
\bibitem [{\citenamefont {Raychowdhury}\ and\ \citenamefont
  {Stryker}(2020)}]{Raychowdhury:2018osk}%
  \BibitemOpen
  \bibfield  {author} {\bibinfo {author} {\bibfnamefont {I.}~\bibnamefont
  {Raychowdhury}}\ and\ \bibinfo {author} {\bibfnamefont {J.~R.}\ \bibnamefont
  {Stryker}},\ }\bibfield  {title} {\bibinfo {title} {Solving {G}auss's law on
  digital quantum computers with loop-string-hadron digitization},\ }\href
  {https://doi.org/10.1103/PhysRevResearch.2.033039} {\bibfield  {journal}
  {\bibinfo  {journal} {Phys. Rev. Res.}\ }\textbf {\bibinfo {volume} {2}},\
  \bibinfo {pages} {033039} (\bibinfo {year} {2020})},\ \Eprint
  {https://arxiv.org/abs/1812.07554} {arXiv:1812.07554 [hep-lat]} \BibitemShut
  {NoStop}%
\bibitem [{\citenamefont {Roggero}\ and\ \citenamefont
  {Carlson}(2018)}]{Roggero:2018hrn}%
  \BibitemOpen
  \bibfield  {author} {\bibinfo {author} {\bibfnamefont {A.}~\bibnamefont
  {Roggero}}\ and\ \bibinfo {author} {\bibfnamefont {J.}~\bibnamefont
  {Carlson}},\ }\href@noop {} {\bibinfo {title} {Linear response on a quantum
  computer}} (\bibinfo {year} {2018}),\ \Eprint
  {https://arxiv.org/abs/1804.01505} {arXiv:1804.01505 [quant-ph]} \BibitemShut
  {NoStop}%
\bibitem [{\citenamefont {Alexandru}\ \emph
  {et~al.}(2019{\natexlab{b}})\citenamefont {Alexandru}, \citenamefont
  {Bedaque}, \citenamefont {Harmalkar}, \citenamefont {Lamm}, \citenamefont
  {Lawrence},\ and\ \citenamefont {Warrington}}]{Alexandru:2019nsa}%
  \BibitemOpen
  \bibfield  {author} {\bibinfo {author} {\bibfnamefont {A.}~\bibnamefont
  {Alexandru}}, \bibinfo {author} {\bibfnamefont {P.~F.}\ \bibnamefont
  {Bedaque}}, \bibinfo {author} {\bibfnamefont {S.}~\bibnamefont {Harmalkar}},
  \bibinfo {author} {\bibfnamefont {H.}~\bibnamefont {Lamm}}, \bibinfo {author}
  {\bibfnamefont {S.}~\bibnamefont {Lawrence}},\ and\ \bibinfo {author}
  {\bibfnamefont {N.~C.}\ \bibnamefont {Warrington}} (\bibinfo {collaboration}
  {NuQS}),\ }\bibfield  {title} {\bibinfo {title} {Gluon field digitization for
  quantum computers},\ }\href {https://doi.org/10.1103/PhysRevD.100.114501}
  {\bibfield  {journal} {\bibinfo  {journal} {Phys. Rev. D}\ }\textbf {\bibinfo
  {volume} {100}},\ \bibinfo {pages} {114501} (\bibinfo {year}
  {2019}{\natexlab{b}})},\ \Eprint {https://arxiv.org/abs/1906.11213}
  {arXiv:1906.11213 [hep-lat]} \BibitemShut {NoStop}%
\bibitem [{\citenamefont {Davoudi}\ \emph {et~al.}(2020)\citenamefont
  {Davoudi}, \citenamefont {Hafezi}, \citenamefont {Monroe}, \citenamefont
  {Pagano}, \citenamefont {Seif},\ and\ \citenamefont
  {Shaw}}]{Davoudi:2019bhy}%
  \BibitemOpen
  \bibfield  {author} {\bibinfo {author} {\bibfnamefont {Z.}~\bibnamefont
  {Davoudi}}, \bibinfo {author} {\bibfnamefont {M.}~\bibnamefont {Hafezi}},
  \bibinfo {author} {\bibfnamefont {C.}~\bibnamefont {Monroe}}, \bibinfo
  {author} {\bibfnamefont {G.}~\bibnamefont {Pagano}}, \bibinfo {author}
  {\bibfnamefont {A.}~\bibnamefont {Seif}},\ and\ \bibinfo {author}
  {\bibfnamefont {A.}~\bibnamefont {Shaw}},\ }\bibfield  {title} {\bibinfo
  {title} {Towards analog quantum simulations of lattice gauge theories with
  trapped ions},\ }\href {https://doi.org/10.1103/PhysRevResearch.2.023015}
  {\bibfield  {journal} {\bibinfo  {journal} {Phys. Rev. Res.}\ }\textbf
  {\bibinfo {volume} {2}},\ \bibinfo {pages} {023015} (\bibinfo {year}
  {2020})},\ \Eprint {https://arxiv.org/abs/1908.03210} {arXiv:1908.03210
  [quant-ph]} \BibitemShut {NoStop}%
\bibitem [{\citenamefont {Davoudi}\ \emph {et~al.}(2021)\citenamefont
  {Davoudi}, \citenamefont {Raychowdhury},\ and\ \citenamefont
  {Shaw}}]{Davoudi:2020yln}%
  \BibitemOpen
  \bibfield  {author} {\bibinfo {author} {\bibfnamefont {Z.}~\bibnamefont
  {Davoudi}}, \bibinfo {author} {\bibfnamefont {I.}~\bibnamefont
  {Raychowdhury}},\ and\ \bibinfo {author} {\bibfnamefont {A.}~\bibnamefont
  {Shaw}},\ }\bibfield  {title} {\bibinfo {title} {Search for efficient
  formulations for {H}amiltonian simulation of non-{A}belian lattice gauge
  theories},\ }\href {https://doi.org/10.1103/PhysRevD.104.074505} {\bibfield
  {journal} {\bibinfo  {journal} {Phys. Rev. D}\ }\textbf {\bibinfo {volume}
  {104}},\ \bibinfo {pages} {074505} (\bibinfo {year} {2021})},\ \Eprint
  {https://arxiv.org/abs/2009.11802} {arXiv:2009.11802 [hep-lat]} \BibitemShut
  {NoStop}%
\bibitem [{\citenamefont {Ciavarella}\ \emph {et~al.}(2021)\citenamefont
  {Ciavarella}, \citenamefont {Klco},\ and\ \citenamefont
  {Savage}}]{Ciavarella:2021nmj}%
  \BibitemOpen
  \bibfield  {author} {\bibinfo {author} {\bibfnamefont {A.}~\bibnamefont
  {Ciavarella}}, \bibinfo {author} {\bibfnamefont {N.}~\bibnamefont {Klco}},\
  and\ \bibinfo {author} {\bibfnamefont {M.~J.}\ \bibnamefont {Savage}},\
  }\bibfield  {title} {\bibinfo {title} {Trailhead for quantum simulation of
  {SU(3)} {Y}ang-{M}ills lattice gauge theory in the local multiplet basis},\
  }\href {https://doi.org/10.1103/PhysRevD.103.094501} {\bibfield  {journal}
  {\bibinfo  {journal} {Phys. Rev. D}\ }\textbf {\bibinfo {volume} {103}},\
  \bibinfo {pages} {094501} (\bibinfo {year} {2021})},\ \Eprint
  {https://arxiv.org/abs/2101.10227} {arXiv:2101.10227 [quant-ph]} \BibitemShut
  {NoStop}%
\bibitem [{\citenamefont {Hall}\ \emph {et~al.}(2021)\citenamefont {Hall},
  \citenamefont {Roggero}, \citenamefont {Baroni},\ and\ \citenamefont
  {Carlson}}]{Hall:2021rbv}%
  \BibitemOpen
  \bibfield  {author} {\bibinfo {author} {\bibfnamefont {B.}~\bibnamefont
  {Hall}}, \bibinfo {author} {\bibfnamefont {A.}~\bibnamefont {Roggero}},
  \bibinfo {author} {\bibfnamefont {A.}~\bibnamefont {Baroni}},\ and\ \bibinfo
  {author} {\bibfnamefont {J.}~\bibnamefont {Carlson}},\ }\bibfield  {title}
  {\bibinfo {title} {Simulation of collective neutrino oscillations on a
  quantum computer},\ }\href {https://doi.org/10.1103/PhysRevD.104.063009}
  {\bibfield  {journal} {\bibinfo  {journal} {Phys. Rev. D}\ }\textbf {\bibinfo
  {volume} {104}},\ \bibinfo {pages} {063009} (\bibinfo {year} {2021})},\
  \Eprint {https://arxiv.org/abs/2102.12556} {arXiv:2102.12556 [quant-ph]}
  \BibitemShut {NoStop}%
\bibitem [{\citenamefont {Meurice}(2021)}]{Meurice:2021pvj}%
  \BibitemOpen
  \bibfield  {author} {\bibinfo {author} {\bibfnamefont {Y.}~\bibnamefont
  {Meurice}},\ }\bibfield  {title} {\bibinfo {title} {Theoretical methods to
  design and test quantum simulators for the compact {A}belian {H}iggs model},\
  }\href {https://doi.org/10.1103/PhysRevD.104.094513} {\bibfield  {journal}
  {\bibinfo  {journal} {Phys. Rev. D}\ }\textbf {\bibinfo {volume} {104}},\
  \bibinfo {pages} {094513} (\bibinfo {year} {2021})},\ \Eprint
  {https://arxiv.org/abs/2107.11366} {arXiv:2107.11366 [quant-ph]} \BibitemShut
  {NoStop}%
\bibitem [{\citenamefont {Troyer}\ and\ \citenamefont
  {Wiese}(2005)}]{Troyer:2004ge}%
  \BibitemOpen
  \bibfield  {author} {\bibinfo {author} {\bibfnamefont {M.}~\bibnamefont
  {Troyer}}\ and\ \bibinfo {author} {\bibfnamefont {U.-J.}\ \bibnamefont
  {Wiese}},\ }\bibfield  {title} {\bibinfo {title} {Computational complexity
  and fundamental limitations to {F}ermionic quantum {M}onte {C}arlo
  simulations},\ }\href {https://doi.org/10.1103/PhysRevLett.94.170201}
  {\bibfield  {journal} {\bibinfo  {journal} {Phys. Rev. Lett.}\ }\textbf
  {\bibinfo {volume} {94}},\ \bibinfo {pages} {170201} (\bibinfo {year}
  {2005})},\ \Eprint {https://arxiv.org/abs/cond-mat/0408370}
  {arXiv:cond-mat/0408370 [cond-mat]} \BibitemShut {NoStop}%
\bibitem [{\citenamefont {Jordan}\ \emph {et~al.}(2012)\citenamefont {Jordan},
  \citenamefont {Lee},\ and\ \citenamefont {Preskill}}]{Jordan:2011ne}%
  \BibitemOpen
  \bibfield  {author} {\bibinfo {author} {\bibfnamefont {S.~P.}\ \bibnamefont
  {Jordan}}, \bibinfo {author} {\bibfnamefont {K.~S.~M.}\ \bibnamefont {Lee}},\
  and\ \bibinfo {author} {\bibfnamefont {J.}~\bibnamefont {Preskill}},\
  }\bibfield  {title} {\bibinfo {title} {Quantum algorithms for quantum field
  theories},\ }\href {https://doi.org/10.1126/science.1217069} {\bibfield
  {journal} {\bibinfo  {journal} {Science}\ }\textbf {\bibinfo {volume}
  {336}},\ \bibinfo {pages} {1130} (\bibinfo {year} {2012})},\ \Eprint
  {https://arxiv.org/abs/1111.3633} {arXiv:1111.3633 [quant-ph]} \BibitemShut
  {NoStop}%
\bibitem [{\citenamefont {Wiese}(2013)}]{Wiese:2013uua}%
  \BibitemOpen
  \bibfield  {author} {\bibinfo {author} {\bibfnamefont {U.-J.}\ \bibnamefont
  {Wiese}},\ }\bibfield  {title} {\bibinfo {title} {Ultracold quantum gases and
  lattice systems: {Q}uantum simulation of lattice gauge theories},\ }\href
  {https://doi.org/10.1002/andp.201300104} {\bibfield  {journal} {\bibinfo
  {journal} {Annalen Phys.}\ }\textbf {\bibinfo {volume} {525}},\ \bibinfo
  {pages} {777} (\bibinfo {year} {2013})},\ \Eprint
  {https://arxiv.org/abs/1305.1602} {arXiv:1305.1602 [quant-ph]} \BibitemShut
  {NoStop}%
\bibitem [{\citenamefont {Ba\~nuls}\ \emph {et~al.}(2020)\citenamefont
  {Ba\~nuls}, \citenamefont {Blatt}, \citenamefont {Catani}, \citenamefont
  {Celi}, \citenamefont {Cirac} \emph {et~al.}}]{Banuls:2019bmf}%
  \BibitemOpen
  \bibfield  {author} {\bibinfo {author} {\bibfnamefont {M.~C.}\ \bibnamefont
  {Ba\~nuls}}, \bibinfo {author} {\bibfnamefont {R.}~\bibnamefont {Blatt}},
  \bibinfo {author} {\bibfnamefont {R.}~\bibnamefont {Catani}}, \bibinfo
  {author} {\bibfnamefont {A.}~\bibnamefont {Celi}}, \bibinfo {author}
  {\bibfnamefont {J.~I.}\ \bibnamefont {Cirac}}, \emph {et~al.},\ }\bibfield
  {title} {\bibinfo {title} {Simulating lattice gauge theories within quantum
  technologies},\ }\href {https://doi.org/10.1140/epjd/e2020-100571-8}
  {\bibfield  {journal} {\bibinfo  {journal} {Eur. Phys. J. D}\ }\textbf
  {\bibinfo {volume} {74}},\ \bibinfo {pages} {165} (\bibinfo {year} {2020})},\
  \Eprint {https://arxiv.org/abs/1911.00003} {arXiv:1911.00003 [quant-ph]}
  \BibitemShut {NoStop}%
\bibitem [{\citenamefont {Bender}\ \emph {et~al.}(2020)\citenamefont {Bender},
  \citenamefont {Emonts}, \citenamefont {Zohar},\ and\ \citenamefont
  {Cirac}}]{Bender:2020jgr}%
  \BibitemOpen
  \bibfield  {author} {\bibinfo {author} {\bibfnamefont {J.}~\bibnamefont
  {Bender}}, \bibinfo {author} {\bibfnamefont {P.}~\bibnamefont {Emonts}},
  \bibinfo {author} {\bibfnamefont {E.}~\bibnamefont {Zohar}},\ and\ \bibinfo
  {author} {\bibfnamefont {J.~I.}\ \bibnamefont {Cirac}},\ }\bibfield  {title}
  {\bibinfo {title} {Real-time dynamics in {2+1d} compact {QED} using complex
  periodic {G}aussian states},\ }\href
  {https://doi.org/10.1103/PhysRevResearch.2.043145} {\bibfield  {journal}
  {\bibinfo  {journal} {Phys. Rev. Res.}\ }\textbf {\bibinfo {volume} {2}},\
  \bibinfo {pages} {043145} (\bibinfo {year} {2020})},\ \Eprint
  {https://arxiv.org/abs/2006.10038} {arXiv:2006.10038 [hep-th]} \BibitemShut
  {NoStop}%
\bibitem [{\citenamefont {Singh}\ and\ \citenamefont
  {Chandrasekharan}(2019)}]{Singh:2019uwd}%
  \BibitemOpen
  \bibfield  {author} {\bibinfo {author} {\bibfnamefont {H.}~\bibnamefont
  {Singh}}\ and\ \bibinfo {author} {\bibfnamefont {S.}~\bibnamefont
  {Chandrasekharan}},\ }\bibfield  {title} {\bibinfo {title} {Qubit
  regularization of the {$O(3)$} sigma model},\ }\href
  {https://doi.org/10.1103/PhysRevD.100.054505} {\bibfield  {journal} {\bibinfo
   {journal} {Phys. Rev.}\ }\textbf {\bibinfo {volume} {D100}},\ \bibinfo
  {pages} {054505} (\bibinfo {year} {2019})},\ \Eprint
  {https://arxiv.org/abs/1905.13204} {arXiv:1905.13204 [hep-lat]} \BibitemShut
  {NoStop}%
\bibitem [{\citenamefont {Hackett}\ \emph {et~al.}(2019)\citenamefont
  {Hackett}, \citenamefont {Howe}, \citenamefont {Hughes}, \citenamefont {Jay},
  \citenamefont {Neil},\ and\ \citenamefont {Simone}}]{Hackett:2018cel}%
  \BibitemOpen
  \bibfield  {author} {\bibinfo {author} {\bibfnamefont {D.~C.}\ \bibnamefont
  {Hackett}}, \bibinfo {author} {\bibfnamefont {K.}~\bibnamefont {Howe}},
  \bibinfo {author} {\bibfnamefont {C.}~\bibnamefont {Hughes}}, \bibinfo
  {author} {\bibfnamefont {W.}~\bibnamefont {Jay}}, \bibinfo {author}
  {\bibfnamefont {E.~T.}\ \bibnamefont {Neil}},\ and\ \bibinfo {author}
  {\bibfnamefont {J.~N.}\ \bibnamefont {Simone}},\ }\bibfield  {title}
  {\bibinfo {title} {Digitizing gauge fields: {L}attice {M}onte {C}arlo results
  for future quantum computers},\ }\href
  {https://doi.org/10.1103/PhysRevA.99.062341} {\bibfield  {journal} {\bibinfo
  {journal} {Phys. Rev. A}\ }\textbf {\bibinfo {volume} {99}},\ \bibinfo
  {pages} {062341} (\bibinfo {year} {2019})},\ \Eprint
  {https://arxiv.org/abs/1811.03629} {arXiv:1811.03629 [quant-ph]} \BibitemShut
  {NoStop}%
\bibitem [{\citenamefont {Carena}\ \emph {et~al.}(2022)\citenamefont {Carena},
  \citenamefont {Lamm}, \citenamefont {Li},\ and\ \citenamefont
  {Liu}}]{Carena:2022kpg}%
  \BibitemOpen
  \bibfield  {author} {\bibinfo {author} {\bibfnamefont {M.}~\bibnamefont
  {Carena}}, \bibinfo {author} {\bibfnamefont {H.}~\bibnamefont {Lamm}},
  \bibinfo {author} {\bibfnamefont {Y.-Y.}\ \bibnamefont {Li}},\ and\ \bibinfo
  {author} {\bibfnamefont {W.}~\bibnamefont {Liu}},\ }\bibfield  {title}
  {\bibinfo {title} {Improved {H}amiltonians for quantum simulations of gauge
  theories},\ }\href {https://doi.org/10.1103/PhysRevLett.129.051601}
  {\bibfield  {journal} {\bibinfo  {journal} {Phys. Rev. Lett.}\ }\textbf
  {\bibinfo {volume} {129}},\ \bibinfo {pages} {051601} (\bibinfo {year}
  {2022})},\ \Eprint {https://arxiv.org/abs/2203.02823} {arXiv:2203.02823
  [hep-lat]} \BibitemShut {NoStop}%
\bibitem [{\citenamefont {Ji}\ \emph {et~al.}(2022)\citenamefont {Ji},
  \citenamefont {Lamm},\ and\ \citenamefont {Zhu}}]{Ji:2022qvr}%
  \BibitemOpen
  \bibfield  {author} {\bibinfo {author} {\bibfnamefont {Y.}~\bibnamefont
  {Ji}}, \bibinfo {author} {\bibfnamefont {H.}~\bibnamefont {Lamm}},\ and\
  \bibinfo {author} {\bibfnamefont {S.}~\bibnamefont {Zhu}},\ }\href@noop {}
  {\bibinfo {title} {Gluon digitization via character expansion for quantum
  computers}} (\bibinfo {year} {2022}),\ \Eprint
  {https://arxiv.org/abs/2203.02330} {arXiv:2203.02330 [hep-lat]} \BibitemShut
  {NoStop}%
\bibitem [{\citenamefont {Liu}\ and\ \citenamefont
  {Chandrasekharan}(2022{\natexlab{a}})}]{Liu:2021tef}%
  \BibitemOpen
  \bibfield  {author} {\bibinfo {author} {\bibfnamefont {H.}~\bibnamefont
  {Liu}}\ and\ \bibinfo {author} {\bibfnamefont {S.}~\bibnamefont
  {Chandrasekharan}},\ }\bibfield  {title} {\bibinfo {title} {Qubit
  regularization and qubit embedding algebras},\ }\href
  {https://doi.org/10.3390/sym14020305} {\bibfield  {journal} {\bibinfo
  {journal} {Symmetry}\ }\textbf {\bibinfo {volume} {14}},\ \bibinfo {pages}
  {305} (\bibinfo {year} {2022}{\natexlab{a}})},\ \Eprint
  {https://arxiv.org/abs/2112.02090} {arXiv:2112.02090 [hep-lat]} \BibitemShut
  {NoStop}%
\bibitem [{\citenamefont {Chandrasekharan}\ and\ \citenamefont
  {Wiese}(1997)}]{Chandrasekharan:1996ih}%
  \BibitemOpen
  \bibfield  {author} {\bibinfo {author} {\bibfnamefont {S.}~\bibnamefont
  {Chandrasekharan}}\ and\ \bibinfo {author} {\bibfnamefont {U.~J.}\
  \bibnamefont {Wiese}},\ }\bibfield  {title} {\bibinfo {title} {Quantum link
  models: {A} discrete approach to gauge theories},\ }\href
  {https://doi.org/10.1016/S0550-3213(97)00006-0} {\bibfield  {journal}
  {\bibinfo  {journal} {Nucl. Phys. B}\ }\textbf {\bibinfo {volume} {492}},\
  \bibinfo {pages} {455} (\bibinfo {year} {1997})},\ \Eprint
  {https://arxiv.org/abs/hep-lat/9609042} {arXiv:hep-lat/9609042} \BibitemShut
  {NoStop}%
\bibitem [{\citenamefont {Brower}\ \emph {et~al.}(1999)\citenamefont {Brower},
  \citenamefont {Chandrasekharan},\ and\ \citenamefont
  {Wiese}}]{PhysRevD.60.094502}%
  \BibitemOpen
  \bibfield  {author} {\bibinfo {author} {\bibfnamefont {R.}~\bibnamefont
  {Brower}}, \bibinfo {author} {\bibfnamefont {S.}~\bibnamefont
  {Chandrasekharan}},\ and\ \bibinfo {author} {\bibfnamefont {U.-J.}\
  \bibnamefont {Wiese}},\ }\bibfield  {title} {\bibinfo {title} {{QCD} as a
  quantum link model},\ }\href {https://doi.org/10.1103/PhysRevD.60.094502}
  {\bibfield  {journal} {\bibinfo  {journal} {Phys. Rev. D}\ }\textbf {\bibinfo
  {volume} {60}},\ \bibinfo {pages} {094502} (\bibinfo {year}
  {1999})}\BibitemShut {NoStop}%
\bibitem [{\citenamefont {Beard}\ \emph {et~al.}(2005)\citenamefont {Beard},
  \citenamefont {Pepe}, \citenamefont {Riederer},\ and\ \citenamefont
  {Wiese}}]{Beard:2004jr}%
  \BibitemOpen
  \bibfield  {author} {\bibinfo {author} {\bibfnamefont {B.}~\bibnamefont
  {Beard}}, \bibinfo {author} {\bibfnamefont {M.}~\bibnamefont {Pepe}},
  \bibinfo {author} {\bibfnamefont {S.}~\bibnamefont {Riederer}},\ and\
  \bibinfo {author} {\bibfnamefont {U.}~\bibnamefont {Wiese}},\ }\bibfield
  {title} {\bibinfo {title} {Study of {CP(N-1)} theta-vacua by
  cluster-simulation of {SU(N)} quantum spin ladders},\ }\href
  {https://doi.org/10.1103/PhysRevLett.94.010603} {\bibfield  {journal}
  {\bibinfo  {journal} {Phys. Rev. Lett.}\ }\textbf {\bibinfo {volume} {94}},\
  \bibinfo {pages} {010603} (\bibinfo {year} {2005})},\ \Eprint
  {https://arxiv.org/abs/hep-lat/0406040} {arXiv:hep-lat/0406040} \BibitemShut
  {NoStop}%
\bibitem [{\citenamefont {Brower}\ \emph {et~al.}(2004)\citenamefont {Brower},
  \citenamefont {Chandrasekharan}, \citenamefont {Riederer},\ and\
  \citenamefont {Wiese}}]{Brower:2003vy}%
  \BibitemOpen
  \bibfield  {author} {\bibinfo {author} {\bibfnamefont {R.}~\bibnamefont
  {Brower}}, \bibinfo {author} {\bibfnamefont {S.}~\bibnamefont
  {Chandrasekharan}}, \bibinfo {author} {\bibfnamefont {S.}~\bibnamefont
  {Riederer}},\ and\ \bibinfo {author} {\bibfnamefont {U.}~\bibnamefont
  {Wiese}},\ }\bibfield  {title} {\bibinfo {title} {{D} theory: {F}ield
  quantization by dimensional reduction of discrete variables},\ }\href
  {https://doi.org/10.1016/j.nuclphysb.2004.06.007} {\bibfield  {journal}
  {\bibinfo  {journal} {Nucl. Phys. B}\ }\textbf {\bibinfo {volume} {693}},\
  \bibinfo {pages} {149} (\bibinfo {year} {2004})},\ \Eprint
  {https://arxiv.org/abs/hep-lat/0309182} {arXiv:hep-lat/0309182} \BibitemShut
  {NoStop}%
\bibitem [{\citenamefont {Wiese}(2006)}]{Wiese:2006kp}%
  \BibitemOpen
  \bibfield  {author} {\bibinfo {author} {\bibfnamefont {U.~J.}\ \bibnamefont
  {Wiese}},\ }\bibfield  {title} {\bibinfo {title} {{D}-theory: A quest for
  nature's regularization},\ }\bibfield  {booktitle} {\emph {\bibinfo
  {booktitle} {{Hadron physics, proceedings of the Workshop on Computational
  Hadron Physics, University of Cyprus, Nicosia, Cyprus, 14-17 September
  2005}}},\ }\href {https://doi.org/10.1016/j.nuclphysbps.2006.01.027}
  {\bibfield  {journal} {\bibinfo  {journal} {Nucl. Phys. Proc. Suppl.}\
  }\textbf {\bibinfo {volume} {153}},\ \bibinfo {pages} {336} (\bibinfo {year}
  {2006})}\BibitemShut {NoStop}%
\bibitem [{\citenamefont {Liu}\ and\ \citenamefont
  {Chandrasekharan}(2022{\natexlab{b}})}]{Liu2022}%
  \BibitemOpen
  \bibfield  {author} {\bibinfo {author} {\bibfnamefont {H.}~\bibnamefont
  {Liu}}\ and\ \bibinfo {author} {\bibfnamefont {S.}~\bibnamefont
  {Chandrasekharan}},\ }\bibfield  {title} {\bibinfo {title} {Qubit
  regularization and qubit embedding algebras},\ }\bibfield  {journal}
  {\bibinfo  {journal} {Symmetry}\ }\textbf {\bibinfo {volume} {14}},\ \href
  {https://doi.org/10.3390/sym14020305} {10.3390/sym14020305} (\bibinfo {year}
  {2022}{\natexlab{b}})\BibitemShut {NoStop}%
\bibitem [{\citenamefont {Affleck}\ \emph {et~al.}(1987)\citenamefont
  {Affleck}, \citenamefont {Kennedy}, \citenamefont {Lieb},\ and\ \citenamefont
  {Tasaki}}]{PhysRevLett.59.799}%
  \BibitemOpen
  \bibfield  {author} {\bibinfo {author} {\bibfnamefont {I.}~\bibnamefont
  {Affleck}}, \bibinfo {author} {\bibfnamefont {T.}~\bibnamefont {Kennedy}},
  \bibinfo {author} {\bibfnamefont {E.~H.}\ \bibnamefont {Lieb}},\ and\
  \bibinfo {author} {\bibfnamefont {H.}~\bibnamefont {Tasaki}},\ }\bibfield
  {title} {\bibinfo {title} {Rigorous results on valence-bond ground states in
  antiferromagnets},\ }\href {https://doi.org/10.1103/PhysRevLett.59.799}
  {\bibfield  {journal} {\bibinfo  {journal} {Phys. Rev. Lett.}\ }\textbf
  {\bibinfo {volume} {59}},\ \bibinfo {pages} {799} (\bibinfo {year}
  {1987})}\BibitemShut {NoStop}%
\bibitem [{\citenamefont {Affleck}\ and\ \citenamefont
  {Haldane}(1987)}]{Affleck:1987ch}%
  \BibitemOpen
  \bibfield  {author} {\bibinfo {author} {\bibfnamefont {I.}~\bibnamefont
  {Affleck}}\ and\ \bibinfo {author} {\bibfnamefont {F.~D.~M.}\ \bibnamefont
  {Haldane}},\ }\bibfield  {title} {\bibinfo {title} {{Critical Theory of
  Quantum Spin Chains}},\ }\href {https://doi.org/10.1103/PhysRevB.36.5291}
  {\bibfield  {journal} {\bibinfo  {journal} {Phys. Rev.}\ }\textbf {\bibinfo
  {volume} {B36}},\ \bibinfo {pages} {5291} (\bibinfo {year}
  {1987})}\BibitemShut {NoStop}%
\bibitem [{\citenamefont {Itoi}\ and\ \citenamefont
  {Kato}(1997)}]{PhysRevB.55.8295}%
  \BibitemOpen
  \bibfield  {author} {\bibinfo {author} {\bibfnamefont {C.}~\bibnamefont
  {Itoi}}\ and\ \bibinfo {author} {\bibfnamefont {M.-H.}\ \bibnamefont
  {Kato}},\ }\bibfield  {title} {\bibinfo {title} {Extended massless phase and
  the haldane phase in a spin-1 isotropic antiferromagnetic chain},\ }\href
  {https://doi.org/10.1103/PhysRevB.55.8295} {\bibfield  {journal} {\bibinfo
  {journal} {Phys. Rev. B}\ }\textbf {\bibinfo {volume} {55}},\ \bibinfo
  {pages} {8295} (\bibinfo {year} {1997})}\BibitemShut {NoStop}%
\bibitem [{\citenamefont {Binder}\ and\ \citenamefont
  {Barthel}(2020)}]{PhysRevB.102.014447}%
  \BibitemOpen
  \bibfield  {author} {\bibinfo {author} {\bibfnamefont {M.}~\bibnamefont
  {Binder}}\ and\ \bibinfo {author} {\bibfnamefont {T.}~\bibnamefont
  {Barthel}},\ }\bibfield  {title} {\bibinfo {title} {Low-energy physics of
  isotropic spin-1 chains in the critical and haldane phases},\ }\href
  {https://doi.org/10.1103/PhysRevB.102.014447} {\bibfield  {journal} {\bibinfo
   {journal} {Phys. Rev. B}\ }\textbf {\bibinfo {volume} {102}},\ \bibinfo
  {pages} {014447} (\bibinfo {year} {2020})}\BibitemShut {NoStop}%
\bibitem [{\citenamefont {Bhattacharya}\ \emph {et~al.}(2021)\citenamefont
  {Bhattacharya}, \citenamefont {Buser}, \citenamefont {Chandrasekharan},
  \citenamefont {Gupta},\ and\ \citenamefont {Singh}}]{Bhattacharya:2020gpm}%
  \BibitemOpen
  \bibfield  {author} {\bibinfo {author} {\bibfnamefont {T.}~\bibnamefont
  {Bhattacharya}}, \bibinfo {author} {\bibfnamefont {A.~J.}\ \bibnamefont
  {Buser}}, \bibinfo {author} {\bibfnamefont {S.}~\bibnamefont
  {Chandrasekharan}}, \bibinfo {author} {\bibfnamefont {R.}~\bibnamefont
  {Gupta}},\ and\ \bibinfo {author} {\bibfnamefont {H.}~\bibnamefont {Singh}},\
  }\bibfield  {title} {\bibinfo {title} {{Qubit regularization of asymptotic
  freedom}},\ }\href {https://doi.org/10.1103/PhysRevLett.126.172001}
  {\bibfield  {journal} {\bibinfo  {journal} {Phys. Rev. Lett.}\ }\textbf
  {\bibinfo {volume} {126}},\ \bibinfo {pages} {172001} (\bibinfo {year}
  {2021})},\ \Eprint {https://arxiv.org/abs/2012.02153} {arXiv:2012.02153
  [hep-lat]} \BibitemShut {NoStop}%
\bibitem [{\citenamefont {Wu}\ and\ \citenamefont {Yang}(1976)}]{Wu:1976ge}%
  \BibitemOpen
  \bibfield  {author} {\bibinfo {author} {\bibfnamefont {T.~T.}\ \bibnamefont
  {Wu}}\ and\ \bibinfo {author} {\bibfnamefont {C.~N.}\ \bibnamefont {Yang}},\
  }\bibfield  {title} {\bibinfo {title} {Dirac monopole without strings:
  {M}onopole harmonics},\ }\href {https://doi.org/10.1016/0550-3213(76)90143-7}
  {\bibfield  {journal} {\bibinfo  {journal} {Nucl. Phys. B}\ }\textbf
  {\bibinfo {volume} {107}},\ \bibinfo {pages} {365} (\bibinfo {year}
  {1976})}\BibitemShut {NoStop}%
\bibitem [{\citenamefont {Haldane}(1983)}]{Haldane:1982rj}%
  \BibitemOpen
  \bibfield  {author} {\bibinfo {author} {\bibfnamefont {F.}~\bibnamefont
  {Haldane}},\ }\bibfield  {title} {\bibinfo {title} {{Continuum dynamics of
  the 1-D Heisenberg antiferromagnetic identification with the O(3) nonlinear
  sigma model}},\ }\href {https://doi.org/10.1016/0375-9601(83)90631-X}
  {\bibfield  {journal} {\bibinfo  {journal} {Phys. Lett. A}\ }\textbf
  {\bibinfo {volume} {93}},\ \bibinfo {pages} {464} (\bibinfo {year}
  {1983})}\BibitemShut {NoStop}%
\bibitem [{\citenamefont {Sierra}(1996)}]{sierra_nonlinear_1996}%
  \BibitemOpen
  \bibfield  {author} {\bibinfo {author} {\bibfnamefont {G.}~\bibnamefont
  {Sierra}},\ }\bibfield  {title} {\bibinfo {title} {The nonlinear sigma model
  and spin ladders},\ }\href {https://doi.org/10.1088/0305-4470/29/12/032}
  {\bibfield  {journal} {\bibinfo  {journal} {Journal of Physics A:
  Mathematical and General}\ }\textbf {\bibinfo {volume} {29}},\ \bibinfo
  {pages} {3299} (\bibinfo {year} {1996})}\BibitemShut {NoStop}%
\bibitem [{\citenamefont {Sierra}(1997)}]{sierra_application_1997}%
  \BibitemOpen
  \bibfield  {author} {\bibinfo {author} {\bibfnamefont {G.}~\bibnamefont
  {Sierra}},\ }\bibfield  {title} {\bibinfo {title} {On the application of the
  nonlinear sigma model to spin chains and spin ladders},\ }\href
  {https://doi.org/10.1007/BFb0104637} {\bibfield  {journal} {\bibinfo
  {journal} {Lect.Notes Phys.}\ }\textbf {\bibinfo {volume} {478}},\ \bibinfo
  {pages} {137} (\bibinfo {year} {1997})}\BibitemShut {NoStop}%
\bibitem [{\citenamefont {{Mart{\'i}n-Delgado}}\ \emph
  {et~al.}(1996)\citenamefont {{Mart{\'i}n-Delgado}}, \citenamefont {Shankar},\
  and\ \citenamefont {Sierra}}]{martin-delgado_phase_1996}%
  \BibitemOpen
  \bibfield  {author} {\bibinfo {author} {\bibfnamefont {M.~A.}\ \bibnamefont
  {{Mart{\'i}n-Delgado}}}, \bibinfo {author} {\bibfnamefont {R.}~\bibnamefont
  {Shankar}},\ and\ \bibinfo {author} {\bibfnamefont {G.}~\bibnamefont
  {Sierra}},\ }\bibfield  {title} {\bibinfo {title} {Phase {{Transitions}} in
  {{Staggered Spin Ladders}}},\ }\href
  {https://doi.org/10.1103/PhysRevLett.77.3443} {\bibfield  {journal} {\bibinfo
   {journal} {Physical Review Letters}\ }\textbf {\bibinfo {volume} {77}},\
  \bibinfo {pages} {3443} (\bibinfo {year} {1996})}\BibitemShut {NoStop}%
\bibitem [{\citenamefont {Caspar}\ and\ \citenamefont
  {Singh}(2022)}]{PhysRevLett.129.022003}%
  \BibitemOpen
  \bibfield  {author} {\bibinfo {author} {\bibfnamefont {S.}~\bibnamefont
  {Caspar}}\ and\ \bibinfo {author} {\bibfnamefont {H.}~\bibnamefont {Singh}},\
  }\bibfield  {title} {\bibinfo {title} {From asymptotic freedom to
  {$\ensuremath{\theta}$} vacua: Qubit embeddings of the {O(3)} nonlinear
  {$\ensuremath{\sigma}$} model},\ }\href
  {https://doi.org/10.1103/PhysRevLett.129.022003} {\bibfield  {journal}
  {\bibinfo  {journal} {Phys. Rev. Lett.}\ }\textbf {\bibinfo {volume} {129}},\
  \bibinfo {pages} {022003} (\bibinfo {year} {2022})}\BibitemShut {NoStop}%
\bibitem [{\citenamefont {Nguyen}\ and\ \citenamefont
  {Singh}(2023)}]{nguyen_lattice_2023a}%
  \BibitemOpen
  \bibfield  {author} {\bibinfo {author} {\bibfnamefont {M.}~\bibnamefont
  {Nguyen}}\ and\ \bibinfo {author} {\bibfnamefont {H.}~\bibnamefont {Singh}},\
  }\bibfield  {title} {\bibinfo {title} {Lattice regularizations of
  \$\textbackslash ensuremath\{\textbackslash theta\}\$ vacua: {{Anomalies}}
  and qubit models},\ }\href {https://doi.org/10.1103/PhysRevD.107.014507}
  {\bibfield  {journal} {\bibinfo  {journal} {Physical Review D}\ }\textbf
  {\bibinfo {volume} {107}},\ \bibinfo {pages} {014507} (\bibinfo {year}
  {2023})}\BibitemShut {NoStop}%
\bibitem [{\citenamefont {Lieb}\ \emph {et~al.}(1961)\citenamefont {Lieb},
  \citenamefont {Schultz},\ and\ \citenamefont {Mattis}}]{LIEB1961407}%
  \BibitemOpen
  \bibfield  {author} {\bibinfo {author} {\bibfnamefont {E.}~\bibnamefont
  {Lieb}}, \bibinfo {author} {\bibfnamefont {T.}~\bibnamefont {Schultz}},\ and\
  \bibinfo {author} {\bibfnamefont {D.}~\bibnamefont {Mattis}},\ }\bibfield
  {title} {\bibinfo {title} {Two soluble models of an antiferromagnetic
  chain},\ }\href
  {https://doi.org/https://doi.org/10.1016/0003-4916(61)90115-4} {\bibfield
  {journal} {\bibinfo  {journal} {Annals of Physics}\ }\textbf {\bibinfo
  {volume} {16}},\ \bibinfo {pages} {407} (\bibinfo {year} {1961})}\BibitemShut
  {NoStop}%
\bibitem [{\citenamefont {Eggert}(1996)}]{PhysRevB.54.R9612}%
  \BibitemOpen
  \bibfield  {author} {\bibinfo {author} {\bibfnamefont {S.}~\bibnamefont
  {Eggert}},\ }\bibfield  {title} {\bibinfo {title} {Numerical evidence for
  multiplicative logarithmic corrections from marginal operators},\ }\href
  {https://doi.org/10.1103/PhysRevB.54.R9612} {\bibfield  {journal} {\bibinfo
  {journal} {Phys. Rev. B}\ }\textbf {\bibinfo {volume} {54}},\ \bibinfo
  {pages} {R9612} (\bibinfo {year} {1996})}\BibitemShut {NoStop}%
\bibitem [{\citenamefont {Liu}\ \emph {et~al.}(2021)\citenamefont {Liu},
  \citenamefont {Chandrasekharan},\ and\ \citenamefont {Kaul}}]{Liu:2020ygc}%
  \BibitemOpen
  \bibfield  {author} {\bibinfo {author} {\bibfnamefont {H.}~\bibnamefont
  {Liu}}, \bibinfo {author} {\bibfnamefont {S.}~\bibnamefont
  {Chandrasekharan}},\ and\ \bibinfo {author} {\bibfnamefont {R.~K.}\
  \bibnamefont {Kaul}},\ }\bibfield  {title} {\bibinfo {title} {{H}amiltonian
  models of lattice {F}ermions solvable by the meron-cluster algorithm},\
  }\href {https://doi.org/10.1103/PhysRevD.103.054033} {\bibfield  {journal}
  {\bibinfo  {journal} {Phys. Rev. D}\ }\textbf {\bibinfo {volume} {103}},\
  \bibinfo {pages} {054033} (\bibinfo {year} {2021})},\ \Eprint
  {https://arxiv.org/abs/2011.13208} {arXiv:2011.13208 [hep-lat]} \BibitemShut
  {NoStop}%
\bibitem [{\citenamefont {Shankar}\ and\ \citenamefont
  {Read}(1990)}]{Shankar:1989ee}%
  \BibitemOpen
  \bibfield  {author} {\bibinfo {author} {\bibfnamefont {R.}~\bibnamefont
  {Shankar}}\ and\ \bibinfo {author} {\bibfnamefont {N.}~\bibnamefont {Read}},\
  }\bibfield  {title} {\bibinfo {title} {{The $\theta = \pi$ Nonlinear $\sigma$
  Model Is Massless}},\ }\href {https://doi.org/10.1016/0550-3213(90)90437-I}
  {\bibfield  {journal} {\bibinfo  {journal} {Nucl. Phys. B}\ }\textbf
  {\bibinfo {volume} {336}},\ \bibinfo {pages} {457} (\bibinfo {year}
  {1990})}\BibitemShut {NoStop}%
\bibitem [{\citenamefont {Adams}\ and\ \citenamefont
  {Chandrasekharan}(2003)}]{Adams:2003cca}%
  \BibitemOpen
  \bibfield  {author} {\bibinfo {author} {\bibfnamefont {D.~H.}\ \bibnamefont
  {Adams}}\ and\ \bibinfo {author} {\bibfnamefont {S.}~\bibnamefont
  {Chandrasekharan}},\ }\bibfield  {title} {\bibinfo {title} {{Chiral limit of
  strongly coupled lattice gauge theories}},\ }\href
  {https://doi.org/10.1016/S0550-3213(03)00350-X} {\bibfield  {journal}
  {\bibinfo  {journal} {Nucl. Phys. B}\ }\textbf {\bibinfo {volume} {662}},\
  \bibinfo {pages} {220} (\bibinfo {year} {2003})},\ \Eprint
  {https://arxiv.org/abs/hep-lat/0303003} {arXiv:hep-lat/0303003} \BibitemShut
  {NoStop}%
\bibitem [{\citenamefont {Chandrasekharan}\ and\ \citenamefont
  {Jiang}(2006)}]{Chandrasekharan:2006tz}%
  \BibitemOpen
  \bibfield  {author} {\bibinfo {author} {\bibfnamefont {S.}~\bibnamefont
  {Chandrasekharan}}\ and\ \bibinfo {author} {\bibfnamefont {F.-J.}\
  \bibnamefont {Jiang}},\ }\bibfield  {title} {\bibinfo {title} {{Phase-diagram
  of two-color lattice QCD in the chiral limit}},\ }\href
  {https://doi.org/10.1103/PhysRevD.74.014506} {\bibfield  {journal} {\bibinfo
  {journal} {Phys. Rev.}\ }\textbf {\bibinfo {volume} {D74}},\ \bibinfo {pages}
  {014506} (\bibinfo {year} {2006})},\ \Eprint
  {https://arxiv.org/abs/hep-lat/0602031} {arXiv:hep-lat/0602031 [hep-lat]}
  \BibitemShut {NoStop}%
\bibitem [{\citenamefont {Caracciolo}\ \emph {et~al.}(1993)\citenamefont
  {Caracciolo}, \citenamefont {Edwards}, \citenamefont {Pelissetto},\ and\
  \citenamefont {Sokal}}]{Caracciolo:1992nh}%
  \BibitemOpen
  \bibfield  {author} {\bibinfo {author} {\bibfnamefont {S.}~\bibnamefont
  {Caracciolo}}, \bibinfo {author} {\bibfnamefont {R.~G.}\ \bibnamefont
  {Edwards}}, \bibinfo {author} {\bibfnamefont {A.}~\bibnamefont
  {Pelissetto}},\ and\ \bibinfo {author} {\bibfnamefont {A.~D.}\ \bibnamefont
  {Sokal}},\ }\bibfield  {title} {\bibinfo {title} {{Wolff type embedding
  algorithms for general nonlinear sigma models}},\ }\href
  {https://doi.org/10.1016/0550-3213(93)90044-P} {\bibfield  {journal}
  {\bibinfo  {journal} {Nucl. Phys. B}\ }\textbf {\bibinfo {volume} {403}},\
  \bibinfo {pages} {475} (\bibinfo {year} {1993})},\ \Eprint
  {https://arxiv.org/abs/hep-lat/9205005} {arXiv:hep-lat/9205005} \BibitemShut
  {NoStop}%
\bibitem [{\citenamefont {Carena}\ \emph {et~al.}(2021)\citenamefont {Carena},
  \citenamefont {Lamm}, \citenamefont {Li},\ and\ \citenamefont
  {Liu}}]{Carena:2021ltu}%
  \BibitemOpen
  \bibfield  {author} {\bibinfo {author} {\bibfnamefont {M.}~\bibnamefont
  {Carena}}, \bibinfo {author} {\bibfnamefont {H.}~\bibnamefont {Lamm}},
  \bibinfo {author} {\bibfnamefont {Y.-Y.}\ \bibnamefont {Li}},\ and\ \bibinfo
  {author} {\bibfnamefont {W.}~\bibnamefont {Liu}},\ }\bibfield  {title}
  {\bibinfo {title} {Lattice renormalization of quantum simulations},\ }\href
  {https://doi.org/10.1103/PhysRevD.104.094519} {\bibfield  {journal} {\bibinfo
   {journal} {Phys. Rev. D}\ }\textbf {\bibinfo {volume} {104}},\ \bibinfo
  {pages} {094519} (\bibinfo {year} {2021})},\ \Eprint
  {https://arxiv.org/abs/2107.01166} {arXiv:2107.01166 [hep-lat]} \BibitemShut
  {NoStop}%
\bibitem [{\citenamefont {{\c{S}}ahino{\u{g}}lu}\ and\ \citenamefont
  {Somma}(2021)}]{_ahino_lu_2021}%
  \BibitemOpen
  \bibfield  {author} {\bibinfo {author} {\bibfnamefont {B.}~\bibnamefont
  {{\c{S}}ahino{\u{g}}lu}}\ and\ \bibinfo {author} {\bibfnamefont {R.~D.}\
  \bibnamefont {Somma}},\ }\bibfield  {title} {\bibinfo {title} {Hamiltonian
  simulation in the low-energy subspace},\ }\bibfield  {journal} {\bibinfo
  {journal} {npj Quantum Information}\ }\textbf {\bibinfo {volume} {7}},\ \href
  {https://doi.org/10.1038/s41534-021-00451-w} {10.1038/s41534-021-00451-w}
  (\bibinfo {year} {2021})\BibitemShut {NoStop}%
\bibitem [{\citenamefont {Gu}\ \emph {et~al.}(2021)\citenamefont {Gu},
  \citenamefont {Somma},\ and\ \citenamefont {\c{S}ahino\u{g}lu}}]{Gu:2021hyo}%
  \BibitemOpen
  \bibfield  {author} {\bibinfo {author} {\bibfnamefont {S.}~\bibnamefont
  {Gu}}, \bibinfo {author} {\bibfnamefont {R.~D.}\ \bibnamefont {Somma}},\ and\
  \bibinfo {author} {\bibfnamefont {B.}~\bibnamefont {\c{S}ahino\u{g}lu}},\
  }\bibfield  {title} {\bibinfo {title} {{Fast-forwarding quantum evolution}},\
  }\href {https://doi.org/10.22331/q-2021-11-15-577} {\bibfield  {journal}
  {\bibinfo  {journal} {Quantum}\ }\textbf {\bibinfo {volume} {5}},\ \bibinfo
  {pages} {577} (\bibinfo {year} {2021})},\ \Eprint
  {https://arxiv.org/abs/2105.07304} {arXiv:2105.07304 [quant-ph]} \BibitemShut
  {NoStop}%
\end{thebibliography}%
\onecolumngrid
\clearpage
\twocolumngrid
\appendix

\section{Exact calculations on small lattices}
\label{app:exact}

\begin{figure}[hb]
\begin{center}
\includegraphics[width=0.4\textwidth]{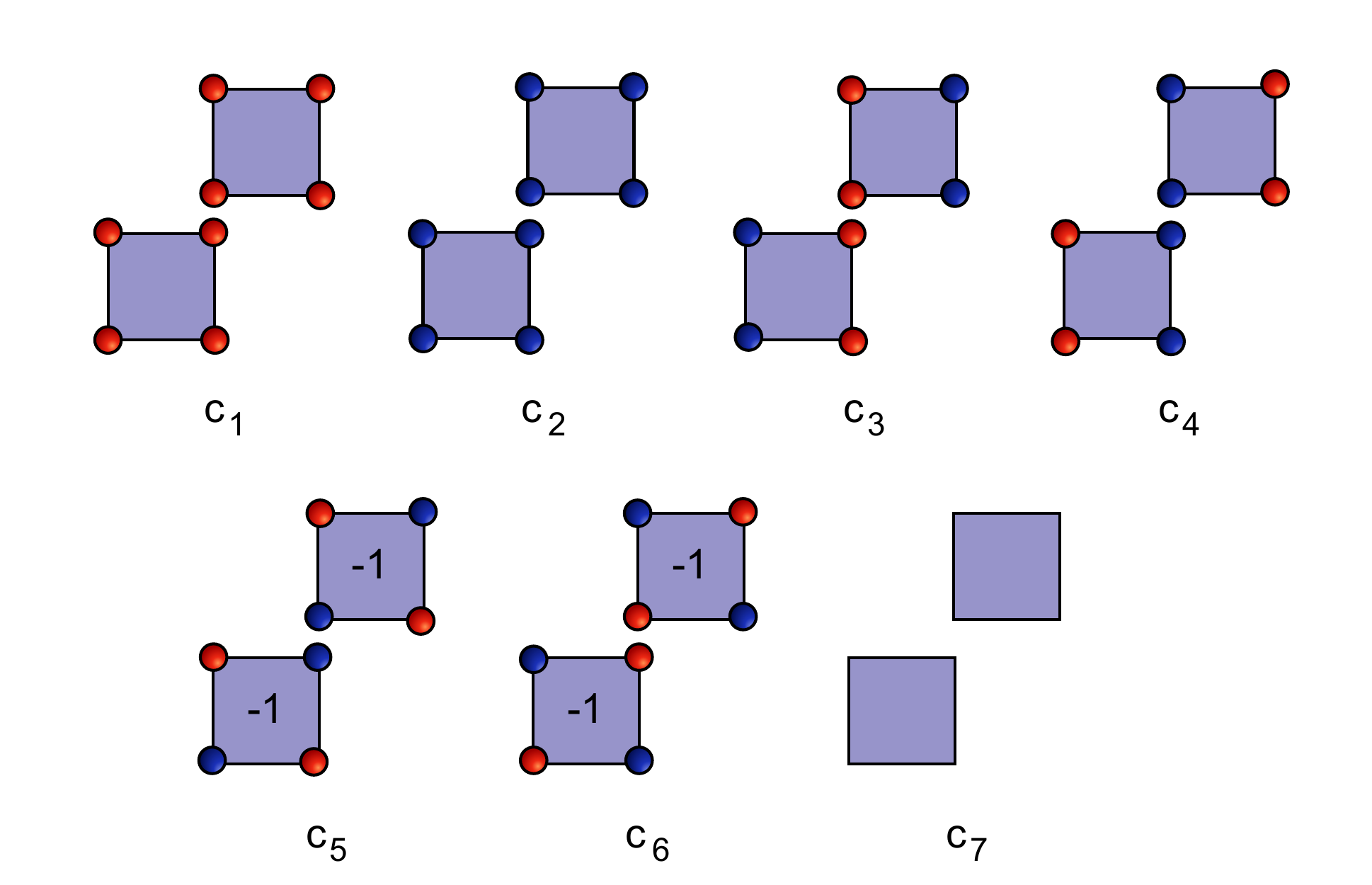}
\vskip0.1in
\includegraphics[width=0.4\textwidth]{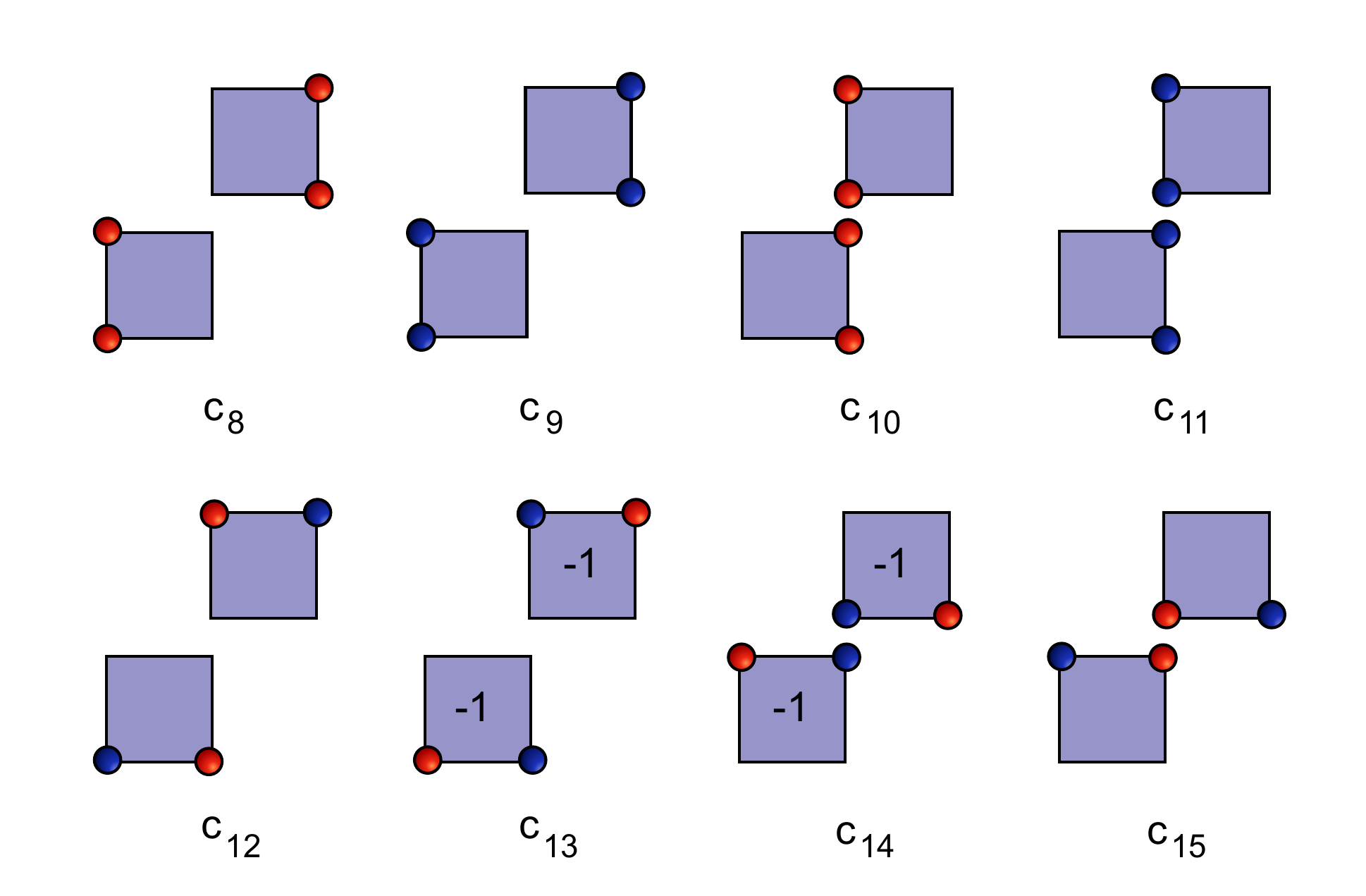}
\end{center}
\caption{\label{fig:2x2Zconf} Spin configurations that contribute to the partition function on a $2\times 2$ periodic lattice. Spatial direction is horizontal and temporal direction is vertical. Using the rules for computing the magnitude of the Boltzmann weight of a configuration we obtain $\Wc(c_1) = \Wc(c_2) = \Wc(c_5) = \Wc(c_6) = 1$, $\Wc(c_3) = \Wc(c_4) = 4$, $\Wc(c_7) = U^4$ and all the remaining eight configurations have weight $U^2$. The sign factors that appear in plaquettes when $\ket{\ua}$ or $\bra{\ua}$ is on an even site are also shown. Note that the sign factors always come in pairs as explained in the text.}
\end{figure}

Here we give some exact results for our observables on a $2\times 2$ lattice and a $2\times 4$ lattice and compare them with our Monte Carlo calculations at a few values of the coupling. The goal here is not only to verify our Monte Carlo method, but also to clarify the definition of the observables defined in the main text. On a $2\times 2$ lattice it is easy to enumerate all configurations $c$ that contribute to the partition function $Z$. There are fifteen configurations which are explicitly shown in \cref{fig:2x2Zconf}.

\begin{figure}[ht]
\begin{center}
\includegraphics[width=0.28\textwidth]{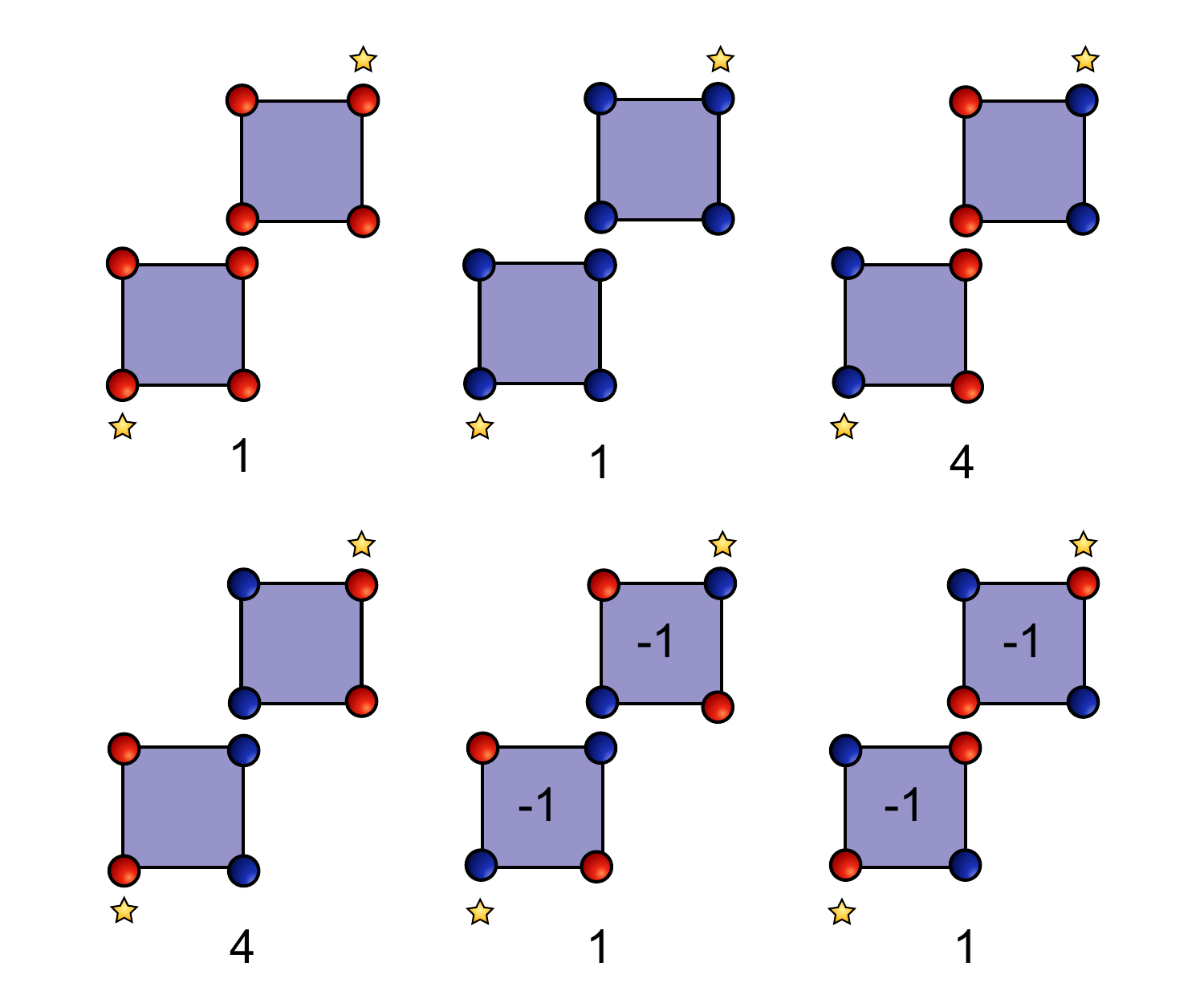}
\includegraphics[width=0.35\textwidth]{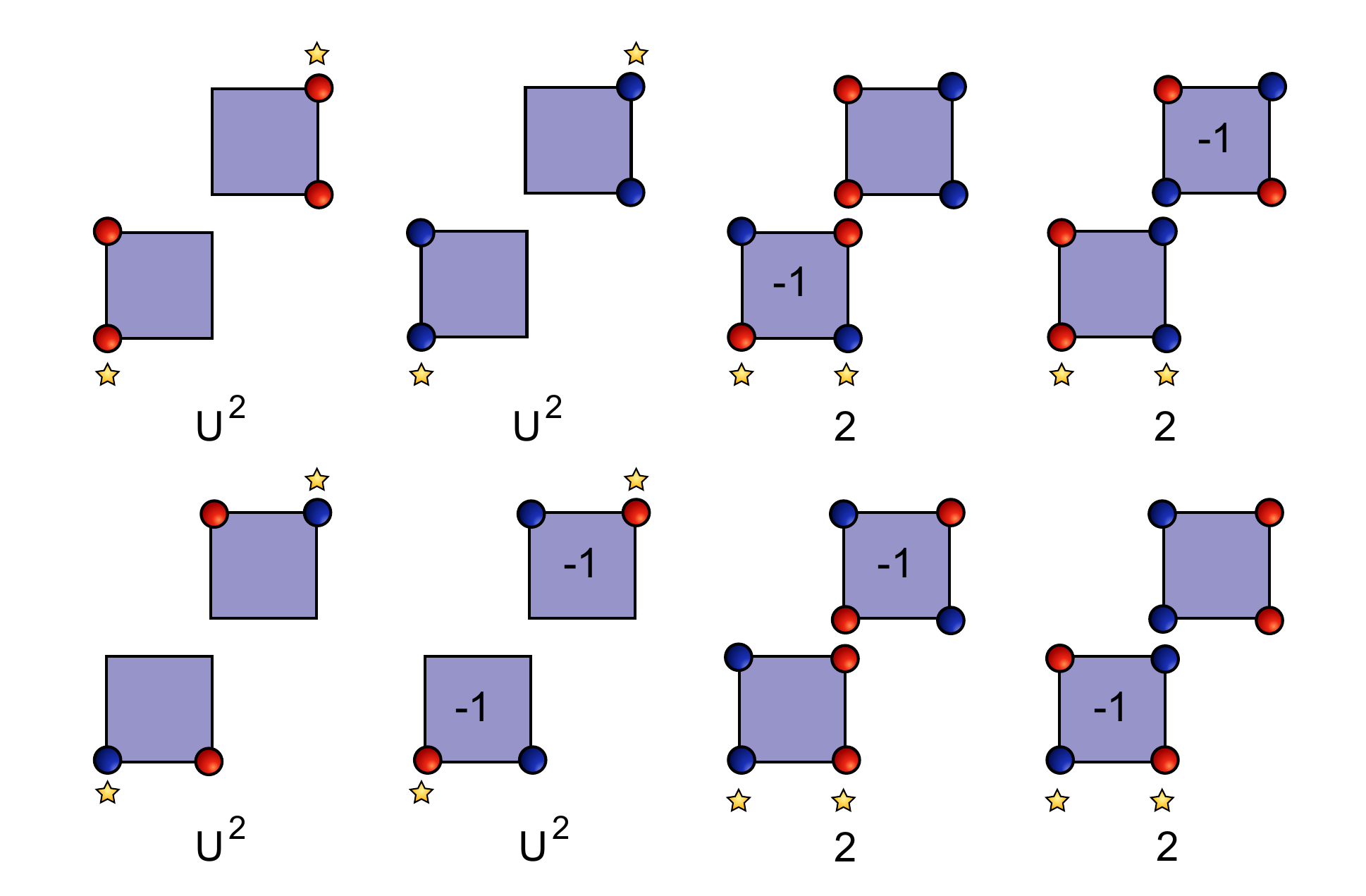}
\includegraphics[width=0.35\textwidth]{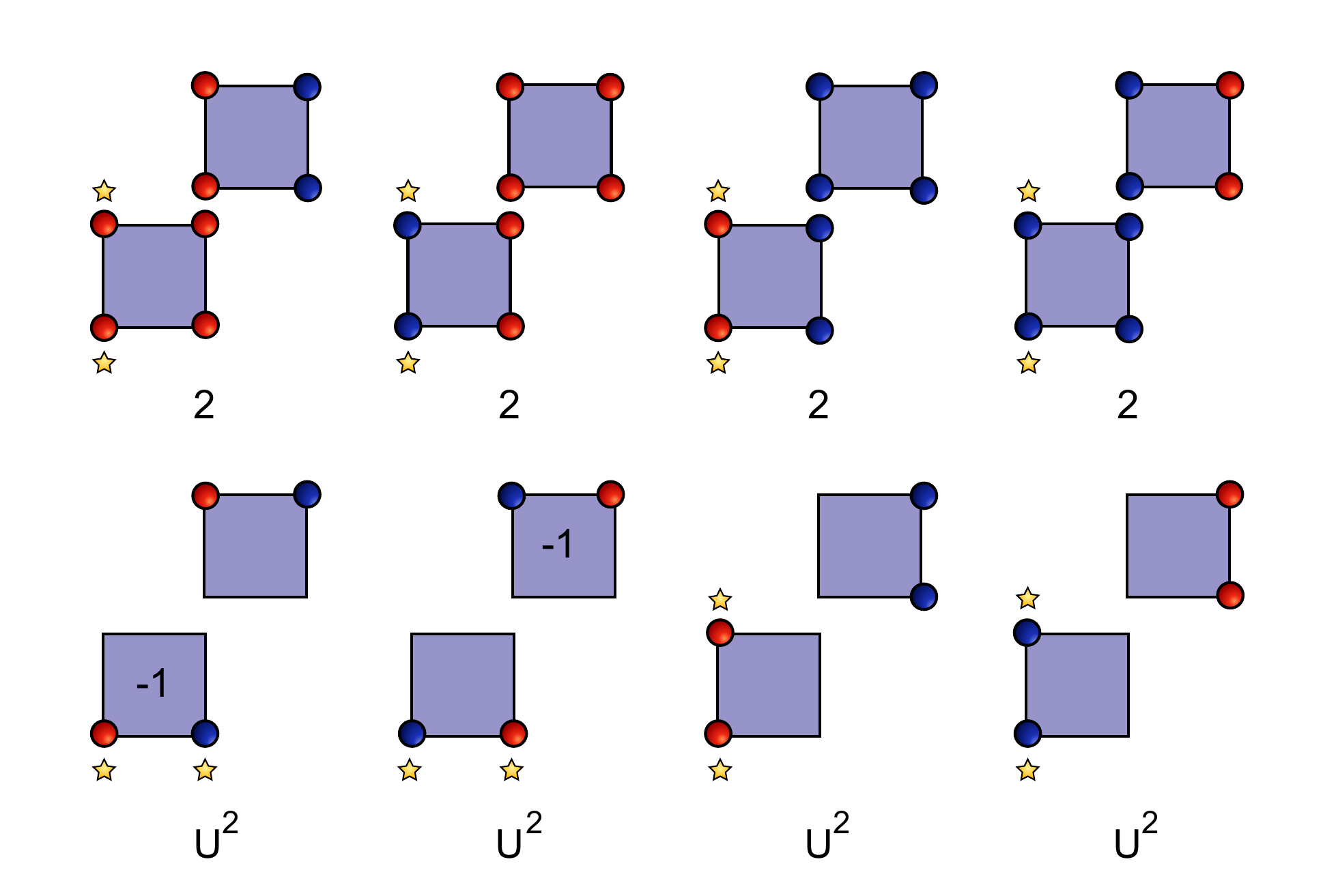}
\includegraphics[width=0.35\textwidth]{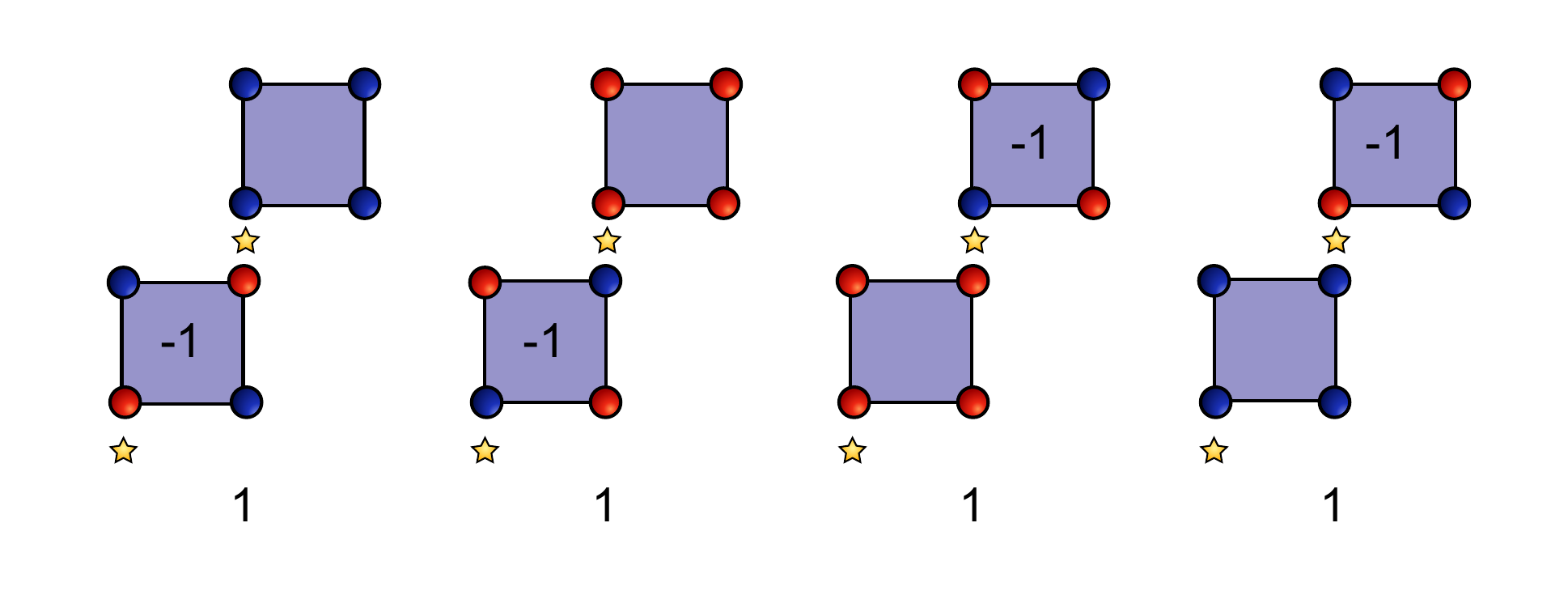}
\end{center}
\caption{\label{fig:2x2G1conf} Spin configurations with insertions of $(-1)^xS^1_{x,t}$  and $S^1_{0,0}$ (shown as yellow stars) that contribute to the calculation of $G_1(x,t)$ defined in \cref{eq:G1}. The magnitude of the Boltzmann weight is shown below each configuration. Note that the plaquette sign factors cancel the $(-1)^x$ signs that come from the sources so every configuration that contributes to $G_1(x,t)$ has a positive weight.}
\end{figure}

\begin{figure*}[ht]
\begin{center}
\includegraphics[width=0.23\textwidth]{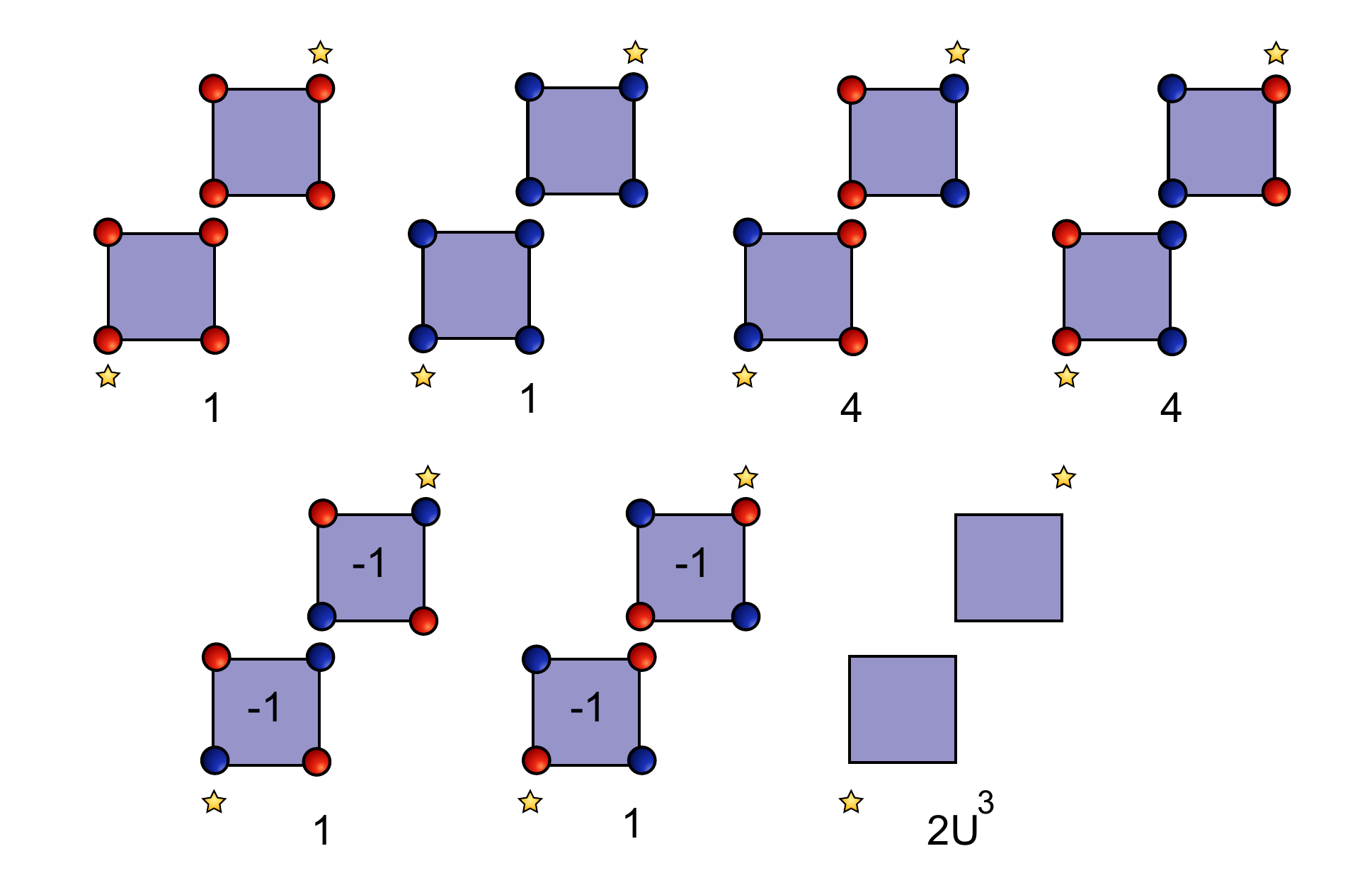}
\includegraphics[width=0.23\textwidth]{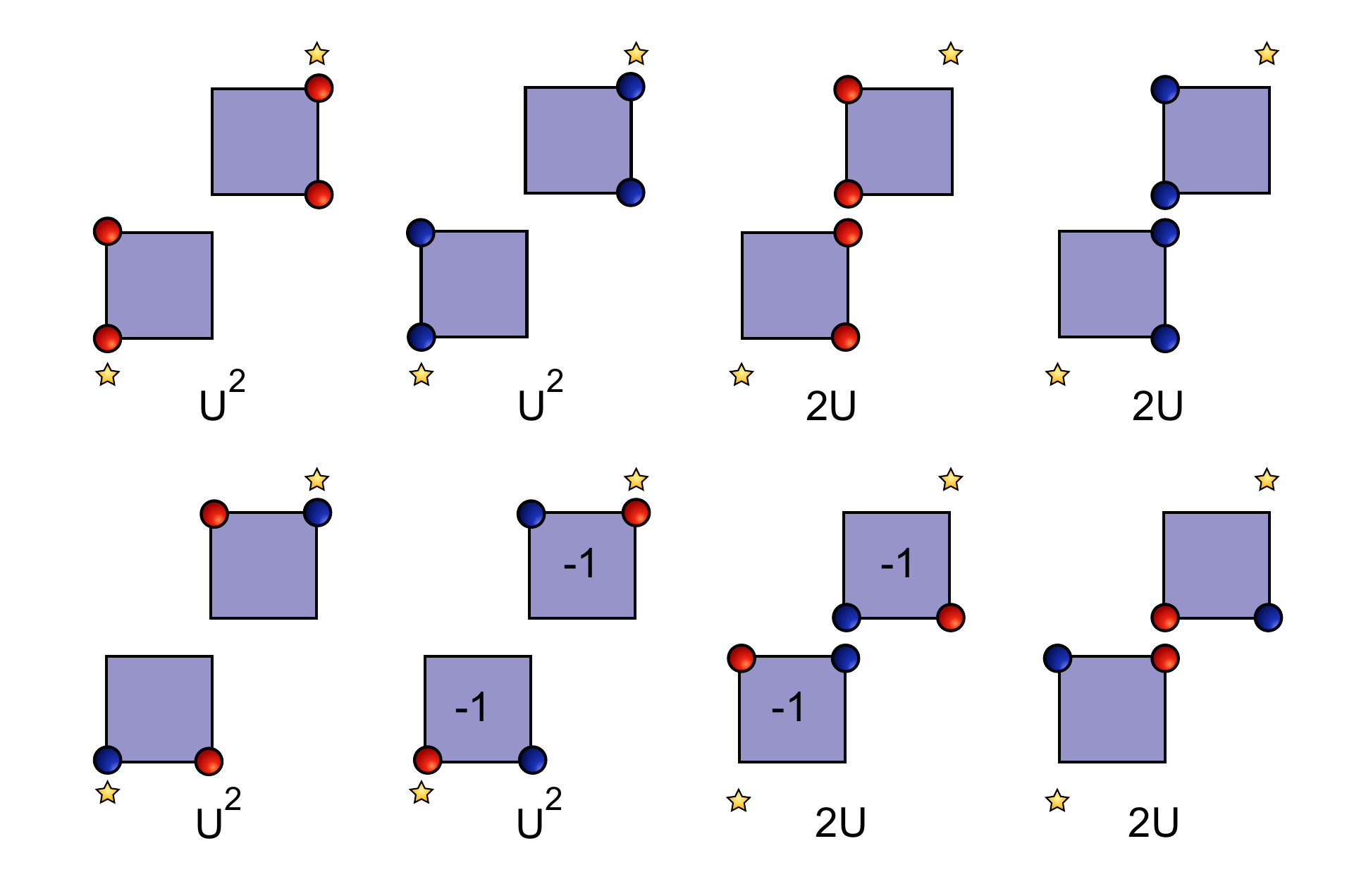}
\includegraphics[width=0.23\textwidth]{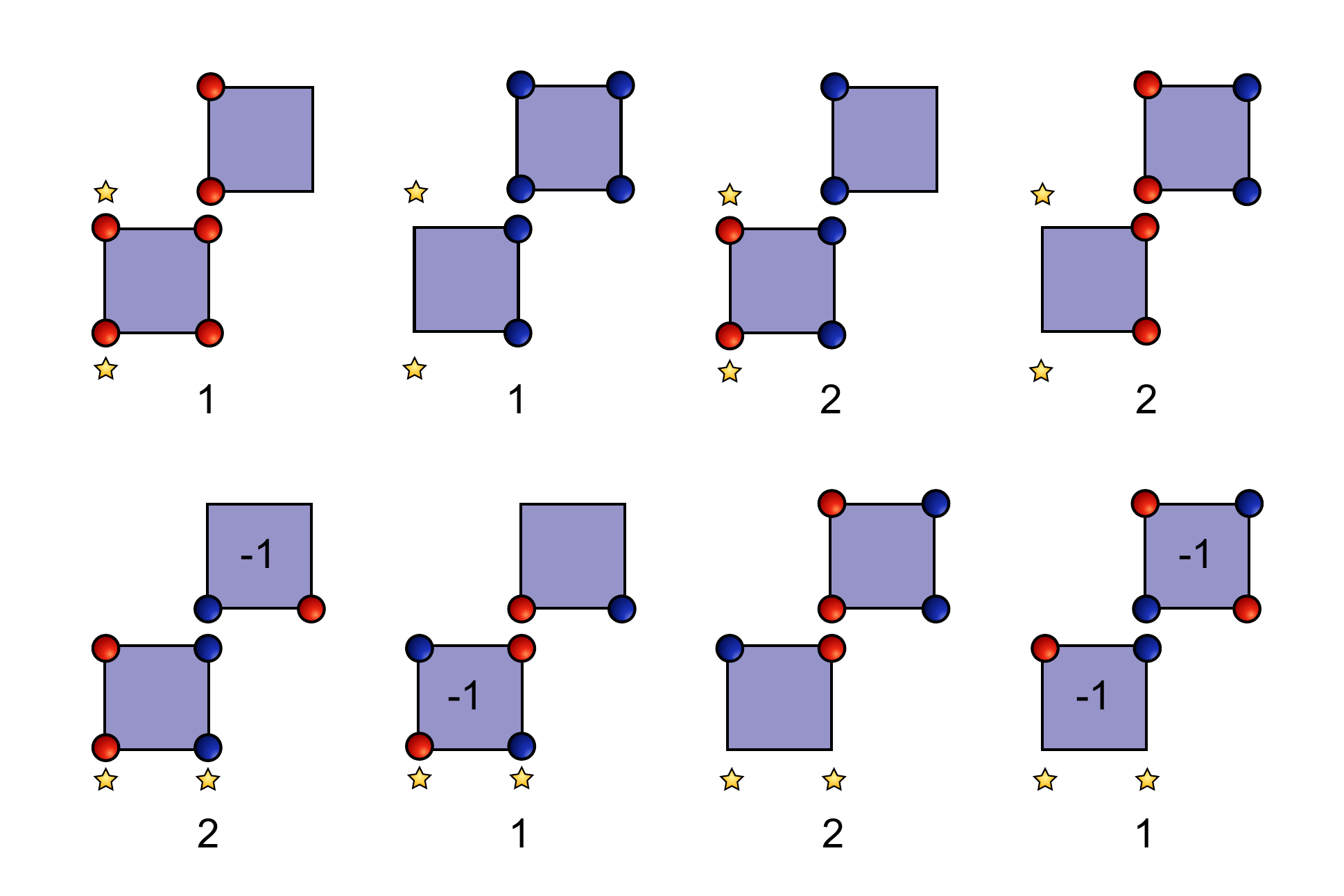}
\includegraphics[width=0.23\textwidth]{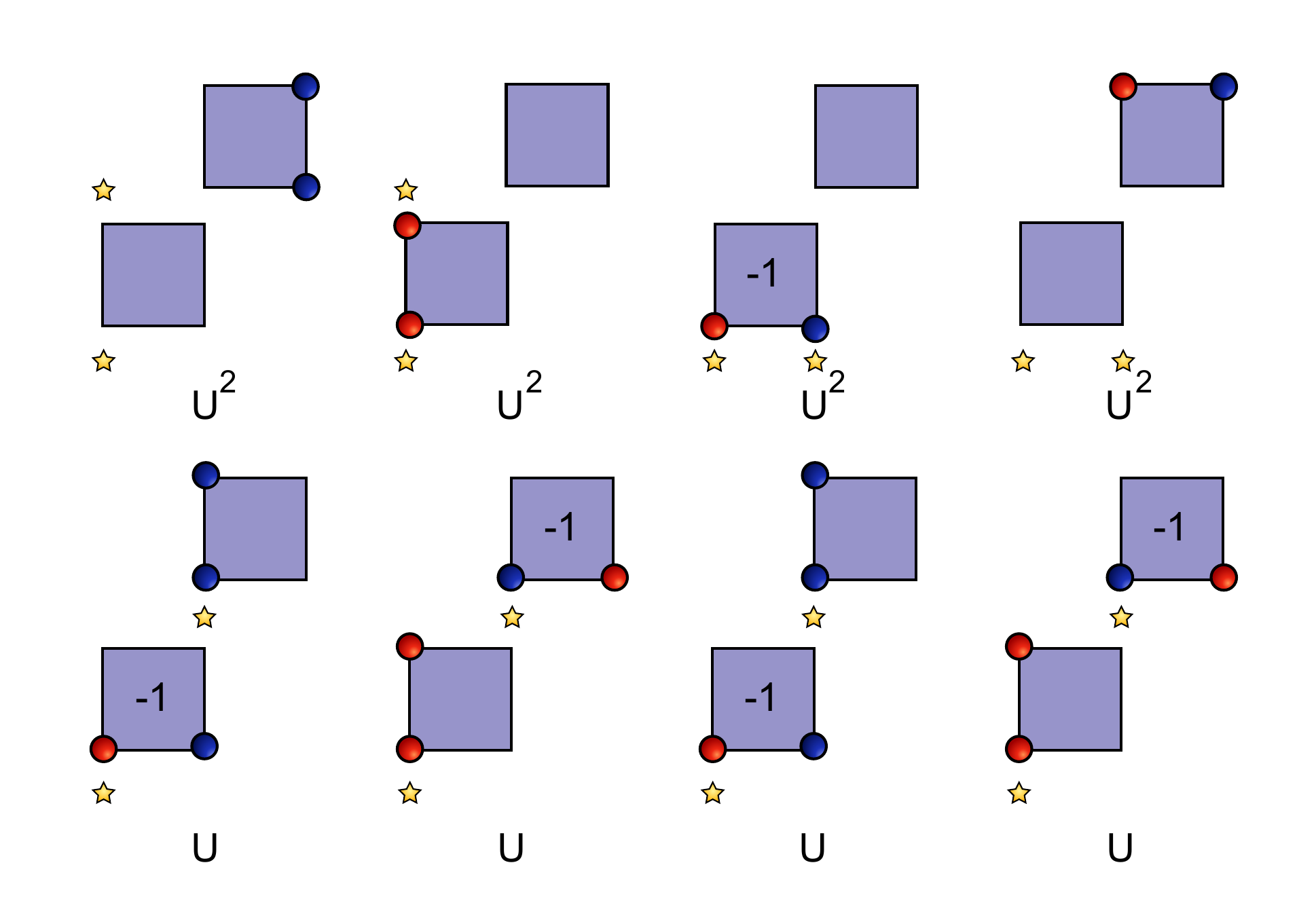}
\end{center}
\caption{\label{fig:2x2G2conf} Spin configurations with insertions of $O^\dagger(x,t)$  and $O(0,0)$ (shown as yellow stars) that contribute to $G_2(x,t)$ defined in \cref{eq:G2}. The magnitude of the Boltzmann weight is shown below each configuration. Note the extra factor of $2$ in the configuration weight if the stars are on the same empty site. This is due to the fact that on those sites we insert $\{O^\dagger,O\}$, which evaluates to $2$ on an empty site. Also note that again the plaquette sign factors cancel the $(-1)^x$ that comes through the definition of source terms so that every configuration that contributes to $G_2(x,t)$ has a positive weight.}
\end{figure*}

\begin{figure*}[ht]
\begin{center}
\includegraphics[width=0.3\textwidth]{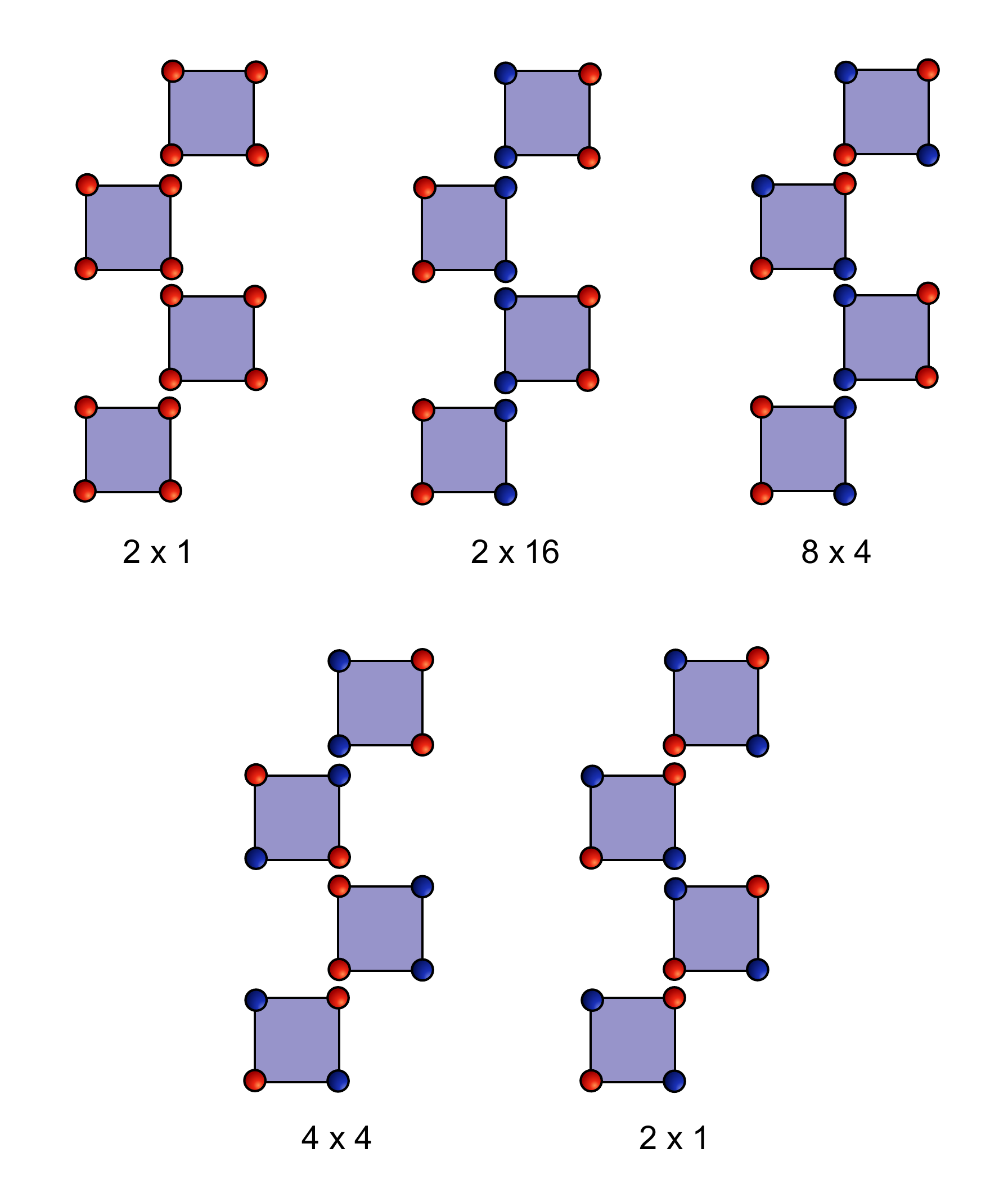}
\includegraphics[width=0.3\textwidth]{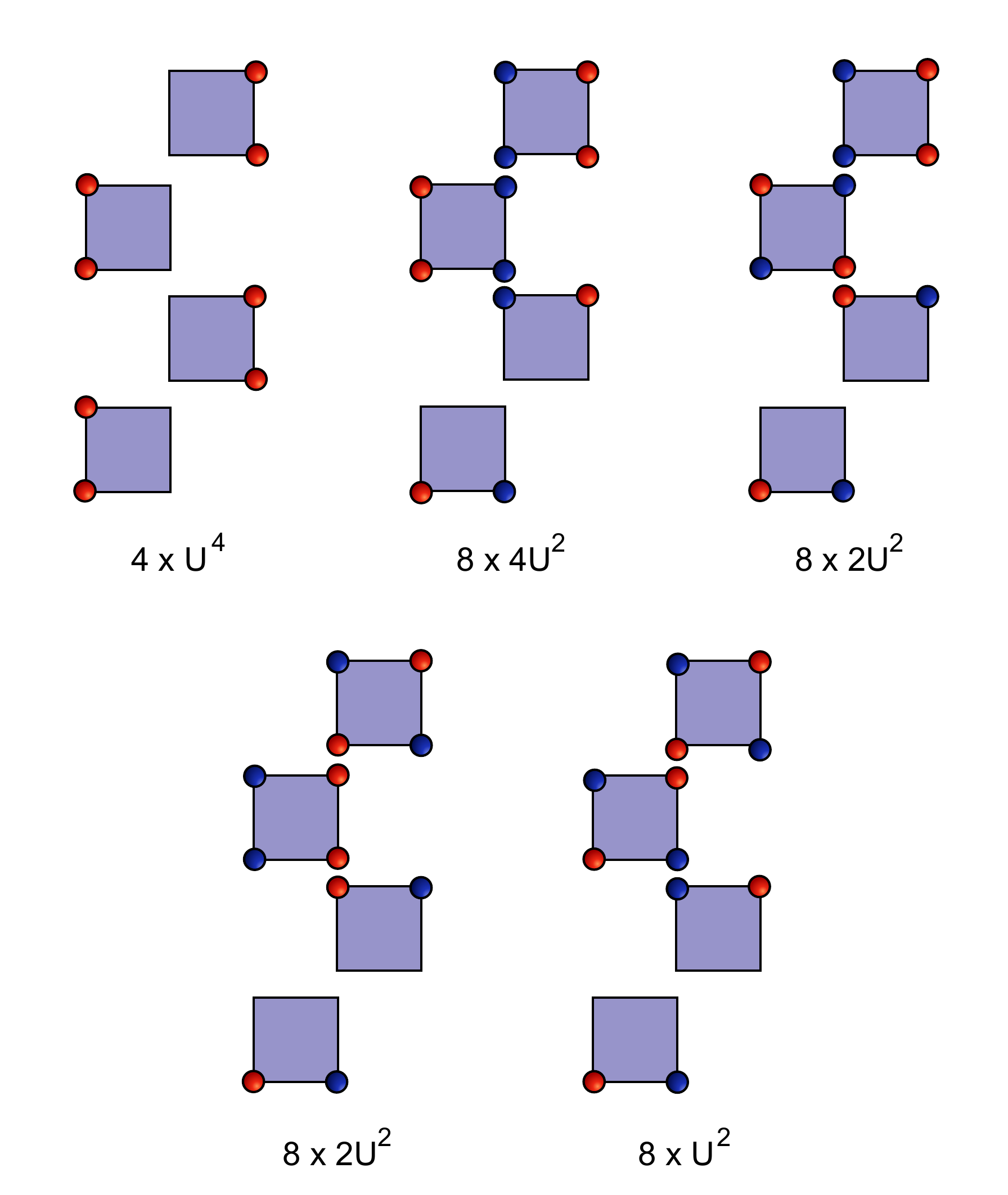}
\includegraphics[width=0.3\textwidth]{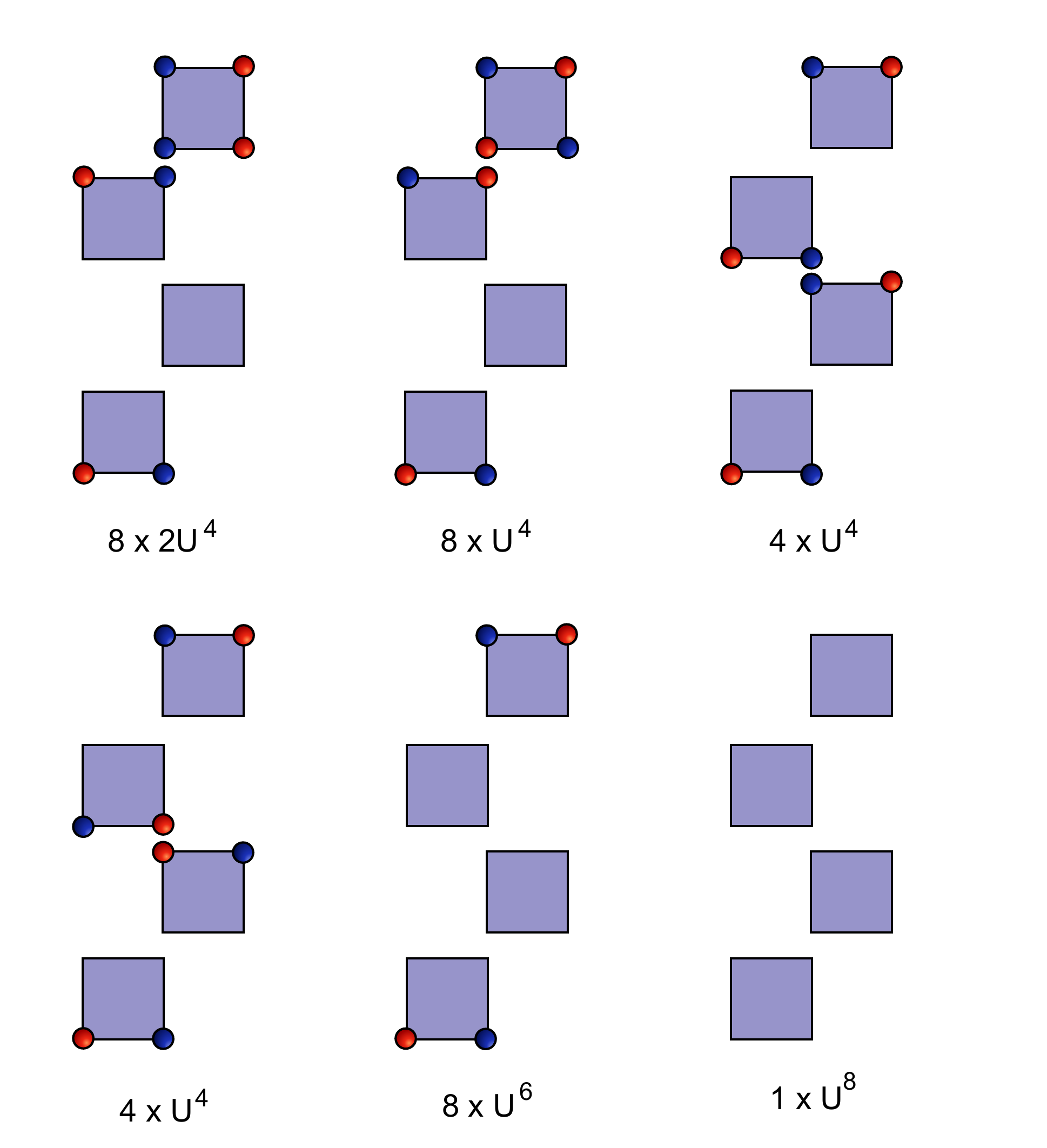}
\end{center}
\caption{\label{fig:2x4Zconf} Spin configurations on a $2\times 4$ lattice. Here we ignore the sign factors on plaquettes since they are expected to cancel in the calculations. We only show one configuration within a class of configurations obtained by translations in space and time and spin-flips. Below each configuration we label the degeneracy $g$ and the weight $W(c)$ as $g \times W(c)$.}
\end{figure*}

\begin{table*}[thb]
\centering
\setlength{\tabcolsep}{4pt}
\renewcommand{\arraystretch}{1.3}
\makegapedcells
\begin{tabular}{r|l|l}
\TopRule
\multicolumn{1}{c|}{} & 
\multicolumn{2}{c}{$U=0$}  \\
\MidRule
\multicolumn{1}{c|}{} & \multicolumn{1}{c|}{Exact} & \multicolumn{1}{c}{Monte Carlo} \\
\MidRule 
 $\rho_0$ & 0 & 0 \\
 $\rho_w$ & 0.66666... & 0.66665(7) \\
$\chi_1$ & 2.66666... & 2.6667(1) \\
 $F_1$ & 0.66666...& 0.6665(1) \\
 $\chi_2$ & 2 & 2 \\
 $F_2$ & 1 & 1.0000(1) \\
\BotRule
\end{tabular}
\begin{tabular}{l|l}
\TopRule
\multicolumn{2}{c}{$U=0.2$}  \\
\MidRule
\multicolumn{1}{c|}{Exact} & \multicolumn{1}{c}{Monte Carlo} \\
\MidRule 
0.013115.. & 0.01311(1) \\
0.662251.. & 0.66234(6) \\
2.623035.. & 2.6229(1) \\
0.662251..& 0.6625(1) \\
2.169848.. & 2.16984(1) \\
1.053109.. & 1.0531(1) \\
\BotRule
\end{tabular}
\begin{tabular}{l|l}
\TopRule
\multicolumn{2}{c}{$U=1.2$}  \\
\MidRule
\multicolumn{1}{c|}{Exact} & \multicolumn{1}{c}{Monte Carlo} \\
\MidRule 
0.306076.. & 0.30613(3) \\
0.537634.. & 0.53760(6) \\
1.700425.. & 1.7003(1) \\
0.537634.. & 0.53761(11) \\
2.085521.. & 2.08545(8) \\
1.016504.. & 1.01642(7) \\
\BotRule
\end{tabular}
\begin{tabular}{l|l}
\TopRule
\multicolumn{2}{c}{$U=2.2$}  \\
\MidRule
\multicolumn{1}{c|}{Exact} & \multicolumn{1}{c}{Monte Carlo} \\
\MidRule 
0.577048.. & 0.57705(3) \\
0.369003.. & 0.36898(5) \\
0.953799.. & 0.9538(1) \\
0.369003.. & 0.36907(8) \\
1.489178.. & 1.48913(5) \\
0.828855.. & 0.82885(6) \\
\BotRule
\end{tabular}
\renewcommand{\arraystretch}{1}
\caption{\label{tab:exact1} Comparison of Monte Carlo results with exact calculations on a $2\times 2$ lattice for various values of $U$. The exact results are obtained using the results discussed in \cref{app:exact}.}
\end{table*}

\begin{table*}[thb]
\centering
\setlength{\tabcolsep}{4pt}
\renewcommand{\arraystretch}{1.3}
\makegapedcells
\begin{tabular}{r|l|l}
\TopRule
\multicolumn{1}{c|}{} & 
\multicolumn{2}{c}{$U=0$}  \\
\MidRule
\multicolumn{1}{c|}{} & \multicolumn{1}{c|}{Exact} & \multicolumn{1}{c}{Monte Carlo} \\
\MidRule 
$\rho_0$ & 0 & 0 \\
$\rho_w$ & 0.5714285.. & 0.57147(10) \\
$\chi_1$ & 3.809523.. & 3.8099(2) \\
$F^x_1$ & 0.1904761.. & 0.19052(12) \\
$F^t_1$ & 1.523809.. & 1.52365(11) \\
$\chi_2$ & 2 & 2 \\
$F^x_2$ & 0.714285.. & 0.71414(11) \\
$F^t_2$ & 1.642857.. & 1.64293(5) \\
\BotRule
\end{tabular}
\begin{tabular}{l|l}
\TopRule
\multicolumn{2}{c}{$U=0.2$}  \\
\MidRule
\multicolumn{1}{c|}{Exact} & \multicolumn{1}{c}{Monte Carlo} \\
\MidRule 
0.008617.. & 0.00861(5) \\
0.568056.. & 0.56801(13) \\
3.757121.. & 3.7572(3) \\
0.184186.. & 0.18428(11) \\
1.518913.. & 1.51897(11) \\
2.196171.. & 2.19619(5) \\
0.733470.. & 0.73340(13) \\
1.684511.. & 1.68457(7) \\
\BotRule
\end{tabular}
\begin{tabular}{l|l}
\TopRule
\multicolumn{2}{c}{$U=1.2$}  \\
\MidRule
\multicolumn{1}{c|}{Exact} & \multicolumn{1}{c}{Monte Carlo} \\
\MidRule 
0.294166.. & 0.29416(3) \\
0.468507.. & 0.46855(7) \\
2.298096.. & 2.2980(2) \\
0.112175.. & 0.11218(8) \\
1.205777.. & 1.20589(11) \\
2.584950.. & 2.58494(13) \\
0.728386.. & 0.72842(8) \\
1.520234.. & 1.52011(9) \\
\BotRule
\end{tabular}
\begin{tabular}{l|l}
\TopRule
\multicolumn{2}{c}{$U=2.2$}  \\
\MidRule
\multicolumn{1}{c|}{Exact} & \multicolumn{1}{c}{Monte Carlo} \\
\MidRule 
0.636190.. & 0.63620(4) \\
0.298041.. & 0.29803(5) \\
0.985396.. & 0.98542(19) \\
0.074463.. & 0.07449(7) \\
0.664496.. & 0.66436(12) \\
1.813853.. & 1.81389(8) \\
0.712170.. & 0.71219(8) \\
1.113073.. & 1.11305(5) \\
\BotRule
\end{tabular}
\renewcommand{\arraystretch}{1}
\caption{\label{tab:exact2} Comparison of Monte Carlo results with exact calculations on a $2\times 4$ lattice for various values of $U$. Since we are no longer working on a square lattice, we define $\rho_w = (\rho^x_w+\rho^t_w)/2$.}
\end{table*}

The magnitude of the weight of each configuration $W(c)$ is defined as the product of plaquette weights. Each plaquette weight without empty sites is $1$ except when the plaquette contains opposite spins that don't hop. These special plaquettes have weight $2$. Plaquettes with empty sites have a weight $(\sqrt{U})^{n_e}$ where $n_e$ is the number of empty sites. Empty sites associated with source terms $O(x,t)$ or $O^\dagger(x,t)$ do not carry these factors. The weights of all the fifteen configurations that contribute to the partition function on the $2\times 2$ lattice are given in the caption of \cref{fig:2x2Zconf}. Also notice that there are extra negative signs associated with plaquettes. These signs always accompany a $\ket{\ua}$ or a $\bra{\ua}$ on an even site. 

We can compute the partition function of our model on a $2\times 2$ lattice by summing over the weights of the fifteen configurations in \cref{fig:2x2Zconf}, and we obtain
\begin{align}
Z(U) \ =\ 12 + 8 U^2 + U^4.
\label{eq:2x2pf}
\end{align}
Using \cref{eq:rho0} we can compute the average monomer density to be
\begin{align}
\rho_0 = \frac{1}{Z(U)}(4 U^2 + U^4). 
\end{align}
For each configuration the spatial winding charge $Q_w(c)$ can be obtained by picking the first spatial site and counting $\ket{\ua\da}$ and $\bra{\da\ua}$ as +1 and $\ket{\da\ua}$ and $\bra{\ua\da}$ as -1 on all forward bonds. Using this rule we note that the non-zero winding charges are $Q_w(c_6)=-Q_w(c_5)=2$ and $Q_w(c_{13})=Q_w(c_{15})=-Q_w(c_{12})=-Q_w(c_{14})=1$. Using \cref{eq:currsus} we compute the current-current susceptibility, which is then given by
\begin{align}
\rho^x_w(U) = \rho^t_w(U) \ =\ \frac{1}{Z(U)}(8 + 4 U^2).
\end{align}
In order to compute $\chi_1$, $F_1^x$ and $F_1^t$ we first enumerate all the configurations that contribute to $G_1(x,t)$ defined in \cref{eq:G1}. These configurations with insertions of $(-1)^x S^1_{x,t}$ and $S^1_{0,0}$ along with their weights are shown in \cref{fig:2x2G1conf}. Using this information we obtain
\begin{align}
\chi_1(U) &= \frac{1}{Z(U)}(32 + 8 U^2) \\
F^x_1(U) &= F^t_1\ =\ \frac{1}{Z(U)}(8 + 4 U^2).
\end{align}
Similarly we can compute $\chi_2$, $F^x_2$ and $F^t_2$ by enumerate all the configurations that contribute to $G_2(x,t)$ defined in \cref{eq:G2}.The configurations with the insertion of $O^\dagger(x,t)$ and $O(0,0)$ and their weights are shown in \cref{fig:2x2G2conf}. Using these weights we obtain
\begin{align}
\chi_2(U) &= \frac{1}{Z(U)}(24 + 12 U + 8 U^2 + 2 U^3) \\
F^x_2(U) &= \ F^t_2(U) \ = \frac{1}{Z(U)}(12 + 4 U + 4 U^2 + 2 U^3)
\end{align}
We have verified that these exact results for our observables are reproduced by our Monte Carlo algorithm. The comparison is shown in \cref{tab:exact1}.

A similar exact analysis is possible for a $2\times 4$ lattice, but the total number of configurations is much larger. Fortunately we can still use symmetries do classify and count the configurations. For example we can use translations in space and time and spin-flip as symmetry transformations. One configuration in each of these symmetry classes is shown in \cref{fig:2x4Zconf}. Below each configuration we show the degeneracy $g$ and the weight $W(c)$ labeled as $g \times W(c)$. Using these results we can easily obtain
\begin{align}
Z(U) &= 84 + 72 U^2 + 36 U^4 + 8 U^6 + U^8 \\
\rho_0(U) &= \frac{1}{Z(U)}(18 U^2 + 18 U^4 + 6 U^6 + U^8) \\
\rho^x(U) &= \frac{1}{Z(U)}(80 + 68 U^2 + 24 U^4 + 4 U^6 ) \\
\rho^t_w(U) &= \frac{1}{Z(U)}(16 + 8 U^4)
\end{align}
In order to obtain the remaining observables it is much easier to work with bond configurations instead of the spin configurations. This gives us
\begin{align}
    \chi_1(U) & = \frac{1}{Z(U)} \left( 320 + 164 U^2 + 48 U^4 + 4 U^6 \right) \\
    F_1^x (U) & = \frac{1}{Z(U)} \left( 16 + 8 U^4 \right) \\
    F_1^t(U) & = \frac{1}{Z(U)} \left( 128 + 100 U^2 + 32 U^4 + 4 U^6 \right) \\
    \chi_2(U) & = \frac{1}{Z(U)} \left(168 + 84 U + 138 U^2 + 66 U^3 \right. \nonumber \\
    & \qquad \qquad \left. +\ 48 U^4 + 18 U^5 + 6 U^6 + 2 U^7 \right) \\
    F_2^x & = \frac{1}{Z(U)} \left( 60 + 12 U + 30 U^2 + 18 U^3 \right. \nonumber \\
    & \qquad \qquad \left. +\ 12 U^4 + 10 U^5 + 2 U^6 + 2 U^7 \right) \\
    F_2^t & = \frac{1}{Z(U)} \left( 138 + 24 U + 84 U^2 + 30 U^3 \right. \nonumber \\
    & \qquad \qquad \left. +\ 28 U^4 + 10 U^5 + 4 U^6 + 2 U^7 \right)
\end{align}
These results have also been verified using an automated mathematica code. In \cref{tab:exact2} we compare the exact results with those obtained using our Monte Carlo method and we find excellent agreement.
\end{document}